\documentclass[aps,twocolumn,linenumbers,showpacs,superscriptaddress,preprintnumbers,amsmath,amssymb,floatfix]{revtex4-2}

\usepackage{dcolumn}
\usepackage{bm}
\usepackage{color}
\usepackage[dvipsnames]{xcolor} 
\usepackage{lineno}
\usepackage{subfigure} 
\usepackage{hyperref}
\usepackage{graphicx}  
 
\linenumbers

\begin{document} 
\nolinenumbers

\begin{titlepage} 

\title{Opportunities for Gas-Phase Science at Short-Wavelength Free-Electron Lasers with Undulator-Based Polarization Control}

\author{Markus Ilchen} \email{markus.ilchen@desy.de}

\affiliation{Institut f\"ur Experimentalphysik, Universit\"at Hamburg, Luruper Chaussee 149, 22761 Hamburg, Germany}
\affiliation{Deutsches Elektronen-Synchrotron DESY, Notkestr. 85, 22607 Hamburg, Germany}
\affiliation{Center for Free-Electron Laser Science CFEL, Deutsches Elektronen-Synchrotron DESY, Notkestr. 85, 22607 Hamburg, Germany}
\affiliation{European XFEL, Holzkoppel 4, 22869 Schenefeld, Germany}

\author{Enrico Allaria}
\affiliation{Elettra Sincrotrone Trieste, Strada Statale 14 km 163.5, 34149 Trieste, Italy}

\author{Primo\v{z} Rebernik Ribi\v{c}}
\affiliation{Elettra Sincrotrone Trieste, Strada Statale 14 km 163.5, 34149 Trieste, Italy}

\author{Heinz-Dieter Nuhn}
\affiliation{SLAC National Accelerator Laboratory, 2575 Sand Hill Road, Menlo Park, CA 94025, USA}

\author{Alberto Lutman}
\affiliation{SLAC National Accelerator Laboratory, 2575 Sand Hill Road, Menlo Park, CA 94025, USA}

\author{Evgeny Schneidmiller}
\affiliation{Deutsches Elektronen-Synchrotron DESY, Notkestr. 85, 22607 Hamburg, Germany}

\author{Markus Tischer}
\affiliation{Deutsches Elektronen-Synchrotron DESY, Notkestr. 85, 22607 Hamburg, Germany}

\author{Mikail Yurkov}
\affiliation{Deutsches Elektronen-Synchrotron DESY, Notkestr. 85, 22607 Hamburg, Germany}

\author{Marco Calvi}
\affiliation{Paul Scherrer Institut, Forschungsstrasse 111, 5232 Villigen PSI, Switzerland}

\author{Eduard Prat}
\affiliation{Paul Scherrer Institut, Forschungsstrasse 111, 5232 Villigen PSI, Switzerland}

\author{Sven Reiche}
\affiliation{Paul Scherrer Institut, Forschungsstrasse 111, 5232 Villigen PSI, Switzerland}

\author{Thomas Schmidt}
\affiliation{Paul Scherrer Institut, Forschungsstrasse 111, 5232 Villigen PSI, Switzerland}

\author{Gianluca Aldo Geloni}
\affiliation{European XFEL, Holzkoppel 4, 22869 Schenefeld, Germany}

\author{Suren Karabekyan}
\affiliation{European XFEL, Holzkoppel 4, 22869 Schenefeld, Germany}

\author{Jiawei Yan}
\affiliation{European XFEL, Holzkoppel 4, 22869 Schenefeld, Germany}

\author{Svitozar Serkez}
\affiliation{European XFEL, Holzkoppel 4, 22869 Schenefeld, Germany}

\author{Zhangfeng Gao}
\affiliation{Shanghai Advanced Research Institute, Chinese Academy of Sciences, China}

\author{Bangjie Deng}
\affiliation{School of Nuclear Science and Technology, Xi’an Jiaotong University, China}
\affiliation{Shanghai Institute of Applied Physics, Chinese Academy of Sciences, China}

\author{Chao Feng}
\affiliation{Shanghai Advanced Research Institute, Chinese Academy of Sciences, China}
\affiliation{Shanghai Institute of Applied Physics, Chinese Academy of Sciences, China}

\author{Haixiao Deng}
\affiliation{Shanghai Advanced Research Institute, Chinese Academy of Sciences, China}
\affiliation{Shanghai Institute of Applied Physics, Chinese Academy of Sciences, China}

\author{Wolfram Helml}
\affiliation{Technical University of Dortmund, Maria-Goeppert-Mayer-Straße 2, 44227 Dortmund, Germany}

\author{Lars Funke}
\affiliation{Technical University of Dortmund, Maria-Goeppert-Mayer-Straße 2, 44227 Dortmund, Germany}

\author{Mats Larsson}
\affiliation{Stockholm University, AlbaNova University Center, 114 21 Stockholm, Sweden}

\author{Vitali Zhaunerchyk}
\affiliation{University of Gothenburg, 405 30 Gothenburg, Sweden}

\author{Michael Meyer}
\affiliation{European XFEL, Holzkoppel 4, 22869 Schenefeld, Germany}

\author{Tommaso Mazza}
\affiliation{European XFEL, Holzkoppel 4, 22869 Schenefeld, Germany}

\author{Till Jahnke}
\affiliation{European XFEL, Holzkoppel 4, 22869 Schenefeld, Germany}
\affiliation{Max-Planck-Institut f\"{u}r Kernphysik, 69117 Heidelberg, Germany.}

\author{Reinhard Dörner}
\affiliation{Institut f\"ur Kernphysik, Goethe-Universit\"at, Max-von-Laue-Stra\ss e 1, 60438 Frankfurt am Main, Germany}

\author{Francesca Calegari}
\affiliation{Institut f\"ur Experimentalphysik, Universit\"at Hamburg, Luruper Chaussee 149, 22761 Hamburg, Germany}
\affiliation{Center for Free-Electron Laser Science CFEL, Deutsches Elektronen-Synchrotron DESY, Notkestr. 85, 22607 Hamburg, Germany}

\author{Olga Smirnova}
\affiliation{Max-Born-Institut, Max-Born-Straße 2A, 12489, Germany}
\affiliation{Technische Universit\"at Berlin, Str. des 17. Juni 135, 10623 Berlin, Germany}
\affiliation{Solid State Institute and Physics Department, Technion – Israel Institute of Technology, Haifa 3200003, Israel}

\author{Caterina Vozzi}
\affiliation{Istituto di Fotonica e Nanotecnologie - CNR, Piazza Leonardo da Vinci 32,
20133 Milano, Italy}

\author{Giovanni De Ninno}
\affiliation{Elettra Sincrotrone Trieste, Strada Statale 14 km 163.5, 34149 Trieste, Italy}
\affiliation{Laboratory of Quantum Optics, University of Nova Gorica, 5001 Nova Gorica, Slovenia}

\author{Jonas W\"atzel}
\affiliation{Institute of Physics, Martin-Luther University Halle-Wittenberg, D-06099 Halle, Germany}

\author{Jamal Berakdar}
\affiliation{Institute of Physics, Martin-Luther University Halle-Wittenberg, D-06099 Halle, Germany}

\author{Sadia Bari}
\affiliation{Deutsches Elektronen-Synchrotron DESY, Notkestr. 85, 22607 Hamburg, Germany}
\affiliation{Zernike Institute for Advanced Materials, University of Groningen, Nijenborgh 4, 9747 AG Groningen, Netherlands}

\author{Lucas Schwob}
\affiliation{Deutsches Elektronen-Synchrotron DESY, Notkestr. 85, 22607 Hamburg, Germany}

\author{J\'er\'emy R. Rouxel}
\affiliation{Chemical Sciences and Engineering Division, Argonne National Laboratory, Lemont, Illinois 60439, United States}

\author{Shaul Mukamel}
\affiliation{Department of Chemistry and Department of Physics and Astronomy, University of California, Irvine, Irvine, California 92697-2025, United States}

\author{Klaus Bartschat}
\affiliation{Department of Physics and Astronomy, Drake University, Des Moines, Iowa, 50311}

\author{Kathryn Hamilton}
\affiliation{Department of Physics, University of Colorado Denver, Denver, Colorado, 80204}

\author{Luca Argenti}
\affiliation{Department of Physics \& CREOL, University of Central Florida, Orlando, Florida, 32816}

\author{Nicolas Douguet}
\affiliation{Department of Physics \& CREOL, University of Central Florida, Orlando, Florida, 32816}

\author{Nikolay M. Novikovskiy}
\affiliation{Institut f\"ur Physik und CINSaT, Universit\"at Kassel, Heinrich-Plett-Stra\ss e 40, 34132 Kassel, Germany}

\author{Philipp V. Demekhin}
\affiliation{Institut f\"ur Physik und CINSaT, Universit\"at Kassel, Heinrich-Plett-Stra\ss e 40, 34132 Kassel, Germany}

\author{Peter Walter} \email{pwalter@slac.stanford.edu}
\affiliation{SLAC National Accelerator Laboratory, 2575 Sand Hill Road, Menlo Park, CA 94025, USA} 
\affiliation{TAU Systems, 201 W 5th Street, Austin, TX 78701, USA}


\begin{abstract}
\begin{center}

\textbf{\large{Abstract}}\\
\end{center}
Free-electron lasers (FELs) are the world’s most brilliant light sources with rapidly evolving technological capabilities in terms of ultrabright and ultrashort pulses over a large range of photon energies. Their revolutionary and innovative developments have opened new fields of science regarding nonlinear light-matter interaction, the investigation of ultrafast processes from specific observer sites, and approaches to imaging matter with atomic resolution. A core aspect of FEL science is the study of isolated and prototypical systems in the gas phase with the possibility of addressing well-defined electronic transitions or particular atomic sites in molecules. Notably for polarization-controlled short-wavelength FELs, the gas phase offers new avenues for investigations of nonlinear and ultrafast phenomena in spin orientated systems, for decoding the function of the chiral building blocks of life as well as steering reactions and particle emission dynamics in otherwise inaccessible ways. This roadmap comprises descriptions of technological capabilities of facilities worldwide, innovative diagnostics and instrumentation, as well as recent scientific highlights, novel methodology and mathematical modeling. The experimental and theoretical landscape of using polarization controllable FELs for dichroic light-matter interaction in the gas phase will be discussed and comprehensively outlined to stimulate and strengthen global collaborative efforts of all disciplines.

\vspace{60pt}
\end{abstract}

\maketitle

\end{titlepage}

\tableofcontents

\onecolumngrid \newpage \twocolumngrid

\section{Introduction}

Free-electron lasers (FELs) with their ultrashort and ultraintense light pulses are covering a broad photon energy range from the Vacuum Ultraviolet (VUV) to the X-ray regime. They have been offering and rapidly developing their unique technical capabilities with large applicability for practically all disciplines of science. 

Demanding goals for studying individual atomic and molecular structures, complex systems, as well as ultrafast and nonlinear light-matter interaction have been identified and successfully pursued, with a variety of groundbreaking achievements in the hard- and soft X-ray regime. As one of the core promises of FELs, their unique pulses allow for taking snapshots of non-crystallizable molecules and clusters via the technique of coherent diffraction imaging. This method allows to take such snapshots of a system before substantial deformation can destroy the contrast of the picture. In synergy to this application, FELs have been enabling access to selectively ionizing electrons up to binding energies of currently about 25 keV. They can hence individually target a large variety of elements at their core-shell electrons. Already in the soft X-ray regime, key constituents of life’s building blocks, such as hydrogen, carbon, nitrogen, oxygen, phosphorous, and sulfur can be fully covered. With the uniquely short and bright FEL pulses, elements in complex systems can thus be interrogated in detail for their view on ultrafast phenomena \cite{bostedt2013ultra, erk2014imaging, young2018roadmap, rolles2023time}, meanwhile with a temporal resolution down to currently hundreds of attoseconds \cite{Duris2020, hartmann2018attosecond, Li2022}. In particular due to their high brilliance, FELs have furthermore been shown to be able to efficiently create and investigate ionic matter up to very high charge states\cite{sorokin2007photoelectric, doumy2011nonlinear, Young2010, Rudek2012, Rudenko2017, boll2022x}, transient core excited states \cite{mazza2020mapping}, and phenomena such as stimulated X-ray Raman scattering \cite{eichmann2020photon, rohringer2019x}, therefore enabling new perspectives for nonlinear X-ray physics and its applications.

\begin{figure}[b]
\includegraphics[width=1\linewidth]{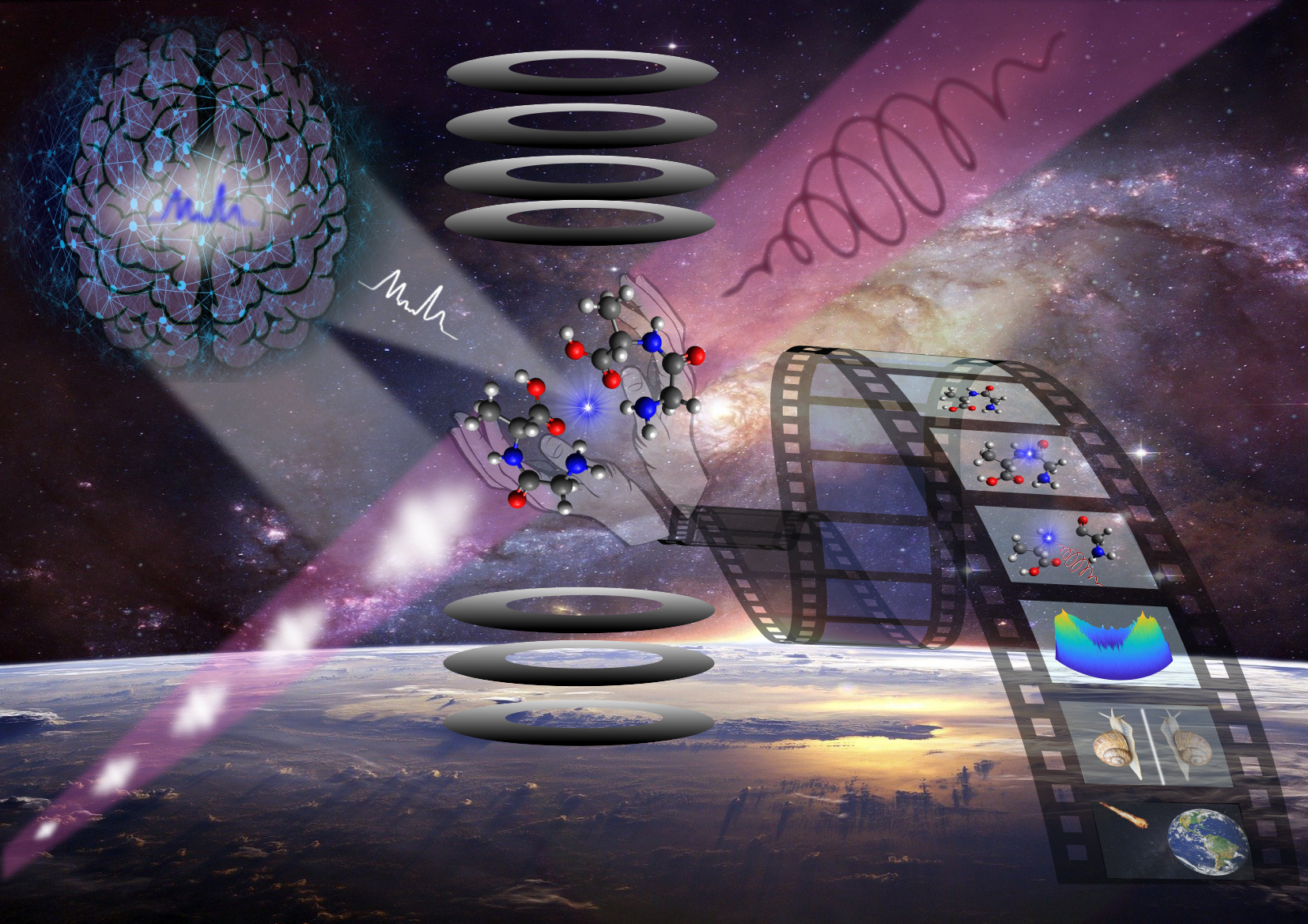}
\caption{Illustration of chirality science with a SASE free-electron
laser. AI-based analysis reveals the time–energy and polarization structure of every ultrashort pulse. The acquried in-depth knowledge allows for site-specific observations in chiral systems and molecules (here: Glycyl-l-alanine).}
\label{fig:GA}
\end{figure}

For a variety of complementary light sources, such as optical lasers, HHG sources, and synchrotrons, it has been demonstrated that circularly polarized photons add a unique tool for studying states and characteristics of matter that are otherwise inaccessible. A prominent example are magnetization studies due to the spin-sensitivity of circularly polarized light \cite{schutz1987absorption, Higley2016}. Here, the different response of a magnetic target to opposing light helicities, i.e. its dichroic behavior, provides otherwise elusive insights about the system. Dichroism studies in general are one particularly powerful method for studying chiral systems and molecular structures. They form all known life and have profound relevance for understanding the very fabric of our universe (see illustration in Fig. \ref{fig:GA}).

In contrast to the condensed phase, gas-phase studies offer access to isolated systems without their environmental interaction and can thus reduce the complexity of a variety of processes to a level where theoretical calculations become feasible and more robust. In fact, as a first step to get access to the complex characteristics of FEL pulses, the gas phase allows for their non-invasive metrology, thus enabling subsequent experiments to benefit from the precise knowledge about each individual pulse. Due to the rapid technological evolution of FELs, metrology of ultrashort pulses in the few femtosecond to attosecond regime, with unprecedented brilliance and versatility in terms of single- and multi-color pump--probe and pulse-shaping schemes can be more robustly harnessed in the gas phase.

Pioneering FEL projects such as FERMI in Italy, LCLS (II) in the USA, European XFEL and FLASH(1 and 2) in Germany, SwissFEL in Switzerland, and SHINE and SXFEL in China have recognized the high importance of undulator-based ultrashort, ultrabright, and fully polarization-controllable light pulses. They have initiated or even finalized the effort to providing the technological basis for a new chapter of polarization-dependent FEL-science. These novel approaches with polarization-controlled FEL pulses will originate new scientific endeavors in e.g. physics, chemistry, and structural biology with broad applicability in the gas phase.

In the following, we will sketch the technological landscape of free-electron laser facilities worldwide and their capabilities for undulator-based polarization control. Based on this technological background, we will give a comprehensive overview over photon-based diagnostic strategies and instrumentation opportunities as well as current and future perspectives for experimenting with undulator-based polarization-controlled FEL pulses. Subsequently, we will sketch the scientific milestones and visions of gas-phase science with such pulses, and the imminent and long-term prospects of the emerging technological era from the perspective of experiment and theory.

\onecolumngrid \newpage \twocolumngrid

\section[Machine Operation - Facilities Worldwide]{Machine Operation - \linebreak Facilities Worldwide}
\label{sec:MachinePart}

\subsection{FERMI@Elettra, Italy}
\label{sec:FERMI}

\begin{center}
Enrico Allaria, Primo\v{z} Rebernik Ribi\v{c}
\end{center}

\subsubsection{Layout of the FERMI facility}
\begin{figure*}
\includegraphics[width=0.9\linewidth]{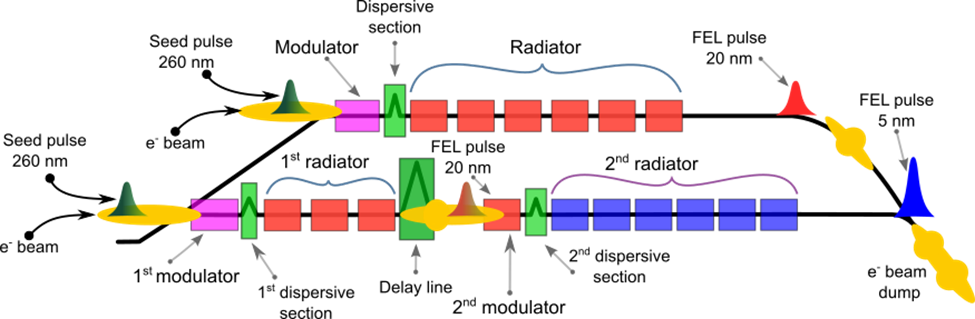}
\caption{Layout of the FERMI FELs. FEL-1 (top) uses a single HGHG stage with one modulator and 6 radiators. FEL-2 (bottom) relies on two stages to reach shorter wavelengths.}
\label{fig:FERMI1}
\end{figure*}

The electron beam produced by the FERMI linear accelerator can be used to emit light in two FEL lines: FEL-1 and FEL-2; see Fig. \ref{fig:FERMI1}. Both use APPLE-2 undulators \cite{Sasaki1994, Kokole2010, Soregaroli2011}, allowing to tune the wavelength and the polarization of the emitted radiation. \\

With an electron-beam energy ranging from 0.9 GeV to 1.5 GeV, the two FEL lines cover the full spectral range from 100 nm down to 4 nm \cite{Allaria2015}. 
FEL-1 \cite{Allaria2012} is designed to emit at long wavelengths ($\sim$100-20 nm) and relies on a single-stage high-gain harmonic-generation scheme (HGHG) \cite{Yu1991}. One planar undulator is used to facilitate the interaction between the electron beam and an external coherent laser, whose wavelength is set according to the required FEL output. FEL radiation with tunable polarization is produced in the long radiator composed of 6 APPLE-2 undulators (see Table \ref{tableFERMI}). \\

\begin{table}[b]
\begin{center}
\centering
\caption{Parameters of the FERMI radiators}
\label{tableFERMI}
\begin{tabular}{lcc}
\hline
FEL parameter                    & FEL-1                  & FEL-2                   \\ \hline
Period (mm)                           & 55           & 35             \\
Number of periods            & 42         & 66        \\
Undulator length (m)                 & 2.31         & 2.31    \\
Number of undulators            & 6        & 6         \\
Minimum gap (mm)                    & 10.5                     & 10.5                     \\
Maximum aw0 (LH)           & 5.3       & 2.4         \\
Maximum aw0 (LV)           & 4.3         & 1.7         \\
Maximum aw0 (CR, CL)            & 5.0          & 2.0        \\
\hline
\end{tabular}
\end{center}
\end{table}

The FEL-2 layout was optimized to allow generation of highly coherent pulses with wavelengths down to 4 nm \cite{Allaria2013}. The nominal operation mode is based on a two stage HGHG scheme. Running FEL-2 using a more efficient seeding scheme called echo-enabled harmonic generation (EEHG) was demonstrated with minimal changes to the layout \cite{Rebernik2019}. Operating FEL-2 in the self-amplified spontaneous emission (SASE) mode, characterized by a large bandwidth and partial coherence, is also possible and done when requested by users \cite{Penco2015}.  
Two undulator sets are available at FEL-2. During normal operations, the first set (with the same parameters as the FEL-1 radiator) generates an FEL pulse with an intermediate wavelength (60-20 nm), which then seeds the second stage. The second stage radiator uses undulators with a smaller period (see Table \ref{tableFERMI}) in order to reach shorter FEL wavelengths with the same electron beam energy. Both stages have the possibility of controlling the polarization state of the emitted radiation.
The FERMI FEL takes full advantage of external seeding that allows producing powerful (GW level), short ($<100$ fs) pulses with a narrow relative bandwidth ($\sim10^{-4}$) and a high degree of coherence (more details on photon properties are available in Ref. \cite{Allaria2015} and at \href{https://www.elettra.eu/lightsources/fermi.html}{FERMI}). 
The two FELs cannot be operated simultaneously and time is shared between users that require long or short wavelengths.   

\subsubsection{Polarization control}
The use of APPLE-2 undulators for all FERMI radiators allows full control of the output polarization of the final FEL pulse \cite{Allaria2014}. For experiments not requiring a specific polarization state, circular polarization is usually chosen in order to take advantage of the stronger FEL amplification \cite{Bonifacio1990}, allowing typically to increase the energy per pulse by a factor of 2 or more with respect to linear polarization. 

Both FEL-1 and FEL-2 lines are mostly operated in linear horizontal (LH) or circular clockwise (C+) polarizations (about 45$\%$ of uptime for each configuration). Circular anticlockwise (C-) and linear vertical (LV) polarizations are chosen only when explicitly required by the user (higher transmission and/or dichroic experiments) and account for about 5$\%$ (each) of the total uptime. 
As a result of different magnet configurations, undulators have a different maximum (normalized) undulator parameter aw0 for each polarization (see Table \ref{tableFERMI}), with a lower value for LV polarization. Adjustments of the electron beam energy are necessary if a broad tuning range is required, especially when the FEL is operated in vertical polarization.\\
At $\sim$GeV electron beam energies used at FERMI, changes of the undulator gap (to adjust the wavelength) and phase (to set the polarization) can have a significant effect on the electron beam focusing. These effects need to be properly compensated for by adjusting the strength of the focusing/defocusing quadrupoles placed between undulators. Both circular polarizations have the same undulator focusing, allowing very fast switching between the two orthogonal polarization states. Typical circular dichroic experiments are performed by switching the polarization every 5 – 10 minutes and the change of the polarization state takes only a few tens of seconds. Due to the required changes of the electron-beam optics, the switching from LH to LV or from LH to circular polarization takes a few minutes.

\subsubsection{Advanced polarization schemes}
\paragraph{Crossed-polarized undulator scheme}\

The possibility to set the polarization of each undulator separately makes FERMI an ideal layout for advanced polarization schemes. One interesting method relies on the superposition of two orthogonally polarized fields with a controlled relative phase \cite{Kim1984, Kim2000}.  The scheme, originally proposed for synchrotrons, benefits from the higher degree of coherence of FELs and has been demonstrated both at FEL-1 and FEL-2 \cite{Ferrari2015, Ferrari2019}.
The possibility to control the polarization state by adjusting only a phase shifter (small magnetic chicane) can be exploited in measurements with continuous or high frequency polarization switching; with the current setup, adjustments of the phase shifter can be done within seconds and with minimal impact on the electron beam dynamic. The cross-polarized undulator scheme enables, e.g.,  a continuous rotation of the direction of linear polarization  \cite{Ferrari2015} and rapid switching between LH and LV states. A much higher frequency of the polarization switching would be possible using an electromagnetic phase shifter that could potentially change the polarization of each FEL shot (50 Hz). \\

\paragraph{Femtosecond polarization shaping of FEL pulses}\label{sec:fermifemtopol}\

Recently, a method to produce FEL pulses with time-dependent polarization has been proposed \cite{Sudar2020, Morgan2021}. During a recent experiment at FERMI, it was demonstrated that such pulses can indeed be generated using an FEL (manuscript under review). At the FEL-1 as well as the FEL-2 line, we produced two sub-pulses with orthogonal polarizations (i.e., C+ and C-) and tunable time delay, which was controlled with either a magnetic chicane or a few undulators tuned to a long wavelength (and not to a harmonic of the seed laser). Such a configuration generates an FEL pulse whose polarization varies from C+ to linear (whose direction can be controlled by varying the phase between the sub-pulses as in the cross-polarized undulator scheme; see above) and then to C- on a timescale of tens of fs. Our preliminary results suggest that it might be possible to extend the applications of such polarization-shaped pulses, originally developed for the visible spectral region (see, e.g., \cite{Kerbstadt2019}), into the extreme-ultraviolet regime.\\

\paragraph{Low harmonic emission}\

FELs with planar or APPLE-2 undulators tuned to linear polarization are known to produce non-negligible on-axis radiation at odd harmonics, with a power typically on the order of a few 10$^{-3}$ of the fundamental emission. This harmonic emission can be used to extend the tuning range of FERMI \cite{Penco2022} but may also be unwanted in some experiments. A low on-axis content of harmonic emission is easily achieved when the undulators are tuned to circular polarization due to the off-axis character of harmonics \cite{Allaria2008}. The use of circularly polarized undulators in the cross-polarized configuration was shown to be the optimal solution for having a low on-axis harmonic content while producing linearly polarized FEL light.\\

\paragraph{Different pointing}\

\begin{figure}
\includegraphics[width=0.9\linewidth]{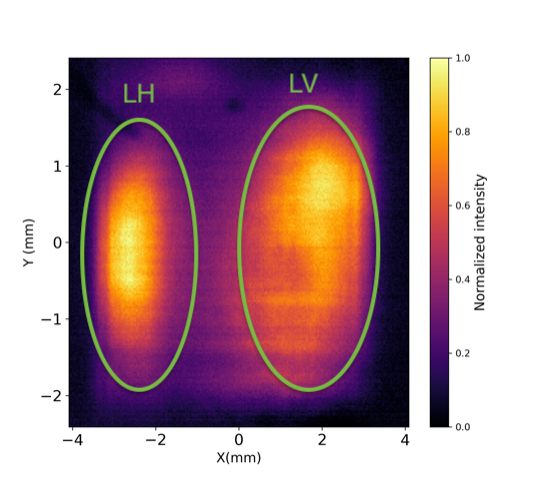}
\caption{Spatial profile of the two pulses produced by the cross-polarized undulator scheme with a kick in the electron-beam trajectory. Data refer to FERMI FEL-1 operated at 20.8 nm (manuscript in preparation).}
\label{fig:FERMI4}
\end{figure}

The cross-polarized undulator scheme in combination with a controlled tilt of the electron-beam trajectory through the radiator \cite{Nuhn2015} is used to produce two orthogonally polarized FEL pulses with different pointing. The two FEL pulses, which are spatially completely separated in the downstream optics (Fig. \ref{fig:FERMI4}), can be used for FEL pump – FEL probe experiments with orthogonal polarizations.\\

\paragraph{Orbital angular momentum}\

\begin{figure}
\centering
\subfigure{\includegraphics[width=0.45\textwidth]{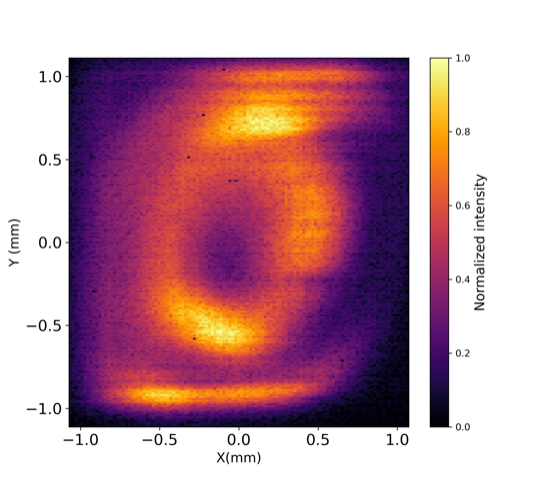}}
\hfill
\subfigure{\includegraphics[width=0.45\textwidth]{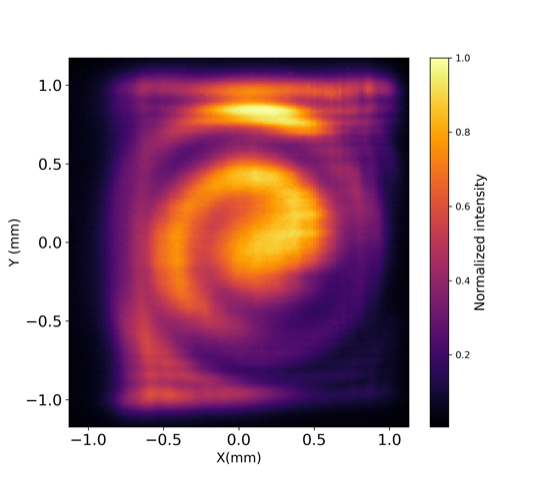}}
\hfill
\caption{FEL intensity distribution corresponding to an OAM optical mode from second harmonic emission (top) and the characteristic spiral interference pattern between the second harmonic OAM mode and the fundamental on-axis emission (bottom).}
\label{fig:FERMI5}
\end{figure}

Besides having full control over the polarization state or equivalently the spin angular momentum of emitted photons, FERMI also allows generating light that carries orbital angular momentum (OAM) \cite{Allen1992}. These so-called optical vortex beams are characterized by a doughnut-shaped intensity profile and a helical phase dependence; see Fig. \ref{fig:FERMI5}. Two approaches are used to generate OAM at FERMI \cite{Ribic2017}: i) vortex beams with a topological charge of $l= 1/-1$ are naturally generated at the second harmonic when the FERMI radiator is set to CR/CL polarization (Fig. \ref{fig:FERMI5}) and ii) high intensity vortex beams with tunable (and independent of polarization) topological charge can be generated at the fundamental wavelength by inserting a spiral zone plate into the beam downstream of the radiator. The latter approach is now routinely used in experiments at FERMI; see chapter \ref{sec:OAM} for details.

\clearpage
\onecolumngrid \newpage \twocolumngrid

\subsection{LCLS@SLAC, USA}
\label{sec:LCLS}

\begin{center}
Heinz-Dieter Nuhn and Alberto Lutman
\end{center}

The LCLS facility has been in operation to produce X-ray free-electron pulses from 2009 \cite{nphoton.2010.176AA} producing X-rays for scientific applications between $270$~eV and $12.8$~keV. The original layout included a Normal Conducting (NC) linac, accelerating electron bunches up to 17~GeV, with a repetition rate up to 120~Hz. Horizontally polarized X-rays were produced in an undulator line equipped with fixed gap undulators. An adjustment of K for each undulator segment was allowed, within a small range, by the canted pole design.  
Operating the NC Linac in nominal conditions, a 250~pC electron bunch is extracted at the cathode, in a first bunch compressor, the bunch charge is reduced to 180~pC~\cite{PhysRevAccelBeams.19.100703} and the bunch is compressed to a current of about $\approx$220~A. A second bunch compressor, operating either at 3 or at 5~GeV compresses the bunch to the final current of few kA, with a typical bunch duration of 35~fs.
Several schemes have been introduced to further control the pulse duration: reducing the bunch charge, using a slotted foil emittance spoiler~\cite{PhysRevLett.92.074801,doi:10.1063/1.4935429}, and Fresh-slice schemes \cite{Lutman2016SLGC, PhysRevLett.120.264801, PhysRevAccelBeams.20.090701, PhysRevLett.120.264802} can reach few femtoseconds duration, while non linear compression and eSASE were able to reach the hundreds of attosecond regime~\cite{PhysRevLett.119.154801, Duris2020}.

\subsubsection{Polarization control}
A variable polarization DELTA Undulator~\cite{lutman2016polarization} was installed as afterburner as last segment of the original LCLS undulator line. The DELTA design presented four parallel longitudinal rows of ${\rm Nd}_2{\rm Fe}_{14}{\rm B}$ magnetic blocks, effectively forming two crossed planar undulators. The polarization was controlled by adjusting the relative longitudinal position of the rows. The mechanical movement required for a full switch of left circular to right circular polarization, can be accomplished in about half a minute.  
To produce FEL radiation, the electron bunch needs to be micro-bunched, at the wavelength of interest, in the upstream segments. The number of upstream segments used is chosen to maximize the microbunching and the output of polarized controllable X-rays emitted by the electrons in the DELTA afterburner.

\paragraph{Beam diverting}
The upstream undulator segments used to micro-bunch the electron beam produce linearly polarized X-rays. The overlap between these unwanted X-rays and the ones produced by the DELTA afterburner compromise the polarization purity of the delivered X-rays. Such X-ray pulses present also a variable and stochastic polarization along the beam, due to the slippage between upstream produced X-rays and the micro-bunched beam. To improve the polarization purity, the power ratio between the X-rays produced by the afterburner and the ones produced by the upstream undulator segments needs to be maximized. A typical ratio of $5$ was achieved using a regular taper configuration. With reverse taper~\cite{Schneidmiller2013}, the ratio was experimentally increased to $\approx15$. To further increase it, the beam diverting configuration has been developed \cite{Lutman2016SLGC}, where the combined effect of a dipole kicker and of a defocusing quadrupole, located between the last linearly polarized segment and the DELTA afterburner, allows to produce the polarized X-ray at an angle with respect to the upstream linearly polarized X-rays. Spatial separation allowed collimating out the unwanted X-rays, increasing the desired power ratio to a factor exceeding $200$, and yielding a degree of circularly polarized X-rays close to 100\%, in pulses of about 10~GW of power. Further studies on the physics of microbunching rotation\cite{PhysRevX.8.041036}, enabling the beam diverting scheme, could further improve its performance and ease the setup of the configuration.

\paragraph{Two-color Two-Polarization}
Split undulator schemes\cite{PhysRevLett.110.134801}, particularly when combined with the Fresh-Slice technique \cite{Lutman2016SLGC}, have proven capable of producing pairs of high power X-ray pulses, where the relative delay can be controlled up to 1~ps, and allowing to scan the delay across time coincidence. Since each X-ray pulse is produced in a different undulator section, properties as wavelength, polarization and pointing can be independently controlled. At the LCLS, a pair of pulses with energy separation of 15~eV, one linearly polarized and one circularly polarized have been produced and characterized in spectrum and polarization with a Time Of Flight (TOF) polarimeter \cite{Ilchen2014} (aslo see Section \ref{sec:InstrumDiag}).

\subsubsection{ LCLS-II --- Upgrading the LCLS FEL facility.}

The SLAC National Accelerator Laboratory (SLAC) is currently commissioning the upgrade project LCLS-II, which has replaced the original LCLS undulator line with two new undulator lines for providing hard X-ray (HXR) and soft X-ray (SXR) pulses, respectively, and added a new source of electrons, based on a new superconduction (SC) linac, in addition to the normal conducting (NC) linac, (the last third of the original SLAC linac), which was already used for the originally LCLS facility. Tab.~\ref{tableLCLS-II} provides a parameter overview for the new facility.

\begin{figure*}
\includegraphics[width=0.9\linewidth]{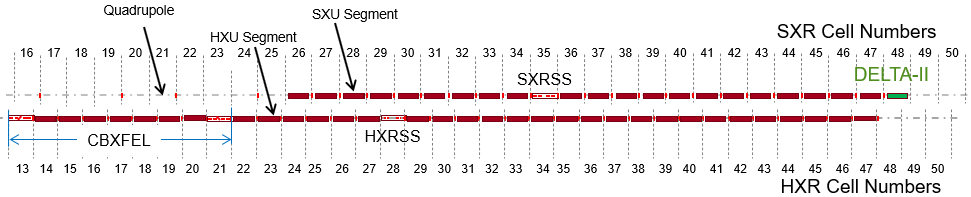}
\caption{LCLS-II main undulator line components layout with the first DELTA-II magnet location indicated in SXR cell 48.}
\label{fig:DELTA-IIa}
\end{figure*}

\begin{figure*}
\includegraphics[width=0.9\linewidth]{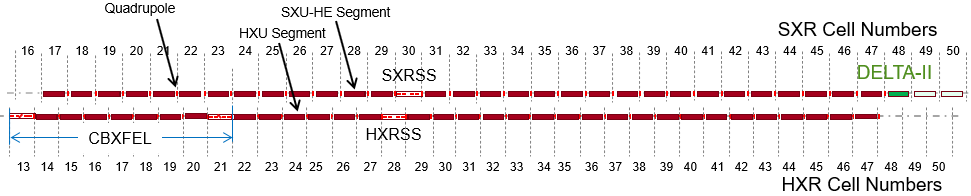}
\caption{LCLS-II-HE main undulator line components layout with the DELTA-II magnets indicated in SXR cells 48 -- 50.}
\label{fig:DELTA-IIb}
\end{figure*}

\begin{table}[]
\begin{center}
\centering
\caption{Parameters of the new LCLS-II facility at SLAC}
\label{tableLCLS-II}
\begin{tabular}{lcccc}
\hline
\rule{0pt}{2.4ex}Undulator line
                                        & \multicolumn{2}{c}{HXR}                 & \multicolumn{2}{c}{SXR}          \\ \hline
\rule{0pt}{2.4ex}Segment type
                                        & \multicolumn{2}{c}{HXU}                 & \multicolumn{2}{c}{SXU}          \\
Undulator period, $\lambda_u$  (mm)      & \multicolumn{2}{c}{26}                   & \multicolumn{2}{c}{39}             \\ 
Undulator minimum gap (mm)                 & \multicolumn{2}{c}{7.2}                 & \multicolumn{2}{c}{7.2}            \\ 
Number of segments                               & \multicolumn{2}{c}{32}                   & \multicolumn{2}{c}{21}             \\ 
Segment  length (m)                               & \multicolumn{2}{c}{3.4}                 & \multicolumn{2}{c}{3.4}            \\ 
Periods per segment                               & \multicolumn{2}{c}{130}                 & \multicolumn{2}{c}{87}             \\ 
Direction of linear polarization                  & \multicolumn{2}{c}{vertical}            & \multicolumn{2}{c}{horizontal} \\ 
Electron source                                      &             NC      &         SC                  &      NC    &   SC                           \\
Max electron energy (GeV)                      &            15        &         4.0                 &    8.0     &  4.0                           \\
Max repetition rate (Hz)                          &           120       &     929,000             &    120     &  929,000                  \\
Max beam power (kW)                            &           0.45      &     120                    &    0.24     &  120                        \\
Min photon energy  (keV)                        &             1.0      &         1.0                &    0.2    &  0.2                             \\
Max photon energy  (keV)                       &            25        &         5.0                &    5.0    &  1.5                             \\
\hline
\end{tabular}
\end{center}
\end{table}

While the LCLS-II project is being completed, another upgrade project, LCLS-II-HE,  is already underway, which will raise the electron beam energy, that is delivered by the superconducting linac, from 4~GeV to 8~GeV, add 9 undulator segments to the SXR line and increase the undulator period of the SXR line undulator segments from $\lambda_{\rm u,SXU}=39$~mm to  $\lambda_{\rm u,HE-SXU}=56$~mm. Operations parameters for the upgraded undulator lines with the LCLS-II-HE linac are summarized  in Tab.~\ref{table:LCLS-II-HE}.

\begin{table}[b]
\begin{center}
\centering
\caption{Parameters of the future LCLS-II-HE facility at SLAC}
\label{table:LCLS-II-HE}
\begin{tabular}{lcccc}
\hline
\rule{0pt}{2.4ex}Undulator line
                                                                     & \multicolumn{2}{c}{HXR}                 & \multicolumn{2}{c}{SXR}          \\ \hline
\rule{0pt}{2.4ex}Segment type
                                                                     & \multicolumn{2}{c}{HXU}                 & \multicolumn{2}{c}{HE-SXU}          \\
Undulator period, $\lambda_u$  (mm)                                   & \multicolumn{2}{c}{26}                   & \multicolumn{2}{c}{56}             \\ 
Undulator minimum gap (mm)                                              & \multicolumn{2}{c}{7.2}                 & \multicolumn{2}{c}{7.2}            \\ 
Number of segments                                                           & \multicolumn{2}{c}{32}                   & \multicolumn{2}{c}{30}             \\ 
Segment  length (m)                                                           & \multicolumn{2}{c}{3.4}                 & \multicolumn{2}{c}{3.4}            \\ 
Periods per segment                                                            & \multicolumn{2}{c}{130}                 & \multicolumn{2}{c}{60}             \\ 
Direction of linear polarization                                               & \multicolumn{2}{c}{vertical}            & \multicolumn{2}{c}{horizontal} \\ 
Electron source                                                                   &             NC      &         SC                  &      NC    &   SC                           \\
Max electron energy (GeV)                                                   &            15        &         8.0                 &   10.0     &  8.0                           \\
Max repetition rate (Hz)                                                       &           120       &     929,000             &    120     &  929,000                   \\
Max beam power (kW)                                                        &           0.45      &     120                    &    0.24     &  120                        \\
Min photon energy (keV)                                                     &             1.0      &       1.0                &    0.2    &  0.2                             \\
Max photon energy (keV)                                                     &            25        &    15.0                &    5.0    &  5.0                             \\
\hline
\end{tabular}
\end{center}
\end{table}

\noindent
In addition to the operation with the SC linac, operation with the NC linac will continue to be provided for both beamlines.

In parallel to upgrading the LCLS facility, SLAC is also developing a new, compact variable gap polarization control undulator, which is called DELTA-II~\cite{Nuhn2019}. Its performance is expected to be comparable to that of the DELTA undulator that was successfully operated with the original LCLS undulator system~\cite{Nuhn2015}\cite{Hartmann2016}\cite{Higley2016}\cite{lutman2016polarization}. The development of this new polarizing undulator started in 2015 and it is scheduled to be available to operation in 2024. The DELTA-II magnet arrangement is similar to that of the APPLE X undulators~\cite{Schmidt2018} recently developed at PSI and, at closed gap, it is similar to the DELTA undulator. Tab.~\ref{table:DELTA-II-HE} contains a summary of the DELTA-II parameters.

\begin{table}[]
\begin{center}
\centering
\caption{Parameters of the DELTA-II operating in afterburner mode with the LCLS-II and LCLS-II-HE beamlines.}
\label{table:DELTA-II-HE}
\begin{tabular}{lcccc}
\hline
\rule{0pt}{2.4ex}Parameter
                                                                      & \multicolumn{4}{c}{DELTA-II}       \\ \hline
\rule{0pt}{2.4ex}Undulator period, $\lambda_{\rm u}$ (mm)        & \multicolumn{4}{c}{49.333}          \\
Maximum gap (mm)                                             & \multicolumn{4}{c}{22.6}              \\
Minimum gap (mm)                                              & \multicolumn{4}{c}{6.6}                 \\
Maximum K$_{\rm DELTA-II, linear}$                 & \multicolumn{4}{c}{6.718}             \\
Minimum K$_{\rm DELTA-II, linear}$                  & \multicolumn{4}{c}{1.727}              \\
Segment  core length (m)                                   & \multicolumn{4}{c}{3.26}                  \\
Core Periods per segment                                     & \multicolumn{4}{c}{66}                   \\
Polarization                                                         & \multicolumn{4}{c}{various}            \\ \hline
\rule{0pt}{2.4ex}Undulator line
                                                     &\multicolumn{2}{c}{LCLS-II} & \multicolumn{2}{c}{LCLS-II-HE }  \\ \hline
\rule{0pt}{2.4ex}Electron source
                                                  & NC        & SC        & NC      & SC        \\ \hline
\rule{0pt}{2.4ex}Minimum electron energy (GeV)
                                                        & 3.7       & 3.7       & 4.9      & 4.9       \\
Maximum electron energy (GeV)                        & 8.0       & 4.0       & 8.0      & 8.0       \\
$\lambda_{\rm u,SXU/SXU-HE}$ (mm)               &  39       & 39        & 56        & 56         \\
Maximum used $K_{\rm SXU/SXU-HE}$                      & 5.560   & 5.560  & 6.287   & 6.287    \\
Minimum used $K_{\rm SXU/SXU-HE}$                       & 2.075   & 2.075   & 1.546  & 1.546    \\
Minimum used $K_{\rm DELTA-II, linear}$                  & 1.727  & 1.727   & 1.727   & 1.727    \\
Maximum used $K_{\rm DELTA-II, linear}$                & 4.901   & 4.901   & 6.718   & 6.718    \\
Minimum photon energy (eV)                             &  0.200  & 0.200   & 0.200   & 0.200    \\
Maximum photon energy (eV)                            &  4.945  & 1.500   & 4.945   & 4.945   \\
\hline
\end{tabular}
\end{center}
\end{table}

The DELTA-II undulator will be operated at the downstream end of the  SXR line (SXR line cell 48, see Fig.\ref{fig:DELTA-IIa}) in afterburner mode, i.e., it will operate on an electron beam that has been micro-bunched by a moderate number ($\approx$~6) of regular SXR undulator segments. The change from the DELTA undulator to the DELTA-II undulator is necessary because of the change of the micro-bunching undulators from the original shorter-(30~mm)-period fixed gap LCLS-type undulator segments to the  new  longer-(39~mm)-period, variable gap  LCLS-II SXR type undulator segments. 
The 39~mm period length of the SXR undulator segments (SXUs) had been chosen to optimized the SXR line performance for operating at the highest electron energy (4~GeV) of the LCLS-II superconducting linac. As mentioned above, the LCLS-II-HE project will increase the highest electron energy to 8~GeV. This requires to increase the period length of the SXR undulator segments to 56~mm (HE-SXUs). The period length of the DELTA-II undulator has been chosen to work with both period lengths. After the DELTA-II undulator has been developed and is operational, SLAC will try to secure funding for two additional DELTA-II undulators to be installed right after the first DELTA-II undulator in SXR line cells 49 an 50 (see Fig.~\ref{fig:DELTA-IIb}).

\clearpage
\onecolumngrid \newpage \twocolumngrid

\subsection{FLASH@DESY, Germany}
\label{sec:DESY}

\begin{center}
Evgeny Schneidmiller, Markus Tischer, \linebreak
and Mikail Yurkov
\end{center}

\begin{figure*}
\includegraphics[width=.9\textwidth]{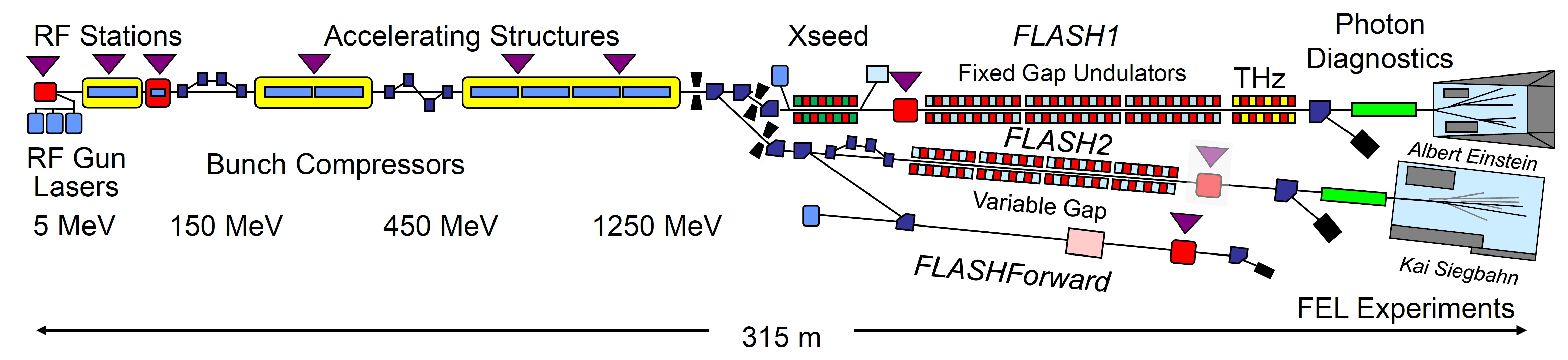}
\caption{Layout of the FLASH facility.}
\label{FL-layout}
\end{figure*}
In the first XUV and soft X-ray FEL user facility
FLASH \cite{flash-nat-phot,tiedtke2009soft}, located at the Deutsches Elektronen-Synchrotron in Hamburg, the electron bunches with maximum energy of 1.25 GeV are distributed between the two undulator lines, FLASH1 and FLASH2 (see Fig.~\ref{FL-layout}). The facility operates in the wavelength range 4 - 60 nm with long pulse trains (several hundred pulses) following with 10 Hz repetition rate. After the current upgrade, the electron energy will reach 1.35 GeV. A part of the upgrade is the installaton of a helical afterburner at FLASH2. Below we briefly describe the afterburner as well as a method to obtain high purity of circularly polarized radiation.\\

The helical afterburner at FLASH2 will be an APPLE-III undulator \cite{schmidt2014magnetic} which is presently developed and constructed at DESY. The APPLE-III configuration 
will be a very efficient magnetic concept for circular beam apertures and well suited to reach the L-edges of the 3d transition metals Fe, Co, Ni despite the only moderate beam energy of 1.35 GeV. Although proposed for already many years, an APPLE-III undulator has not been realized up to now. Like in the well-known APPLE-II device, the pure permanent-magnet arrays are split longitudinally to create an elliptical magnetic field depending on the relative shift to each other. However, the vertically magnetized magnets of this Halbach structure are tilted by 45$^{\circ}$ pointing directly towards the beam. We combine this idea with a force compensation scheme to get a compact magnet structure which can be used together with gap mechanics like the already existing at FLASH2. This also allows for full opening of the gap which will result in a better radiation protection during machine commissioning.\\

\begin{figure}
\includegraphics[width=1\linewidth]{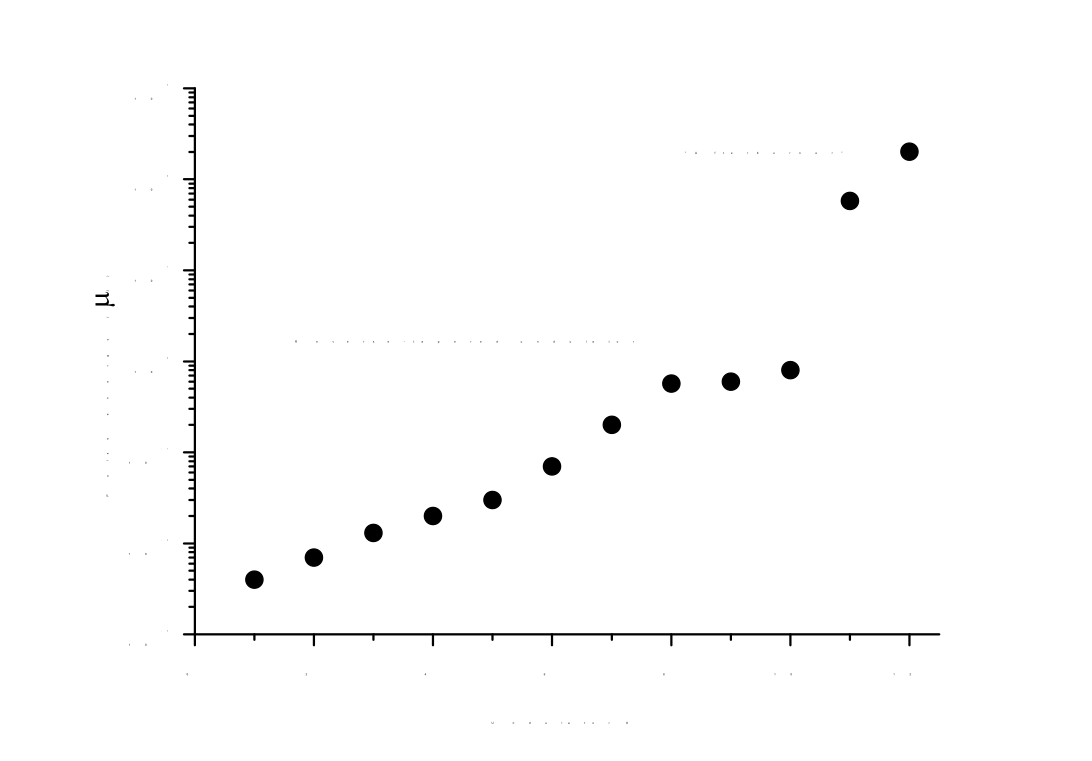}
\caption{FEL pulse energy versus undulator number. First ten undulators are reverse-tapered, last two sections are tuned
to the resonance with the incoming microbunched beam.}
\label{gain-curve}
\end{figure}

The APPLE-III magnet structure will have a period length of 17.5 mm with a minimum magnet gap of 8 mm providing a maximum K-value of 0.92. At closed gap, the circular vacuum chamber (7 mm outer diameter) is mostly surrounded by the magnets leaving a small slit at the sides for its support. The total length will be 2.5 m. A short test structure had been built initially to investigate the force compensation scheme, mechanical performance, the tuning concept, and had also been characterized magnetically. Due the small size and period length, glued pairs of magnets corresponding to half a period will be used which simplifies the keeper design significantly. Meanwhile all parts of the magnet structure are in the fabrication process.\\

A method for suppression of the linearly polarized background from the main undulator was proposed in \cite{Schneidmiller2013}:
an application of the reverse undulator taper.
It was shown that in some range of the taper strength
the bunching factor at saturation is practically the same as in the reference case of the non-tapered undulator,
the saturation length increases moderately while the saturation power is
suppressed by orders of magnitude. Therefore, the proposed scheme is conceptually very simple:
in a tapered main (planar) undulator
the saturation is achieved with a strong microbunching and a suppressed radiation power, then the modulated beam radiates at full
power in a helical afterburner, tuned to the resonance.
This method (in combination with the spatial separation) was used at LCLS to obtain a high degree of circular polarization
\cite{lutman2016polarization, hartmann2016circular} and is routinely used now in user operation.
Obviously, the afterburner (helical or planar) can be tuned to a harmonic of the main undulator. In this case,
the harmonics can be efficiently
generated with a low background at the fundamental.\\ 

The reverse tapering concept was successfully tested at FLASH.
The gap-tunable undulator of FLASH2 consists of twelve 2.5 m long sections with the undulator period of 3.14 cm and
the maximum rms K-value about 1.9.\\

In the experiment in 2016 \cite{schneidmiller2017reverse},
we used the first ten undulator sections as a main undulator with reverse tapering, and the last two sections played the role of the afterburner, i.e. they could be tuned to a resonance with the incoming microbunched beam. 
The electron energy during the measurements was 715 MeV, and the FEL wavelength was 17 nm. 
We did a K-scan of the afterburner, showing the resonance, and also measured the FEL gain curve in this configuration, as shown in Fig.~\ref{gain-curve}. One can see again that
the high contrast (in excess of 200) between the radiation intensity from the afterburner and from the reverse-tapered
undulator is measured. \\

We also demonstrated an efficient
generation of the 2nd and the 3rd harmonics in the afterburner \cite{schneidmiller2017reverse}. The electron energy was 852 MeV,
and the wavelength for the untapered case was set to 25.5 nm. with the rms K parameter of 1.9. The first nine undulator
sections were reverse-tapered, and the following two sections played the role of the afterburner. When the
afterburner sections were completely opened, the background from the reverse-tapered main undulator was measured at the
level of 0.9 $\mu$J. When we tuned the afterburner sections to 26.2 nm, the pulse energy was 132 $\mu$J, i.e. the
contrast above 100 was measured.  Then we tuned the afterburner to the second and the third harmonics of the main
undulator and did the wavelength scan (or, K-scan) of the undulator around the corresponding resonances.
The pulse
energy reached 41 $\mu$J when the afterburner was tuned to 13.2 nm, and 10 $\mu$J for the 8.8 nm tune. 
Thus, we have demonstrated that the reverse tapering in the main undulator can be used for an efficient,
background-free generation of harmonics in the afterburner.

This method will be used for production of high purity circular polarization at FLASH as soon as the APLLE-III afterburner will be installed, as anticipated to be done in mid 2023. To further improve the purity, we also consider using the beam diverting technique in the same way as it was done at LCLS \cite{lutman2016polarization}.

\onecolumngrid \newpage \twocolumngrid

\subsection{SwissFEL@PSI, Switzerland}
\label{sec:PSI}

\begin{center}
Marco Calvi, Thomas Schmidt \linebreak
Sven Reiche, and Eduard Prat
\end{center}

\begin{figure*}[b!]
\centering
    \includegraphics[width=\textwidth]{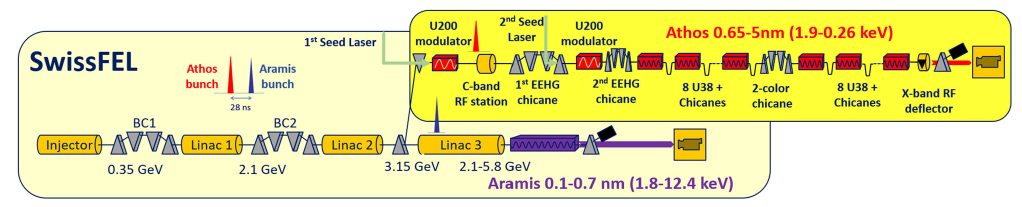}
    \caption{The SwissFEL layout, where the injector and the three linacs are schematically represented together with the details of the two beamlines, Aramis and Athos.} \label{fig:SwissFEL_Layout}
\end{figure*}

\begin{figure}
    \centering
    \includegraphics[scale=0.26]{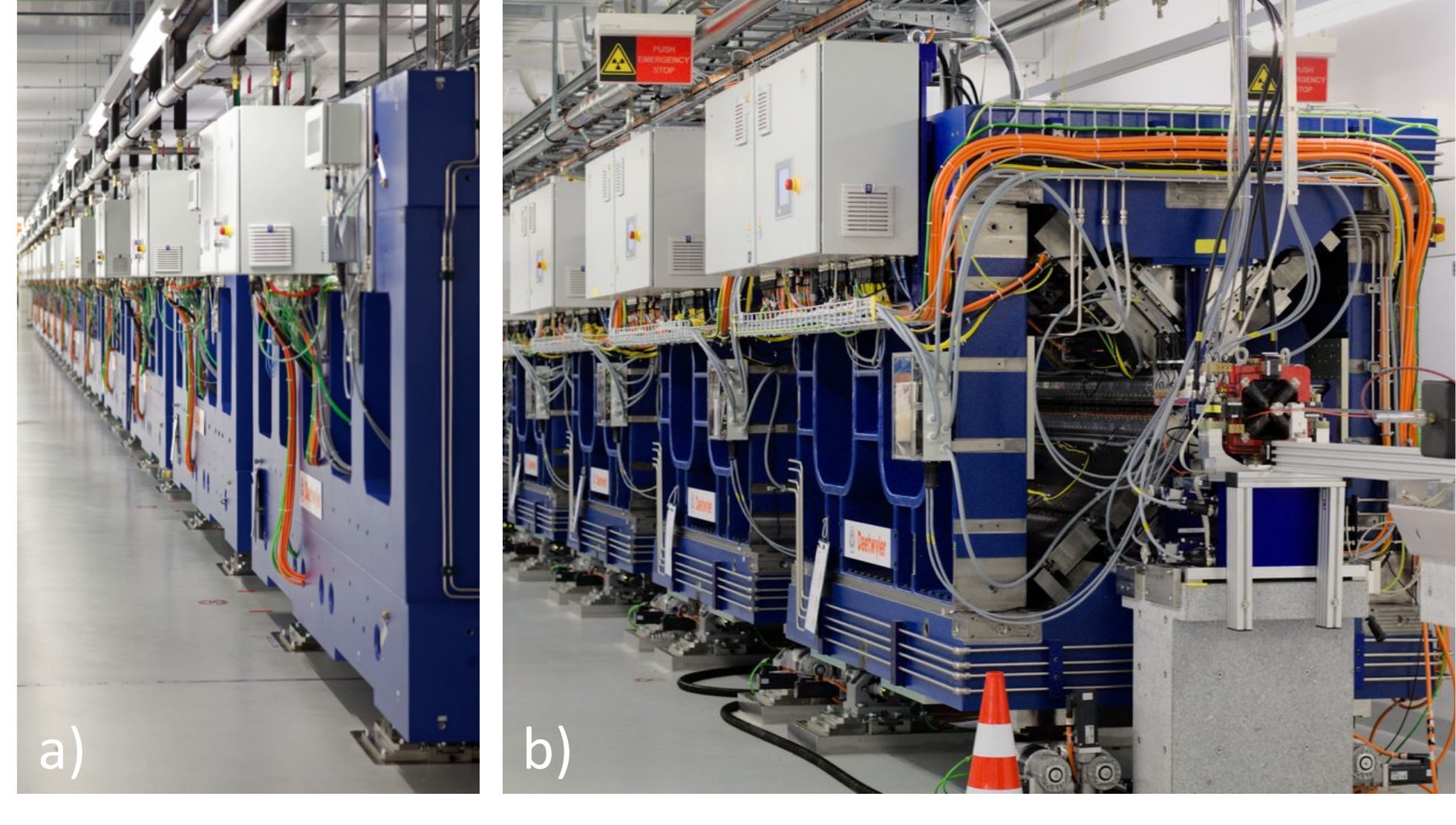}
    \caption{On the left (a), the hard x-ray beamline of SwissFEL, Aramis, with its 13 in-vacuum permanent magnet undulators; on the right (b), the soft x-ray beamline Athos with its 16 Apple X undulators (in the picture only the last four are visible).}
    \label{fig:SwissFEL_Aramis_Athos}
\end{figure}

\begin{figure}
    \centering
    \includegraphics[scale=0.5]{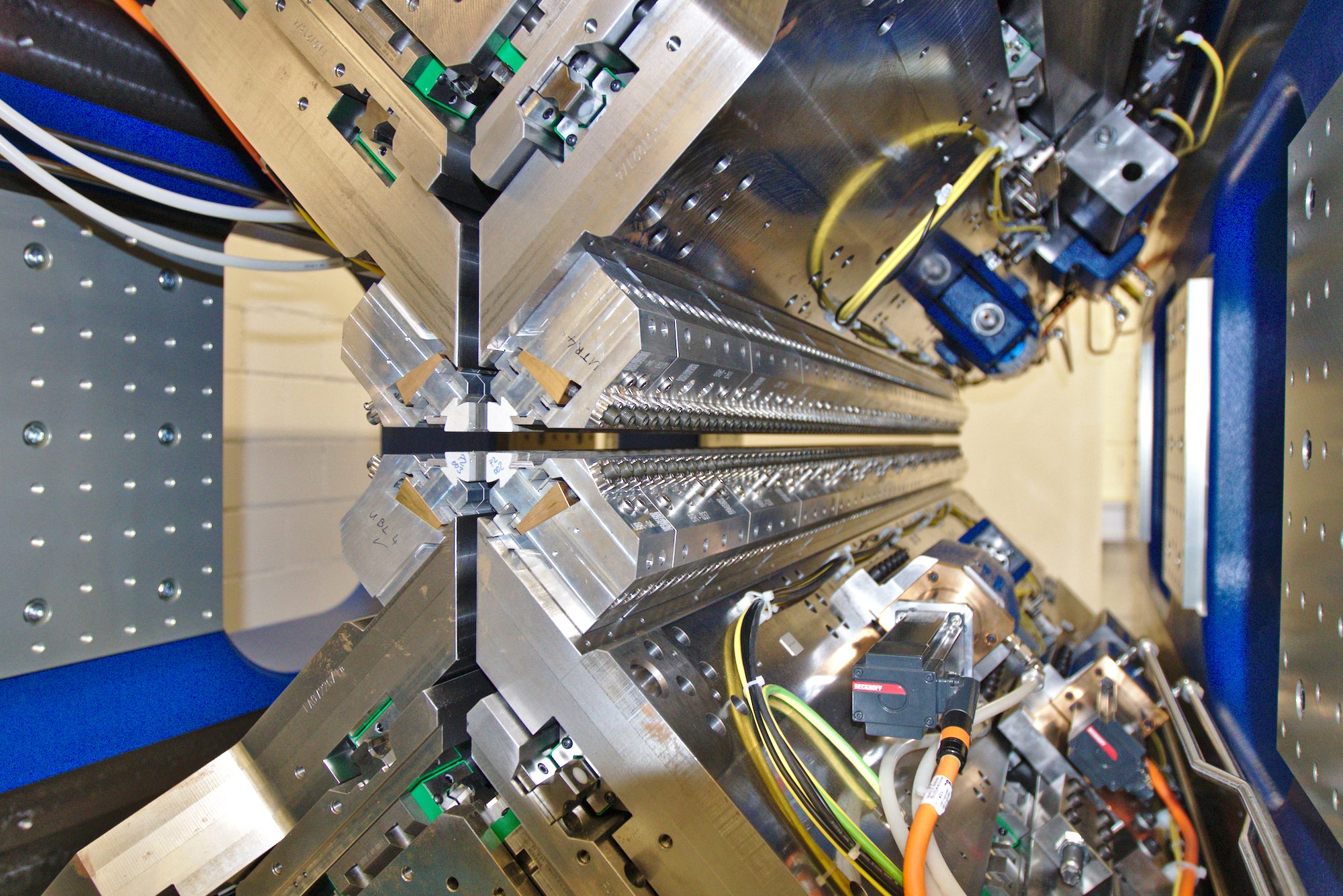}
    \caption{The magnetic structure of an Apple X undulator. The four Halbach rows are visible as well as the stages and motors used both for their radial and longitudinal displacement, to tune the strength and the polarization respectively.}
    \label{fig:Apple_X}
\end{figure}

The FEL presently in user operation at the Paul Scherrer Institute in Switzerland, SwissFEL, is dedicated both to hard x-ray and soft x-ray. The layout of the facility is shown in Fig.\,\ref{fig:SwissFEL_Layout} where the injector, the three linacs and the two stages of compression are schematically presented together with two beamlines. The first beamline commissioned in 2017 was Aramis (Fig.\ref{fig:SwissFEL_Aramis_Athos}a), the SwissFEL hard x-ray branch, equipped with 13 planar in-vacuum undulators, each 4\,m long and with a period length of 15\,mm. With an electron beam energy up to 6.0\,GeV, Aramis delivers photon pulses from 0.7\,nm down to 0.08\,nm in linear horizontal polarization. A second hard x-ray line called Porthos is planned for 2028 were a variable control of the polarization will be possible either with in-vacuum cryogenics APPLE devices \cite{Bahrdt_IPAC2018_AppleVacuum} or with SCAPE \cite{Ivanyushenkov_IPAC2017_SCAPE}, based on superconducting coils made of NbTi. The soft x-ray beamline is called Athos (Fig.\ref{fig:SwissFEL_Aramis_Athos}b) and operates between 7\,nm and 0.7\,nm with Apple\,X undulators \cite{schmidt2018apple} (Fig.\ref{fig:Apple_X}) to deliver polarized photon both in elliptical and linear polarization with arbitrary angle. Within a percentage changes (due to the residual susceptibility of the permanent magnets, geometrical tolerance of the frame and the variation of the forces) the deflection parameter, $K$, does not vary in an Apple X while operated in elliptical mode, which allows to reach the same photon energies both in LV \& LH and in C+ \& C-. The other advantage of the Apple\,X is the larger degrees of freedom with respect to the previous APPLE devices, which allow to move radially and independently the four magnetic rows. This enables to generate Transverse Gradients (TGs) in a controlled way and in any polarization while in an APPLE with four independently longitudinally movable rows it is only possible for elliptical and the relation between $K$ and $\partial K/\partial x(y)$ is fixed \cite{Calvi:vv5164}. \\

 TGs can be used to control the FEL bandwidth \cite{Prat:vv5134} and to increase it above 10\%. This has straight forward applications for instance in absorption spectroscopy which could be executed in single-shot mode, without the need of scanning the $K$. TGs could be used as well to minimize the impact of the slice energy spread if properly associated with the required dispersion as proposed in \cite{Lindberg_FEL2013_TG} for storage ring FEL oscillators. 
The Athos beamline was designed with compact chicanes (later on referred as CHICs) in between each 2 meter long Apple\,X module. This length was the result of the optimization of this novel layout based on the compact chicanes as presented in \cite{Prat:vv5135}. With the help of the CHICs, a series of new operation modes is possible: starting from the optical Klystron which makes the saturation length shorter \cite{Prat.119.15} and pave the way for more compact FEL sources; the high brightness SASE which reduces the bandwidth (enhancing the longitudinal coherence) of the SASE-FEL pulses by delaying the electrons with respect to the photons between undulator modules, thus increasing the cooperation length between electrons and photons \cite{Thompson_IPAC2010_TUPE050,Wu_FEL2012_TUPD07,Wu_IPAC2013_WEODB101,PhysRevLett.110.134802}; and the short and high power FEL pulses based on superradiance \cite{BONIFACIO1984373,PhysRevA.44.R3441} and implemented with a horizontally tilted beam \cite{PhysRevSTAB.18.100701} where only the on axis slice contribute to the lasing process and is replaced with a fresh one any time by means of the CHICs (which are designed to introduce offsets up to $\pm$0.3\,mm if required).

Upstream the Athos beamline, two modulators separated by two magnetic chicanes have been recently (2022) installed to couple in two external lasers, ranging from 260-1600\,nm wavelengths. This equipment is designed for multiple purposes and several options will be tested in the coming years. Nevertheless, the main application will be to seed Athos using the EEHG technique \cite{PhysRevLett.102.074801}, today available for photon energies up to about 400 eV. The 0.2\,m period length modulators are designed with high flexibility to fulfill novel requirements. The possibility to individually tune the strength of each of the 18 poles allow for extreme taper configurations, tune a subset of poles to a different wavelength and even to triple the period, reaching 0.6\,m.  

The first lasing results of Athos have been recently reported in \cite{Prat_nature_2023}. In particular, the Athos team has demonstrated the capacity of the Athos beamline to reduce the saturation length by about 35\% when using the optical klystron and helical undulator configuration (in comparison to the standard planar undulators and no optical klystron),  the production of short pulses with higher power than in standard configurations following the superradiance regime, and the generation of variable polarization FEL radiation using the APPLE-X devices for photon energies at the keV level.  \\

\onecolumngrid \newpage \twocolumngrid

\subsection{European XFEL, Germany}
\label{sec:EuXFEL}

\begin{center}
Gianluca Aldo Geloni, Suren Karabekyan, Jiawei Yan, \linebreak
and Svitozar Serkez    
\end{center}

\subsubsection{Implementing polarization control capabilities at the SASE3 FEL line}
At the European XFEL, two undulator systems for hard X-rays – SASE1 and SASE2, and one for soft X-rays – SASE3 have been successfully in operation since 2018\cite{Abeghyan2019}. All systems were designed and equipped with variable gap planar undulators that could generate linearly polarized radiation. The SASE3 undulator system consists, in particular, of 21 planar U68 variable-gap undulators, each 5 m long with a period of 68 mm. Depending on the energy of the electron beam (8.5 GeV - 17.5 GeV), it can generate radiation in the range from 0.24 keV to 4.6 keV. Due to high demand from users, a project to generate soft X-ray radiation with variable polarization was initiated in 2019 that consists in equipping SASE3 with an afterburner based on the APPLE-X undulator design developed at PSI \cite{Abela2019, Schmidt2018, Liang2019} (see also Section \ref{sec:PSI}).
The idea is to use a micro bunched electron beam after a system of planar undulators and direct it into a system of helical undulators tuned to the resonant frequency, where it emits powerful coherent radiation with controlled polarization. To achieve a high degree of, for instance, circular polarization, it is necessary to suppress or cut off the linear polarization generated by the planar undulators. For this purpose, the SASE3 planar undulators are used in reverse tapering mode \cite{Schneidmiller2013}, where the bunching factor is approximately the same as for nontapered undulators, but the radiation power decreases by orders of magnitude. The results of a numerical 3D simulation using the FAST code \cite{Saldin1999} show that, for example, at a wavelength of 1.5 nm (0.83 keV), using the last 11 planar undulator cells with a total length of 55 meters, in the 2.1$\%$ reverse taper mode, the linearly polarized power reaches a value of 0.4 GW, while the bunching factor increases continuously. The power of circularly polarized radiation generated by a 10 m long helical undulator, located directly behind the system of linear undulators, increases very rapidly and reaches up to 155 GW. Hence, the degree of circular polarization reaches a value of ~99.9$\%$. Similar results were obtained using a numerical three-dimensional SIMPLEX simulation \cite{Tanaka2015}. Table 1 shows the simulation results for four wavelengths considering a helical afterburner of eight- or twelve-meters composition length \cite{Wei2017}.
These studies were the basis for calculating the electron beam optics in the installation area of the helical afterburner system. An APPLE X-type of a helical undulator was chosen to create a system of helical undulators for SASE3 \cite{Karabekyan2021}. The magnetic structure of helical undulators was designed to overlap the working range of planar undulator U68. The full length of the UE-90 magnet structure is 1.98 m, the undulator period length is 90 mm, the gap variation range is 12.5 -31.6 mm, the longitudinal shift range of each magnet array is $\pm$45 mm, and the permanent magnets material is NdFeB with the nominal remanent field of 1.26 T. The frame length of this undulator is slightly over two meters. It was decided to use the existing design of the intersection and its components, with the afterburner section starts after the last SASE3 undulator cell.

\begin{figure}
\centering
\subfigure[]{\includegraphics[width=0.45\textwidth]{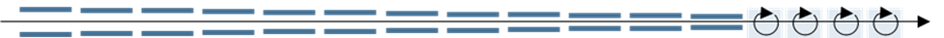}}
\hfill
\subfigure[]{\includegraphics[width=0.45\textwidth]{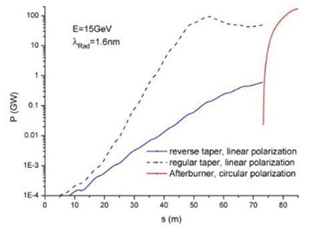}}
\hfill
\caption{Schematic of the project and the evolution of the radiation power from regular taper and reverse taper and afterburner \cite{Karabekyan2021}, with (annexed table) the photon energy ranges generated by the UE-90 undulator by varying only the gap \cite{Karabekyan2022}.}
\label{fig:XFEL1}
\end{figure}

\begin{table}[]
\begin{center}
\centering
\caption{XFEL Parameter}
\label{table:2}
\begin{tabular}{lcc}
\hline
Polarizationmode                  & LH/LV/C+C-                 & Linear 45$^{\circ}$/125$^{\circ}$                  \\ 
K-Range                           & 9.40 - 3.37          & 6.62 - 2.36             \\ \hline
Photon Energy Range (keV)                   &                       &                       \\
@8.5 GeV                & 0.169 - 1.141        & 0.33 - 2.012    \\
@11.5 GeV            & 0.309 - 2.088         & 0.608 - 3.684        \\
@14 GeV             & 0.457 - 3.095        & 0.902 - 5.459         \\
@16.5 GeV           & 0.635 - 4.299       & 1.252 - 7.583         \\
@17.5 GeV         & 0.715 - 4.835        & 1.409 - 8.530         \\
\hline
\end{tabular}
\end{center}
\end{table}

The distinct advantage of the APPLE X undulator is the ability to change the radiation energy by adjusting the gap. This results in a significantly higher field uniformity compared to a fixed-gap delta undulator, where the field can only be controlled by shifting the magnetic structures relative to each other, resulting in large field gradients. This requires extremely small tolerances for the alignment of the fixed-gap undulator system and makes an alignment of extended systems almost impossible. Figure \ref{fig:XFEL1} shows a schematic of the project and the evolution of the radiation power from regular taper and reverse taper plus the afterburner. The table annexed to the figure presents the photon energy ranges generated by the UE-90 undulator by varying only the gap. The magnetic field simulations for LH, LV, C+, C-, and 45° linear polarizations have been performed using the Radia program \cite{Chubar1998}. The actual length of the installed afterburner is 8m (4 undulators)

\subsubsection{Current status of the project}
At the time of writing, the project is at the stage of its completion and commissioning. During the winter shutdown 2021/2022, all four APPLE-X undulators were installed in the tunnel. The helical undulator system was connected to the control system and made operational together with all related components.
During commissioning, calibration measurements were made to compensate for the magnetic fields in order to minimize the deflection of the electron beam from the optimized orbit. Measurements were also made to optimize the values of the phase shifters located between the helical undulators. After alignment of the electron beam and setting up SASE3 in reverse taper configuration, lasing was immediately obtained when all four Apple-X undulators were brought into resonance. This was achieved at a beam energy of 900 eV and a clockwise circular polarization mode (C+). Optimization of the phase shifter gap resulted in a 30-fold gain of the circularly polarized radiation over the linearly polarized radiation of the planar undulators. Later, lasing was also obtained for counterclockwise circular (C-), linear 45$^{\circ}$  (45$^{\circ}$ ), linear vertical (LV), and linear horizontal (LH) modes. The photon beam energy was changed to 700 eV where lasing was obtained in LH, C+ and C- polarization modes.
The availability of the APPLE-X afterburner will allow for further development of advanced polarization schemes, which will be based on flexible control of the electron bunching by means of the SASE3 undulator. Here we will discuss, in particular, two possible future options.

\subsubsection{Possible future developments I: FEL radiation with Orbital angular Momentum}
Undulator radiation pulses produced by a helical undulator at the h-th harmonic of the fundamental carry Orbital Angular Momentum (OAM) with topological charge $\pm$(h-1) \cite{Sasaki2008}. In \cite{Hemsing2020}, it was proposed to exploit this fact by preparing an electron beam bunched at the h-th harmonic of the fundamental of a short helical radiator using the FEL process in a longer undulator preceding the radiator. Minimization of the background emission is obtained by reverse tapering. Due to the availability of the APPLE-X afterburner, this kind of setup is readily available at the SASE3 FEL, where tests will be carried out following the finalization of the afterburner commissioning. For further perspectives, see also Section \ref{sec:FERMI}.
\begin{figure}
\centering
\subfigure[]{\includegraphics[width=0.45\columnwidth]{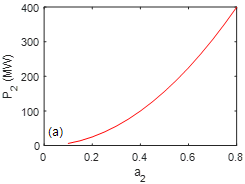}}
\hfill
\subfigure[]{\includegraphics[width=0.45\columnwidth]{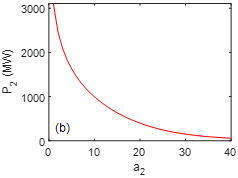}}
\hfill
\subfigure[]{\includegraphics[width=0.45\columnwidth]{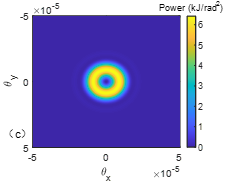}}
\hfill
\subfigure[]{\includegraphics[width=0.45\columnwidth]{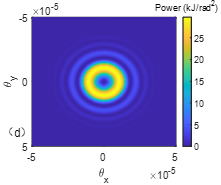}}
\hfill
\caption{Left. Emitted power from a 14 GeV electron beam entering the U-90 Apple-X afterburner after having been bunched in the main SASE3 undulator for a fixed transverse rms size of 20 $\mu$m (a) or a fixed bunching factor of 0.8 (b). Right. Output comparison for the beam with a bunching factor of 0.5 and a transverse size of 20 $\mu$m (c) and the beam with a bunching factor of 0.8 and a transverse size of 10 $\mu$m (d) }
\label{fig:XFEL2}
\end{figure}
Schematically, the setup looks like Figure \ref{fig:XFEL2} 1, with the possible addition of an electron pulse kick between SASE3 and the helical afterburner to decrease the radiation background from SASE3 even more \cite{Serkez2016}. However, now SASE3 and the afterburner are tuned at different frequencies, for example 1000 eV and 500 eV respectively. In this way, the beam entering the afterburner is bunched at 1000 eV and thus emits strongly at the second harmonic of the APPLE-X, producing radiation with OAM.  Estimations of the output can be carried out semi-analytically by computing the bunching at 1000 eV via FEL simulations and subsequently calculating analytically the coherent emission in a short afterburner (assuming no gain). The energy (power) generated strongly depends on the electron beam transverse size.

This means that a magnetic optics solution with small betatron functions at the afterburner location is to be preferred. Figure \ref{fig:XFEL2} shows the dependence on the bunching factor at fixed transverse rms size and the dependence of the transverse rms size for a fixed bunching factor in the case of a normalized emittance of 0.5 $\mu$m and an electron energy of 14 GeV, assuming a flat-top, 20 fs-long profile with 5 kA peak current. The figure also shows two output profiles for two different bunching levels. This method is expected to produce GW-level pulses carrying OAM.

\subsubsection{Possible future developments II: Polarization shaping}
A second possible development concerns polarization-shaping technique, proposed in \cite{Serkez2019}. Consider two overlapping, longitudinally coherent pulses of radiation emitted in two separate parts of the APPLE-X radiators with orthogonal polarization states. If the two pulses have the same frequency, their superposition results in a different polarization state, which depends on the relative phase between the initial two. The resulting polarization state will be located somewhere on the large circle of the Poincar\'{e} sphere between the original polarization states. For example, if the two initial pulses are circular left- and right-polarized, by overlapping them one obtains a linearly polarized pulse, with the polarization plane depending on the phase difference. When the difference between the intensities and phases changes along the direction of propagation or across the transverse coordinates (or both), the degree of polarization, averaged over time or across the transverse direction, degrades. One can, however, control and maximize this effect so that the resulting polarization depends on the location on the sample and/or on the arrival time of the radiation within the pulse. 

\begin{figure}
\centering
{\includegraphics[width=0.95\columnwidth]{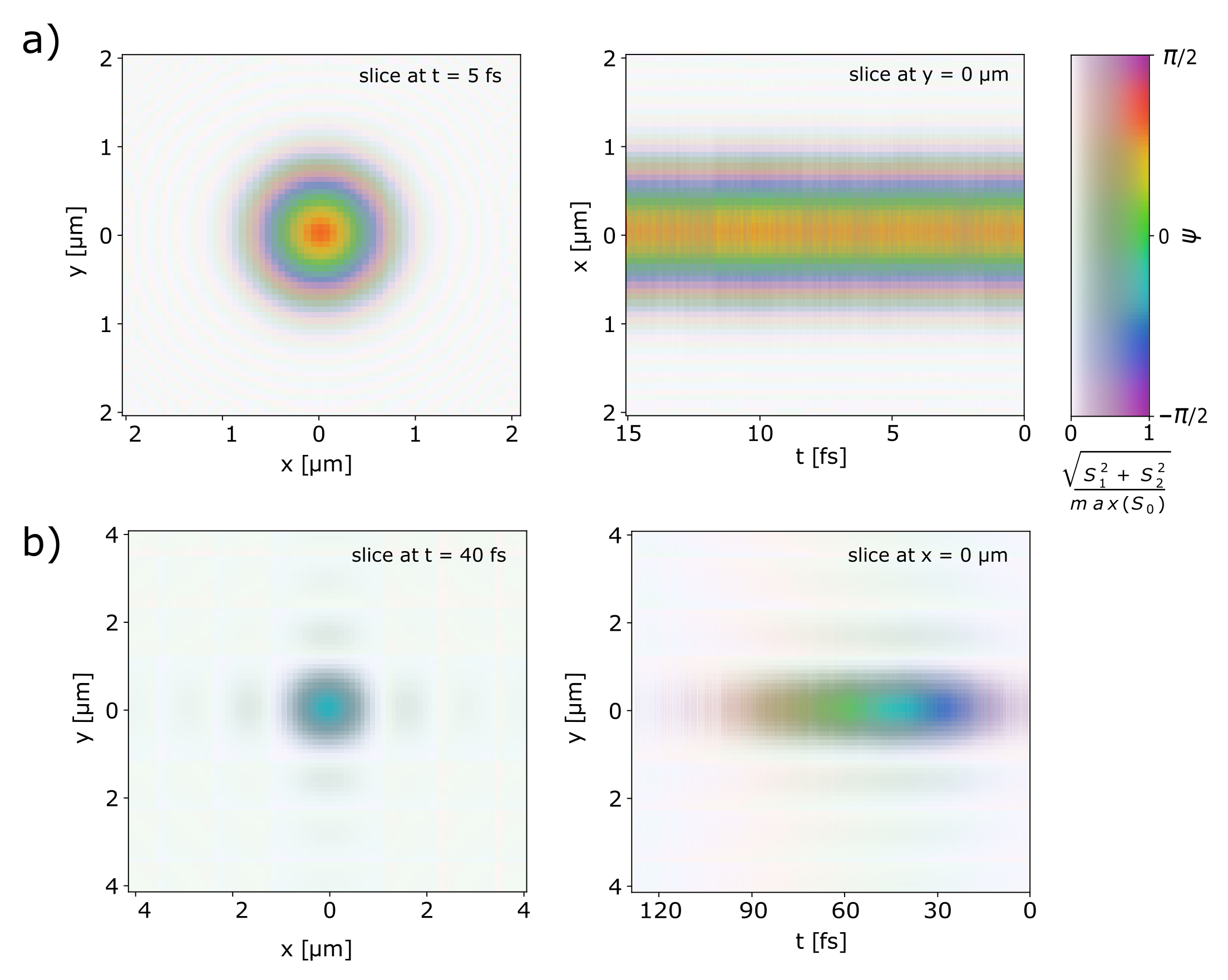}}
\caption{Adapted from Figures 9 and 10 of~\cite{Serkez2019}: Radiation polarization properties calculated for the SASE3 beamline of the European XFEL. 
Subplots depict radiation intensity and orientation of the
polarization plane in false colors as a function of transverse coordinates and time using setup geometry, optimized for \textit{spatial}, row (a) and \textit{temporal} polarization shaping, row (b).}
\label{fig:XFEL3}
\end{figure}

To obtain radiation pulses with shaped polarization one may prepare a background-free electron beam with a temporally coherent bunching by combining soft X-ray self-seeding~\cite{Ratner2015} and inverse tapering schemes~\cite{Schneidmiller2016}. 
This electron beam may then sequentially radiate in two out of four consequent radiators - APPLE-X undulators, set to orthogonal polarizations.

To maximize transverse polarization shaping effect, one should maximize the distance between the active radiators. Once the sample is located between the waists of the radiator images, difference of wavefront curvatures of both polarizations would result in phase difference (hence the resulting polarization) depending on the transverse coordinate, see Figure~\ref{fig:XFEL3}, row (a). With this method polarization may be shaped on down to sub-$\mu m$ scale.

To maximize temporal polarization shaping (similar to that discussed in Section~\ref{sec:fermifemtopol}), one can make use of an electron beam with a linear energy chirp. The passage through the radiators results in a stretching of the beam, which affects its bunching frequency. Thus the consequent raidators would emit pulses with slightly different frequencies. In other words, phase difference between the emitted polarized pulses will change over time. As a result, the resulting combined pulse would be polarized, but its polarization state would change as a function of time. The rate of this polarization shaping may be controlled with electron beam chirp and may reach frequency of THz scale, as illustrated on Figure~\ref{fig:XFEL3}, row (b).\\


\onecolumngrid \newpage \twocolumngrid

\subsection{SHINE and SXFEL}
\label{sec:SHINESXFEL}

\begin{center}
Zhangfeng Gao, Bangjie Deng, Chao Feng, \linebreak
and Haixiao Deng
\end{center}

The Shanghai Deep-Ultraviolet Free-Electron Laser (SDUV-FEL)\cite{zhao2010progress} served as an integrated multi-purpose test facility for FEL principles, laying a solid foundation for the Shanghai Soft X-ray Free-Electron Laser (SXFEL)\cite{zhao2017status} and Shanghai High-Repetition-Rate XFEL and Extreme Light Facility (SHINE)\cite{zhu2017sclf}. Among those FEL experiments carried out at SDUV-FEL, the FEL polarization switching demonstration was successfully performed in 2013, by means of the crossed-planar undulators technique\cite{deng2014polarization}. The first practical success of this technique at a seeded FEL stands a good chance of fast polarization switching of short-wavelength FEL radiation. The principle is based on the superposition of the horizontally and vertically polarized radiation generated from two different types of planar undulators (the afterburner), which are perpendicular to each other. A pulse phase shifter is installed between the crossed planar undulators to delay the electron beam in the vertical planar undulator, thus controlling the phase difference between the horizontally and vertically polarized FEL radiation, which will directly lead to different final polarization states.

\begin{figure}
\includegraphics[width=0.9\linewidth]{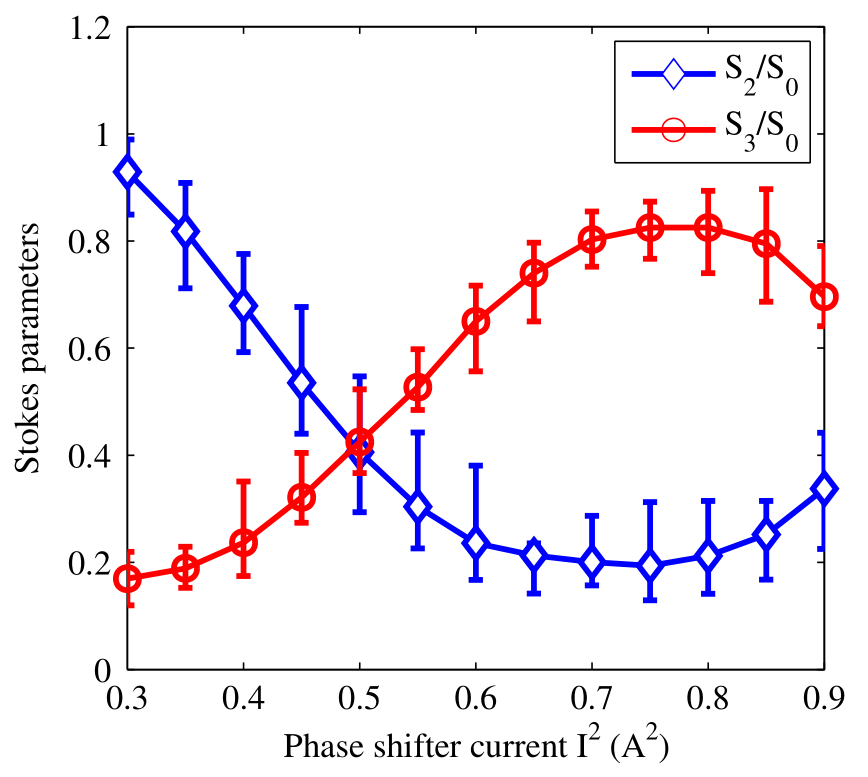}
\caption{The normalized Stokes parameters of coherent 523 nm radiation measured at SDUV-FEL.}
\label{SHINEfig2}
\end{figure}

The 1047 nm seed laser was used in the modulator, and the 523 nm radiation (second harmonic) was generated from the afterburner. The EMU consisted of 10 periods with a period length of 65 mm, while the PMU-H and PMU-V both consisted of 10 periods with the period length of 50 mm. A Division-Of-Amplitude Photo-polarimeter (DOAP), containing one optical lens, three beam splitters, four polarizers, one quarter-wave plate, and four photo-detectors, was constructed to characterize the polarization degree of the FEL radiation. By dividing the incident laser into four separate beams, all the Stokes parameters can be measured at the same time, thus determining the final polarization degree in a single shot\cite{feng2015single}. The Stokes parameters varies with the magnetic field intensity of the phase shifter, just as represented in Fig. \ref{SHINEfig2}, and the maximum polarization degree approaches 90\% as long as the two FEL pulses are perfectly superimposed. The spatial overlap is also highly significant for the final polarization, which is mainly determined by the beam orbit and envelope in the afterburner, as well as the propagation of the FEL pulses between the last undulator and the detector.

\begin{figure}
\includegraphics[width=0.9\linewidth]{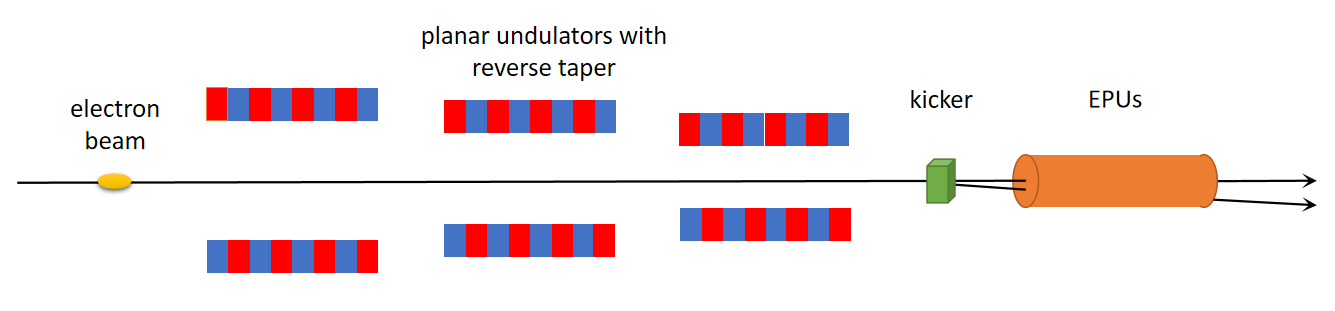}
\caption{Schematic of the polarization control experiment at SXFEL-UF. The kicker serves to bend the prebunched electron beam from the reverse tapered undulators so as to separate the circularly polarized radiation from the linearly polarized radiation.}
\label{SHINEfig3}
\end{figure}

SXFEL is the first soft X-ray FEL facility in China, and the project proceeds in two stages, the test facility (SXFEL-TF) and the user facility (SXFEL-UF). SXFEL-UF consists of two FEL lines, one SASE line and one seeding line, both of which are supposed to generate polarization controllable FEL radiation. Numerous simulations and several experiments on polarization have been performed at the seeding line (under SASE mode)
. Fig. \ref{SHINEfig3} shows the schematic of the polarization control scheme at SXFEL-UF. The upstream planar undulators with a step-gap reverse taper provide well-bunched electron beams with relatively low energy spread, as well as linearly polarized radiation with much lower pulse energy. The kicker serves as a switch to determine the orbit of the electron beams bunch by bunch, which makes it possible to control the position and intensity of the elliptically polarized undulator (EPU) radiation. 

\begin{figure}
\centering
\subfigure[]{ 
\label{fig5.1}
\includegraphics[width=0.45\linewidth]{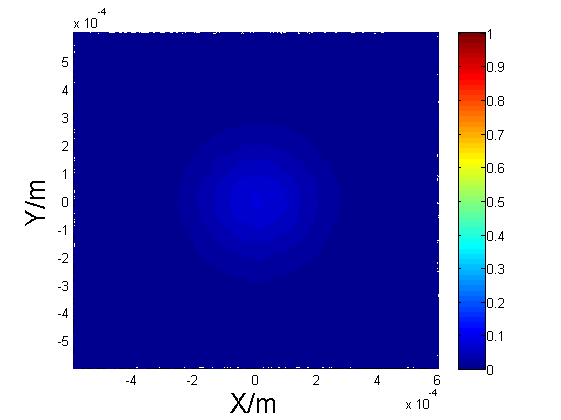}}  
\subfigure[]{
\label{fig5.2}
\includegraphics[width=0.45\linewidth]{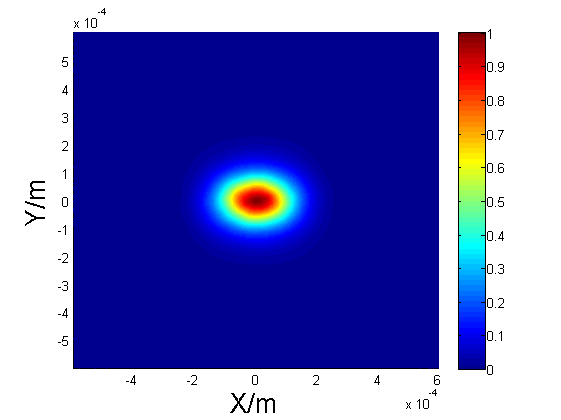}}
\subfigure[]{ 
\label{fig6.1}
\includegraphics[width=0.45\linewidth]{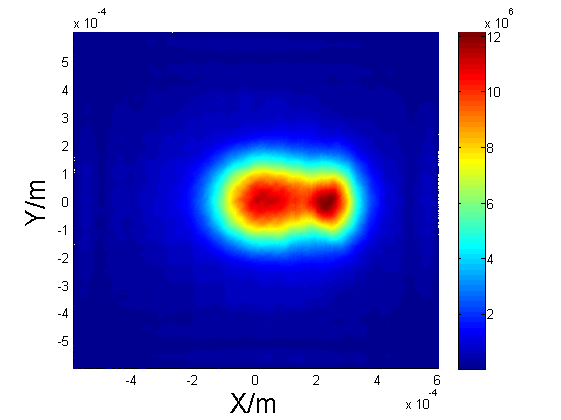}}  \subfigure[]{
\label{fig6.2}
\includegraphics[width=0.45\linewidth]{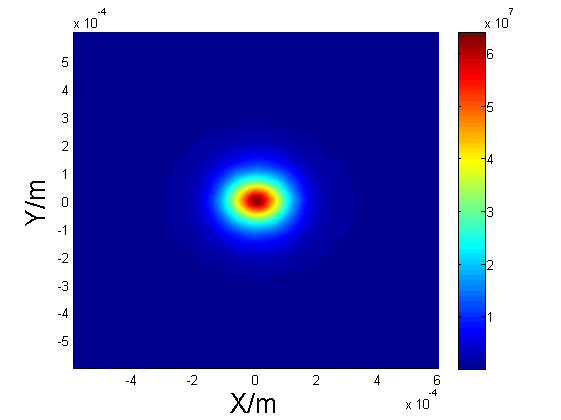}}
\caption{The normalized simulated transverse profiles of linearly (a) and circularly (b) polarized radiation, and the simulated transverse profiles of radiation when the kicker is working (c) or not (d).}
\label{SHINEfig5}
\end{figure}

Due to lack of time for experiments, so far they are carried out with all the undulators tuned to generate 6.75 nm radiation. 8 planar undulators (length of 3 m, period length of 23.5 mm) and 2 EPUs (length of 3 m, period length of 30 mm) are used in the experiments, with a beam energy of 930 MeV. Fig. \ref{SHINEfig5} shows the normalized simulated faculae of linearly and circularly polarized radiation with proper reverse taper. The pulse energy of circularly polarized laser is over one order of magnitude higher than the upstream linearly polarized laser, which means the theoretical polarization degree reaches as high as 95\%. In addition, by adjusting the reverse taper intensity, both linearly and circularly polarized radiation with similar pulse energy can be obtained in simulation at the same time as long as the kicker is working, as shown in Fig. \ref{SHINEfig5}. The experimental results match well with the simulated ones apart from the radiation intensity, which needs more time and attempts to improve. The pulse energy of circularly polarized radiation is around 15 $\mu$J, 7 times higher than the linearly polarized one. Besides, the experimental transverse profiles of radiation is also similar to the simulations when the kicker is on. Moreover, one EPU (length of 4 m, period length of 20 mm) is also going to be installed at the end of the SASE line, providing circularly polarized radiation at a shorter wavelength (2 nm). It is supposed to provide 70 $\mu$J circularly polarized radiation with over 90\% polarization degree.

\begin{figure}
\centering
\subfigure[]{ 
\label{fig7.1}
\includegraphics[width=0.45\linewidth]{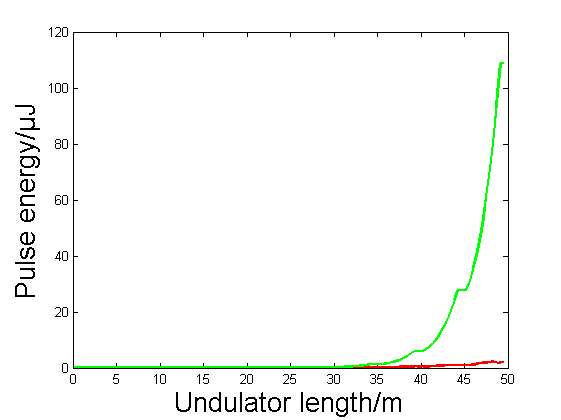}}  \subfigure[]{
\label{fig7.2}
\includegraphics[width=0.45\linewidth]{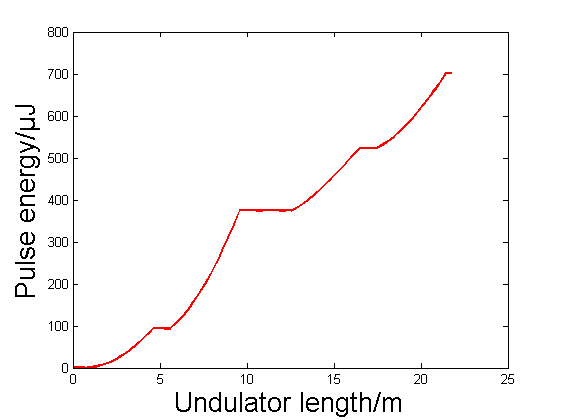}}
\caption{The pulse energy of the linearly polarized radiation with (red) or without (green) reverse taper (a), and the pulse energy of circularly polarized radiation (b).}
\label{SHINEfig7}
\end{figure}

SHINE is the first hard X-ray FEL facility in China, and it is still under construction. So far SHINE consists of 3 FEL lines, in which FEL-II covers the lowest photon energy (down to soft X-ray region) and contains 4 EPUs for polarization control. Numerous simulations have been done on the performance of the polarization control scheme, in which the 1 nm radiation case will be taken as an example. With a beam energy of 8 GeV, the period length of planar undulators is supposed to be 55 mm, the same as that of the EPUs. Fig. \ref{SHINEfig7} shows the pulse energy of the linearly polarized radiation, depending on whether a reverse taper is used, and nearly two orders of magnitude difference in radiation intensity can be observed. Since the pulse energy of circularly polarized radiation is as high as 700 $\mu$J, the theoretical polarization degree is calculated to be over 99\%. Moreover, FEL-III is expected to generate linearly polarized laser with different polarization directions, which still needs more theoretic and simulation supports.

Currently, inspired by the 'CookieBox' instrument developed at DESY\cite{Ilchen2014, hartmann2018attosecond} , a series of angularly resolving polarimeters (called 'ARPolar' instruments) based on measuring photoelectron angular distributionsare proposed, constructed or being offline tested at SXFEL and SHINE. The ARPolar instrument contains 16 $mu$-metal-shielded detection assemblies equipped with a electrostatic optical system and fast MCP detectors for both FELs' energy-spectrum monitoring and the polarity degree \& angle measurement. The former is based on photo-ionized electrons' time-of-flight and the latter by means of measuring their angular distribution. All ARPolar instruments are designed to achieve a linear polarity resolution better than 2\% for fully polarized XFELs. For soft X-ray FEL polarization diagnosis, two machines (ARPolar-CORE and ARPolar-SXFEL) are scheduled for the SXFEL facility. The prototype (ARPolar-CORE) is manufactured firstly for offline experimental validation of mechanics, vacuum and electronics system\cite{Liu2022}, and it is featured by the capability of high-resolution multi-channel charge measurement with MHz repetition rates based on the non-gated charge-to-time conversion principles using low-cost time-to-amplitude converters (TDCs)\cite{Liu2022}. The ARPolar-CORE instrument  will be installed   at the seeding line of SXFEL facility.  Meanwhile, the ARPolar-SXFEL instrument is supposed to be the successor of the ARPolar-CORE and featured by its design of error compensation because of FEL pulses' transverse positions for polarization diagnosis\cite{liu2021}. It is currently installed at the SASE line of SXFEL facility. For hard X-ray FEL pulses, the ARPolar-SHINE instrument is designed for the commission of polarization diagnosis on the FEL-II undulator line at the SHINE facility, covering an energy range from 0.4 keV to 3 keV.



\onecolumngrid \newpage \twocolumngrid

\section{Experimental Perspectives for Polarization-Controlled FELs}

\subsection{Instrumentation}
\label{sec:InstrumDiag}

\begin{center}
Markus Ilchen, Wolfram Helml, and Peter Walter
\end{center}

Free-electron laser facilities provide a unique variety of instrumentation for covering the interdisciplinary demands from diverse user communities. They typically offer a versatile and well-established instrument portfolio for users  that is optimized and calibrated for the specific experimental endstations. The advantage for new users can be that, typically, an expert team of scientists and technical staff aids them in all steps of planning, experiment conduction, and even data analysis. Thus, externally provided equipment by users is not the common case but can also be adapted for specific experiments. In this section, we will sketch a selection of relevant instrumentation for investigations of stereochemistry and dichroic light-matter interaction at FELs. This selection aims to give an orientation for planning of new experimental campaigns within the roadmap's topic, stimulate collaborations and attract users to join the effort of gas-phase-based studies with polarization-controlled FELs.

For the initial demonstration of a stereochemically sensitive experimental scheme at an XFEL as a step towards time-resolved and nonlinear investigations of building blocks of life, the method of monitoring the forward-backward asymmetry of photoelectrons (PECD) \cite{ritchie1975theoretical, bowering2001asymmetry} has been identified to yield large potential \cite{Beaulieu2016, ilchen2017emitter, ilchen2021site, PhysRevX.13.011044} (see also Section \ref{sec:PECD} and \ref{sec:PECD2}). 
Photoelectron yields and their angular distributions have furthermore been used to investigate nonlinearly dichroic effects in atoms and molecules, as outlined in Section \ref{sec:NLCD}. Another recently emerged perspective is realized via exploiting the interplay of photonic spin and orbital angular momentum as indicated in Section \ref{sec:FERMI} and further detailed in Section \ref{sec:OAM}. With the aim to derive highly differential information about specific structures and processes, schemes like (partial) covariance analysis \cite{allum2022localized} and coincidence spectroscopy \cite{Dorner2000, Ullrich2003} have been demonstrated to yield promising potential, also and in particular for investigations with polarization control (see Section \ref{sec:COLTRIMS}). 
The latter technique is even offering complementary access to absolute configurations of molecular structures via Coulomb explosion imaging \cite{Pitzer2013}.
Currently, it is under development to complement or even combine these techniques with fluorescence spectroscopy \cite{Knie2014}, coherent-diffraction imaging \cite{Chapman2006, Chapman2011}, and X-ray scattering spectroscopy \cite{Weninger2013}. 
Some of the commonly employed instruments for gas-phase studies with polarization-controlled FELs are: \\

\paragraph{Velocity Map Imaging (VMI)}\

The VMI technique images a two-dimensional projection of the full three-dimensional velocity distribution of charged particles, here originated via photoionization, which provides both angular and velocity information over a 4$\pi$ collection angle. It is based on the technique of ion imaging, introduced by Chandler and Huston \cite{Chandler1987}, with the addition of the 2D imaging capability of a standard time-of-flight spectrometer with homogeneous fields, initially reported by Eppink and Parker \cite{Eppink1997}. Crucially, the latter authors have introduced an Einzel lens into the setup. Such a lens focuses charged particles with similar velocities to a distinct location on a position-sensitive detector such as a micro-channel-plate assembly. The inherent 'blurring' due to large interaction regions can be mitigated via a finite molecular source and/or ionizing beam \cite{parker_eppink_2003, Wester2014}. The readout of the MCPs can be used for time-of-flight spectroscopy, or realized via phosphor screens that convert the charge-avalanche of the MCPs to visible light dots or via delay-line anodes \cite{Lampton1987, Hanold1999, Debrah2020}. In the first case, cameras can then record images of either accumulated 'hits' or in the case of FELs, record single-shot spectra, even in a time-resolved fashion \cite{DR_FLASH2018} using fast-decaying phosphor screens. A substantial limitation is the readout speed of the camera which poses a challenge to high-repetition-rate single-shot operation, which is an important development at FELs as discussed in Section \ref{sec:MachinePart}.
\begin{figure}
\includegraphics[width=1\linewidth]{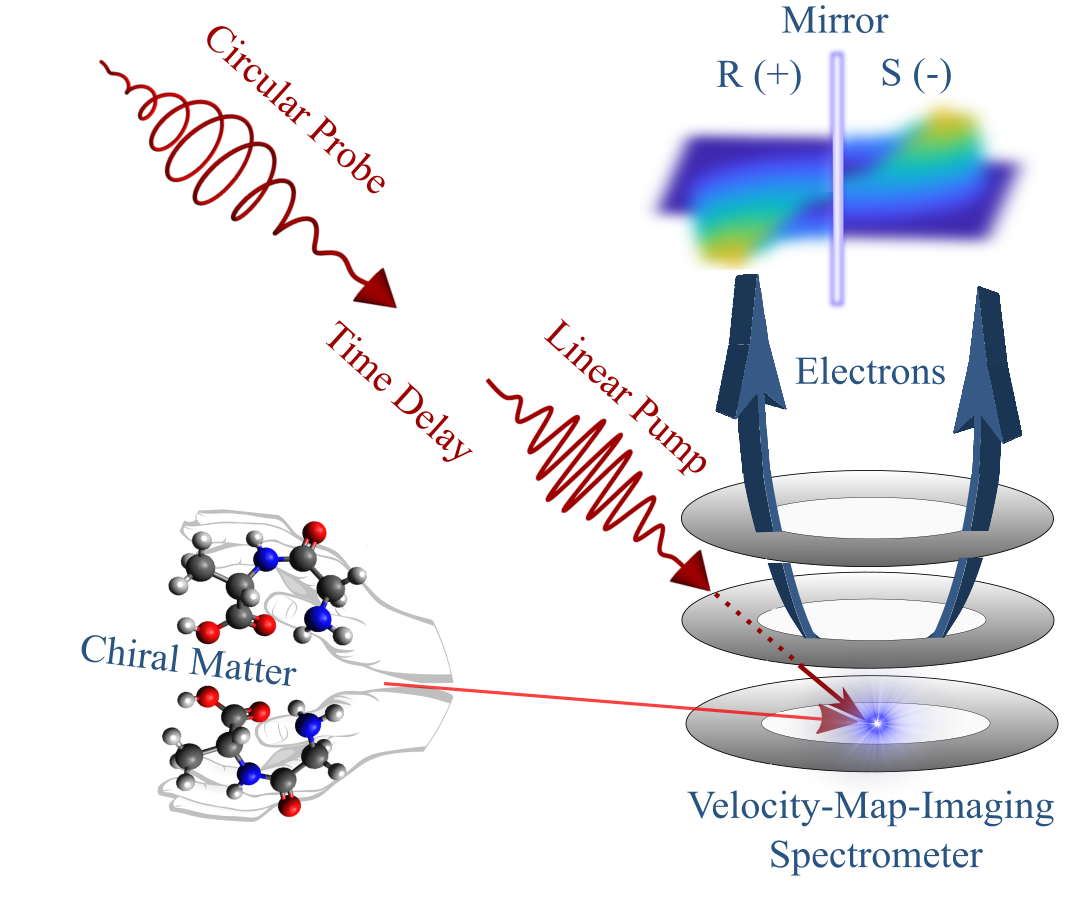}
\caption{Adapted figure from \cite{ilchen2021site} illustrating a possible experimental scheme for pump-probe investigations of chiral systems (here glycyl-l-alanine) by VMI spectroscopy.}
\label{fig:VMI}
\end{figure}
In general, VMIs have enabled new approaches to ion- and photoelectron spectroscopy for molecular photodissociation, collision and reaction-dynamics investigations \cite{Wester2014}, attosecond spectroscopy and stereochemistry, to name a few topically relevant fields of interest. While historically, charged particle imaging has been applied separately for the detection of positive ions and electrons, another important extension is to image the ions and electrons in coincidence or covariance in order to isolate particles from specific ionization channels. Nowadays, such double-sided VMIs are state-of-the-art setups at XFELs \cite{Osipov2018, Erk2018, FERMI_VMI}. 
It is notable that a so called coaxial VMI, where the beam passes through the detector in order to enable a perpendicular projection axis to the standard case, has been used at LCLS for retrieving temporal X-ray pulse characteristics via angular streaking \cite{coaxVMI}.\\

\paragraph{COLTRIMS/REMI}\

A conceptually similar instrumental setup to VMIs can be used as Cold Target Recoil Ion Momentum Spectrometer or COLTRIMS \cite{Dorner2000, Ullrich2003, Jahnke2004} as will be discussed in detail in Section \ref{sec:COLTRIMS} of D\"orner and Jahnke. To give a peek into this powerful technique, we briefly introduce the technique here. One of the decisive differences between a VMI and the COLTRIMS approach is its capability of imaging the full 3-dim. momentum space instead of only a 2-dim. projection. In addition, COLTRIMS is intrinsically a coincidence measurement scheme. 
These advantages of full coincidence come at the price that single-molecule conditions must be met during the measurement, i.e., only one atom or molecule may interact with the ionizing projectile beam inside the interaction volume at a time. This implies that one of the most significant challenges lies in maintaining stringent vacuum conditions and employing a highly dilute molecular jet as the method for sample delivery. The COLTRIMS technique has been developed to give access to kinematically complete experiments on atomic and molecular-fragmentation processes by coincident and momentum-resolved detection of recoiling target ions and emitted electrons, and has been shown at XFELs \cite{Kastirke2020, Li2021}. Through the momentum vectors of all emitted electrons and ionic fragments in the same event the entangled many-particle dynamics in the molecular system can be explored via Coulomb-explosion imaging, which, at FELs, allows to image single molecules in the gas phase \cite{Jahnke2021PRX, PhysRevX.10.021052}. Where it recently has been shown at European XFEL that a molecule size limitation can be advanced to larger systems via the absorption of many photons before substantial molecular restructuring can occur \cite{boll2022x}.\\

\paragraph{Time-of-flight (TOF) spectroscopy}\

Time-of-Flight (ToF) spectroscopy converts 'flight' times of charged particles from an ionization event in the interaction region between the investigated target and the photon pulse, to kinetic energies via their simple physical relation. For the common case of using microchannel plates (MCPs) as fast-response detectors, the electron and ion-mass spectra can be generated in two ways; either by counting the events and accumulating a histogram over many pulses or by recording the current through the MCPs. The latter is a prerequisite for some diagnostic methods at FELs, as outlined below, and also decisively extends the methodology for FEL applications. Notably, such detector operation poses a variety of new challenges that are under development but also yields large potential due to the unique capability of accommodating single-shot spectroscopy up to high repetition rates from kHz to MHz, meeting the full range of currently available machine operation modes \cite{Walter2021}.
ToF-spectrometers can be tailored in a way that charged-particle optics, i.e. electrostatic sections with independently adjustable potentials, decelerate the particles for retrieving maximum energy resolution also for high kinetic energies. These lenses can also be used for acceleration and shaping of the trajectories for increasing the collection efficiency. 
Electron ToF-spectroscopy can provide relatively high energy resolution, moderate collection efficiency and/or angle-resolved spectroscopic insights into a broad variety of scientific questions at pulsed sources like FELs, also in different sample environments \cite{Ilchen2014, lutman2016polarization, ilchen2018symmetry, Walter2021, de2022high}. Collection efficiency and energy resolution are in a trade-off balance and can be adapted to specific experimental needs. 
It is noteworthy that angle-resolving time-of-flight spectrometer setups have expanded their application towards FEL-based investigations and diagnostics \cite{Allaria2014,hartmann2016circular, hartmann2019recovery,Walter2021,de2022high} even in non-dipole geometry for retrieving asymmetric angular distribution patterns which are manifested through forward-backward asymmetries and thus potentially causing differentiation issues with PECD signatures \cite{ilchen2018symmetry}.
For experiments requiring high collection efficiency, magnetic-bottle spectrometers can be employed for ions as well as electrons at the cost of angular resolution \cite{Rijs2000, Matsuda2011}. In the light of stereochemistry, such spectrometers can be used for tracking small changes in photoionization yields due to swapping the helicity of polarization.\\

\paragraph{Hemispherical analyzers}\

Another type of electron detector is the so-called hemispherical electron energy analyzer or hemispherical deflection analyzer. It is usually used where very high energy resolution is needed and was also the instrumental choice for the first discovery of a PECD from the liquid phase \cite{Pohl2022}. The analyzer maps kinetic energies to positions on a detector by two concentric conductive hemispheres that serve as electrodes which bend the trajectories of the electrons. This bending separates the different kinetic energies electrons in space. Selecting specific trajectory radii via a narrow slit at one end enables energy resolutions down to fractions of meV. The resulting disadvantage is a directly correlated energy window as well as reduced collection efficiency. 
Performing two independent measurements with left- and right-handed polarized light while keeping the angular orientation, one can for example measure the forward-backward asymmetry of chiral molecules site-specifically with very high resolution \cite{ulrich2008giant, Pohl2022}. The superior energy resolution can furthermore be important for investigations of energetically similar resonances and -fine structures in chiral and also transient matter. 

\newpage

\subsection{Diagnostics}
\label{sec:Diag}

\begin{center}
Markus Ilchen, Lars Funke, Wolfram Helml, and Peter Walter
\end{center}

Accurate and robust diagnostics at FELs need to give immediate feedback in real time for advanced experimental schemes and methodology. Ideally, they provide online, non-invasive characterization of every incoming FEL pulse and cover a broad variety of machine operation parameters. Knowledge about all relevant X-ray parameters including the intensity, spectrum and chirp, and ideally the full time--energy structure of the pulses, in combination with their arrival time, wavefront, and focus size and shape, as well as state and degree of polarization thus enables access to a variety of new scientific areas and otherwise inaccessible states of matter. The available diagnostic details may even crucially influence the choice and optimization of instrumentation for experimental FEL studies. 

In a very contemporary effort, machine-learning approaches have been identified and initially demonstrated to add new perspectives to the growing toolbox of photon metrology \cite{sanchez2017accurate}. They thrive towards enabling active experimentation, i.e. actively steering the interplay between diagnostics for photon characterization, instrumentation and machine operation, launching an endeavor that has been gaining increasing attention by the FEL community \cite{dingel2022artificial}.

To date, several of these diagnostic challenges have already been met, providing detailed information about the ultrashort and ultrabright photon pulses of free-electron lasers. Successful operation of beam-position \cite{juranic2018swissfel, tiedtke2009soft, moeller2011photon, zangrando2012photon, grunert2019x, zhao2017status, tono2017overview}, spectral \cite{li2022ghost, zhao2017status, brenner2011first}, and intensity \cite{tiedtke2008gas} diagnostics as well as wavefront \cite{keitel2016hartmann, zangrando2015recent, seaberg2019wavefront, kayser2014wavefront} and focal-spot characterization have been important pillars for machine, beamline, and experiment commissioning at (X)FELs. 

Furthermore, correlation of photon- to electron-beam-based diagnostics has been a vital asset and important cross check, as many derivable properties of the electron bunch are imprinted on the photon pulses as well \cite{craievich2018status, vogt2011status, behrens2014few}.


In the light of the topical orientation of this roadmap, it is noteworthy that the commissioning of undulator-based polarization control for ultrabright and ultrashort pulses was pioneered at FERMI and has spear-headed the effort towards polarization diagnostics at short-wavelength FELs \cite{Allaria2014,ferrari2015single}. Here, a multi-experiment approach was used that combined electron-bunch diagnostics, photoionization of simple atomic targets in the gas phase for non-invasive polarization diagnostics at the experimental endstation and a VUV optical polarimeter in simultaneous operation. In addition, a fluorescence polarimeter at a neighboring beamline was employed for sequential diagnostic validation \cite{Allaria2014}. 

Since the approach of non-invasively measuring the dipole patterns of electron emission from direct photoionization yields information about the degree of linear polarization, it was thus possible to obtain an estimation for the degree of circular polarization under the assumption that the amount of randomly or unpolarized light is negligible. This assumption of absent unpolarized light requires validation via an experimental scheme that can directly access the actual amount of circular polarization. As first demonstration, we have chosen to pursue this validation via measuring the circular dichroism in the formation of sidebands \cite{mazza2014determining}. Here, gaseous helium was ionized with circularly polarized XUV photons and the ionization region was spatially overlapped with a temporally synchronized optical laser of either co-rotating or counter-rotating field vectors. The ionized helium can absorb or emit these quanta of energy that correspond to the energy of an optical laser photon. The yield of the individual sideband’s formation crucially depends on the degree of polarization and provides the required information of an absolute degree of circular polarization that in turn can validate the assumption of absent randomly polarized light. The sketched experiment of Mazza et al. did not only demonstrate the first FEL-based CD in the gas phase, it also gave access to a more complex nonlinear CD formation in an oriented atomic system \cite{ilchen2017circular} as will be discussed in more detail in the dedicated section below (see Section \ref{sec:NLCD}).

The first FEL entering the X-ray regime, i.e. the LCLS at SLAC in the Unites States, capitalized on another approach to provide undulator-based polarization control as sketched in Section \ref{sec:LCLS}. The Delta afterburner undulator was a pioneering approach in several regards. Using an afterburner scheme to produce highly intense pulses of several hundreds of $\mu$J requires a pre-bunched electron beam that has not significantly lased before entering the quadrupole Delta undulator of 3.4 m length. The residual amount of linear polarization that was produced upstream of the Delta undulator was blocked by adjustable jaws, thus only propagating the off-axis circular-polarization beam. As discussed in Section \ref{sec:LCLS} and by Lutman et al. \cite{lutman2016polarization} in detail, the upstream-produced linear polarization can also be used for a variety of pump--probe schemes, which ultimately enabled the very first stereochemically sensitive experiment with a chiral molecule at a short wavelength FEL \cite{ilchen2021site} (see also Section \ref{sec:PECD}). The polarization diagnostics for the commissioning of the Delta undulator, in this case provided in almost real-time, were again realized via angle-resolving photoelectron spectroscopy in non-invasive operation. However, in this scheme, it was even more crucial to validate the methodological approach of dipole-based electron spectroscopy since the approximation of negligible randomly polarized light was less substantiated compared to FERMI, where all undulators are quadrupole-magnet structures. We thus performed the first sideband-based CD experiment with an FEL on a molecule, in this case oxygen, in order to determine the absolute degree of circular polarization \cite{hartmann2016circular}. It is noteworthy that the observed indications of very high degrees of circular polarization from the dipole-geometry diagnostic campaign were supported by the first CD measurement, which, for the LCLS, unexpectedly revealed an achievable degree of polarization close to unity. This joint effort of machine scientists and photon diagnostics uniquely enabled first access of an XFEL to studying nonlinear and transient processes in a chiral molecule with full site specificity \cite{ilchen2021site}, as indicated above.

In order to accommodate non-invasive diagnostics, typically a gaseous target of limited complexity is employed, i.e. atoms or small molecules. More recently, the challenge has risen to not only provide online polarimetry but the whole pulse structure including its time--energy distribution, i.e. the substructure in time together with the FEL pulse spectrum, in addition to the degree of polarization and the pulse energy. For FELs based on the principle of self-amplification of spontaneous emission (SASE), the pulse intensity substructure is given by a sequence of very short spikes, whose number, temporal widths, amplitudes and phases start from shot noise and change stochastically from shot to shot \cite{Krinsky2003}. With these demanding requirements in mind the need for advanced diagnostics and instrumentation becomes obvious.

The baseline hardware for uniting all of these functions in a single apparatus was again provided by the already described angle-resolving photoelectron spectroscopy device mentioned above \cite{Allaria2014}. However, for the inclusion of time--energy structure characterization at a SASE FEL with online feedback capabilities, a polarization controllable optical laser with specifically chosen wavelength adapted to the pulse duration window of interest is required for the implementation of angle-resolved photoelectron streaking \cite{Itatani2002, Kienberger2002, Eckle2008, hartmann2018attosecond}.

Furthermore, a promising perspective for online data analysis in this regard is artificial-intelligence processing for keeping up with the high repetition rates of state-of-the-art FELs in real time \cite{dingel2022artificial}.
In the light of the globally emerging high-repetition rate FELs, not only the associated data load and the demanding reconstruction of SASE pulse substructures pose challenges to non-invasive online diagnostics. A crucial factor is the choice of the employed experimental setup since single-shot spectroscopy at high-repetition rates is still an elusive development. MCP-based electron ToF-spectroscopy has shown promising potential to comfortably enter the kHz regime. However, burst-mode schemes with 10 Hz bunch trains and intra-train spacings down to the order of 200 ns like at European XFEL and FLASH as well as the planned continuous MHz operation at LCLS II in the USA and SHINE in China require new developments in order to meet the correspondingly high replenishing rates for the MCP detectors as well as accommodating available readout schemes.

For the currently evolving XFEL-based measurements with attosecond resolution, not only the relative arrival time between the pump and the probe pulse, but also the temporal structure of the pulses becomes important, especially for the stochastic SASE pulses, which are strongly modulated on the (sub-)femtosecond time scale, depending on a variety of parameters such as the photon energy. Importantly, the full reconstruction of FEL pulses is crucial for a variety of experimental schemes and scientific investigations.

In the angular streaking scheme, a similar detection geometry can be used as for the measurement of the polarization state described in \cite{mazza2014determining, Allaria2014, hartmann2018attosecond}. The X-rays ionize a gaseous target and promote photoelectrons from the bound state to the continuum with a given excess kinetic energy. In our previous efforts \cite{hartmann2018attosecond}, the photoelectrons were measured by 16 dipole-encircling ToF-spectrometers, but complementary schemes have been developed as well at LCLS \cite{li2018characterizing}. In the case of angular streaking experiments, the copropagating IR laser is circularly polarized, in contrast to standard attosecond-streaking setups with linear streaking laser polarization. The effect of the IR laser interacting with the emitted electrons can be understood classically as a driving force exerted by its instantaneous electric field, thus changing the electrons’ momenta dependent on the moment when the electrons are set free. This momentum change translates to an angularly varying electron kinetic energy in the presence of a circularly polarized streaking field. The direction of the IR polarization vector determines the coordinates along which electrons, ionized at this specific time, are losing or gaining energy on their way to the detector. Thus, the complete distribution of electron energies in the dipole plane over the whole FEL pulse mimics the duration of the pulse shape, while the relative electron count rates per energy bin of the measured traces correspond to the intensity structure of the pulse.

We have successfully shown at the LCLS \cite{hartmann2018attosecond} that with this scheme it is possible to measure the full time--energy structure of SASE FEL pulses in a single shot down to, currently, the few-hundred-attosecond level. Following this, we implemented the angular streaking technique at the European XFEL and combined the measurement with an online analysis, built on pre-trained machine-learning networks for characterizing various important FEL pulse parameters on the fly \cite{dingel2022artificial}. The employed algorithms demonstrated the promising potential of ML methods to cope with complex pulse characterization schemes at (non-continuous) MHz rate.

\begin{figure}
\includegraphics[width=1\linewidth]{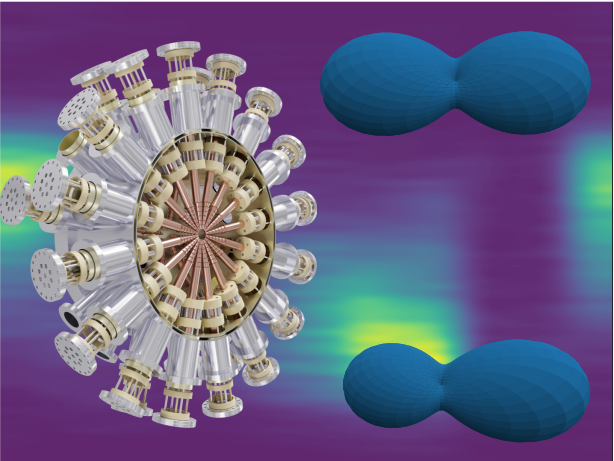}
\caption{Illustration of polarization characterization from streaked electron spectra. Left: CAD drawing of a new dual-interaction-plane time-of-flight apparatus for atto-streaking. The background shows angle-resolved data from the first angular streaking campaign at European XFEL from June 2022. The insets on the right illustrate the 3-dimensional angular distribution patterns for (top) linear FEL polarization of unstreaked electrons versus (bottom) electrons streaked by a circularly polarized laser.}
\label{fig:cpl_streaked}
\end{figure}

We are currently developing a dual-interaction plane chamber called 'Spectrometer with Angular Resolution' 'SpeAR', in which the angular-streaking measurement for FEL pulse characterization can be conducted in line with a second dedicated experiment on a separate target (see Fig. \ref{fig:cpl_streaked}).
The sketched technological advances are not only the key to approaching site-specific electron dynamics in chiral molecules and studying their interplay with nuclear dynamics, they also open the door for combining gas- and liquid-phase experiments. \\

For the successful integration of ultrafast time-resolved and polarization-dependent measurements at the attosecond frontier, we here demonstrate how to simultaneously characterize the FEL polarization via the angular electron intensity distribution in the presence of a circularly polarized streaking field. 
The insets on the right of Fig. \ref{fig:cpl_streaked} show a clearly observable signature of the X-ray polarization imprinted on the measured PAD in the streaked angle-dependent electron intensity distribution. In this case, we compare the effect for linear FEL polarization and circular streaking field with zero (upper right) and relatively strong streaking field (bottom right) and photoelectrons with an maximum anisotropy parameter, i.e. $\beta = 2$. 
When the photoelectrons are subject to the additional field of the circularly polarized streaking laser these angular distributions are changed. The deformed electron distributions can be mathematically reproduced, clearly showing that the deformation in the electron distribution pattern from the original one produced by the linear FEL polarization is the more pronounced the higher the streaking effect is chosen. Thus the effect of streaking can be disentangled from the original PAD by calculating the absolute square of the momentum probability amplitude in the strong field approximation \cite{Kitzler2002}
\begin{equation}
\begin{aligned}
    b(\mathbf{p})\ =\ i\int_{-\infty}^{\infty} &\mathbf{E}(t') \mathbf{d}(\mathbf p - \mathbf A(t')) \\
    &\cdot \exp\left(-i \int_{t'}^{\infty} \frac12 (\mathbf p - \mathbf A(t'') \mathrm d t'' + iI_p t'\right) \mathrm dt'.
\end{aligned}
\end{equation}
Here, $\mathbf E$ is the X-ray electric field, $\mathbf d$ the transition dipole moment, $\mathbf p$ the photoelectron momentum, $\mathbf A$ the streaking laser vector potential and $I_p$ the ponderomotive potential.

A variety of additional promising developments in terms of single-photon-counting and time-resolving cameras, as well as X-ray and fluorescence spectrometers in combination with a variety of methodological and diagnostic novelties (see also below) are set to enable new experimental pathways for a new user community around chirality and dichroism science at FELs. \\
A sketch of the relevant instrumentation and diagnostics in the accelerator tunnels can be found in the respective sections above and references therein.\\
Furthermore, it shall be emphasized that advanced sample-delivery methods (see e.g. Section \ref{sec:ESI}) for gas-phase applications are an important prerequisite for the here focused kind of FEL experiments, and that gas-phase applications will be synergistically complemented by liquid-and solid-phase experiments. A detailed description is, however, not within the scope of this paper.

\onecolumngrid \newpage \twocolumngrid

\subsection{Opportunities for exploring ultrafast und nonlinear stereochemistry} 
\label{sec:PECD}

\begin{center}
Mats Larsson and Vitali Zhaunerchyk    
\end{center}

Ammonia is the textbook example of a molecule that oscillates between two different structures. The nitrogen atom can either be located above the three hydrogen atoms located in a plane, or below the same plane at an equal distance from the plane as when the nitrogen atom is above. From a quantum mechanical point of view all other degrees of freedom (rotation, vibration) can be neglected, and the system can be treated as a two-state system. The two forms correspond to different structural forms, and the barrier for interconversion between them is sufficiently low so that the nitrogen can tunnel between the two structures at a high rate (about $10^{10}$ s$^{-1}$). 

When the molecules get more complicated they can no longer invert themselves either by quantum mechanical tunneling or thermal agitation. As expressed by Phil Anderson in his famous article ``\textit{More is Different}'' \cite{Anderson_1972}: ``\textit{At this point we must forget about the possibility of inversion and ignore the parity symmetry: the symmetry laws have been, not repealed, but broken}''. In fact, chirality was the first example of spontaneous symmetry breaking, later followed by other type of symmetry breaking in both chemistry and physics. Thus, we have been used in thinking about the molecules of life having only one of the two enantiomeric forms; amino acids are L-chiral whereas sugars are D-chiral. On the other hand, if we synthesize sugar by chemical reaction, a racemic mixture with equal amounts of the two enantiomers is obtained. Nature's maintenance of homochirality is surprising since it requires energy. The entropy is lower in a chiral system than in racemic mixtures.

A traditional spectroscopic method to detect chirality is based on circular dichroism (CD). In this method an investigated sample is irradiated with circularly polarized light (CPL) with photons from infrared to the near-ultraviolet regions and chirality is inferred from differences in absorptions of right- and left-handed CPL. It has been suggested that CD can be enhanced by using azimuthally and radially polarized vector beams or by using light with orbital angular momentum \cite{Mukamel_2019,Mukamel_2021}. CD in the X-ray regime allows, in particular, selecting a molecular site being core-excited or -ionized. For example, as was shown in Ref. \cite{Mukamel_2017} for core-resonant CD, the magnitude of the latter depends on the location of the initially excited site and chiral center, \textit{i.e.}, the site-specificity capability inherent to X-rays enables manipulating the strength of CD. The main disadvantage of the CD method is that CD signals are rather weak. Photoelectron CD (PECD) is another approach to sensing chirality and it is usually few orders of magnitude more sensitive to chiral configurations than the conventional CD (see also Section \ref{sec:Demekhin}). 

Ever since Louis Pasteur discovered molecular chirality in the 19$^{th}$ century, and it was realized that the origin of chirality is the tetrahedral asymmetric carbon atom, chirality has been viewed to a large extent as a static phenomenon. A molecule is either achiral, and if it is chiral it can occur in only two possible enantiomeric forms. The development of free-electron lasers (FELs) has opened new perspectives for time-resolved PECD (TR-PECD): apart from providing access to the femtosecond time-window, the X-ray FELs offer the element- and site-specificity.

A new generation of experiments to probe ultrafast chiral dynamics is still in its infancy. One of the envisioned experiments investigates the so-called transient chirality. It requires an amplified, ultrashort optical laser to photo-induce some change in the molecule, and an X-ray probe with circularly polarized light. The optical pulse can for example induce a transition in a molecule from an achiral ground state to a chiral excited state where the molecules switches from one enantiomeric form to the other on a short time scale. The X-ray can then probe this oscillatory motion as a function of time. For a simple molecule with only a single carbon atom this situation is very similar to the ammonia case, with the exception that the evolving chirality takes place in an excited state and not the ground state. For more complex molecules, one can expect the signals from different cores to be different, with complex dynamics on different timescales. Feasibility of such experiments has theoretically been demonstrated in Ref. \cite{Mukamel_formamide}. The case of formamide was considered as the molecule has a planar achiral structure in the ground state but upon electronic excitation becomes chiral with a low inversion barrier between two enantiomers. The authors showed that if CPL is used as a pump with a pulse duration of tens of fs, an enantiomeric excess is created that bounces back and forth between two enantiomers on a hundred fs-timescale. Such an oscillatory behavior can be probed by fs-short X-ray pulses from FELs with photon energies at and exceeding the C, N and O K-edges.

\begin{figure}
\includegraphics[width=1\linewidth]{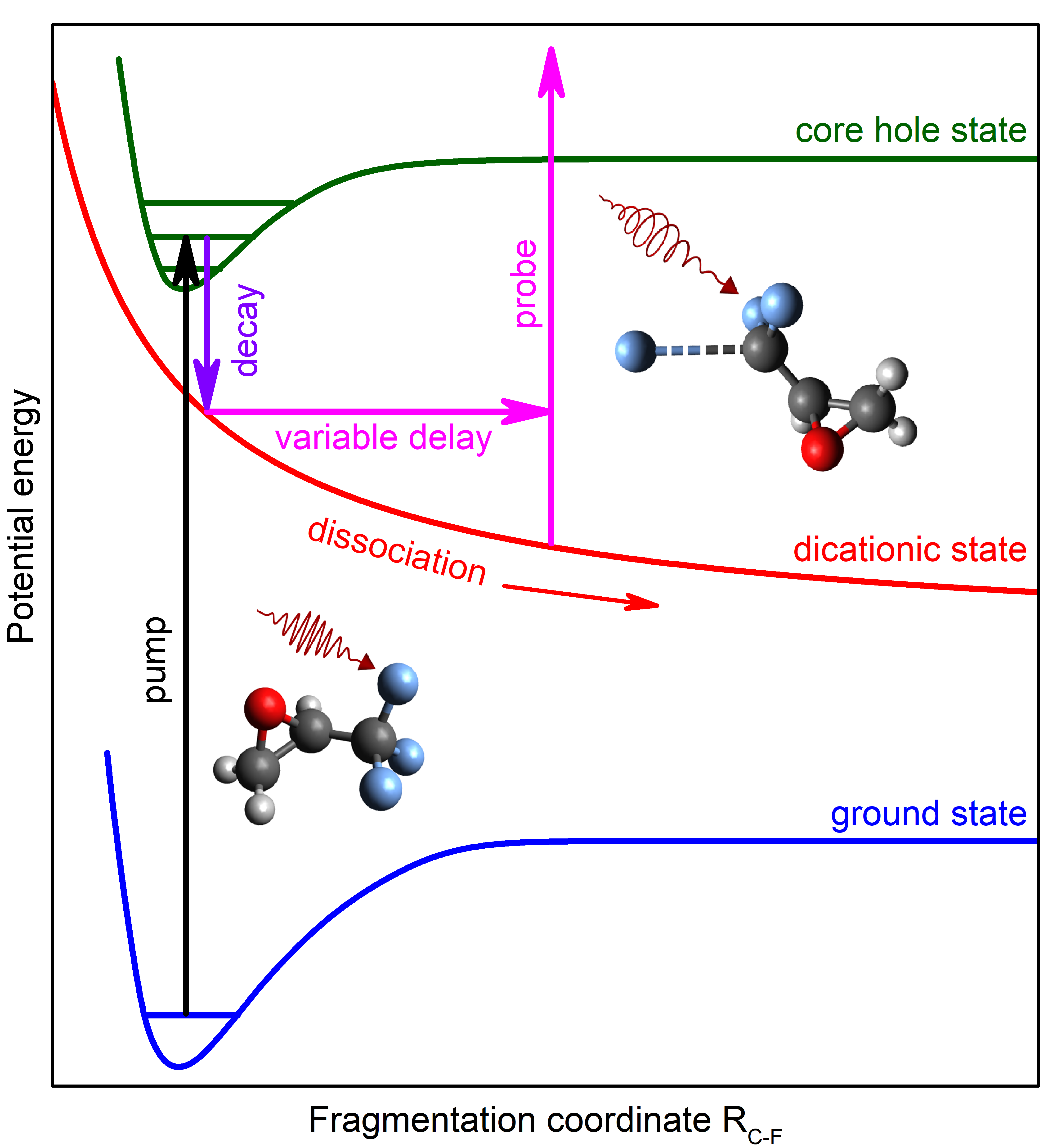}
\caption{A possible pump-probe experiment to investigate ultrafast chiral dynamics. The pump pulse core-excites or -ionizes a specific molecular site leaving the molecule unstable towards dissociation. Chirality of the dissociating molecule is probed with CPL via PECD. The figure has been adapted from \cite{ilchen2021site}.}
\label{fig:PP}
\end{figure}

Another interesting research topic related to ultrafast chiral dynamics is fragmentation of chiral molecules into achiral fragments or, vice versa, fragmentation of achiral molecules into chiral fragments. Such experimental studies will bring insight into the question whether transition from chirality to achirality is similar to a ``quantum jump'' in quantum mechanics (\textit{i.e.}, a process that can only be described as ``before'' or ``after'') or it is a continuous process for which the degree of chirality can be quantified as a function of time (see also Section \ref{sec:Demekhin}). In these envisioned experiments photolysis can be initiated by an ultrafast optical or X-ray pulse, products of which are probed with circularly polarized X-ray photons (Fig. \ref{fig:PP}). Such experiments have been initiated \cite{ilchen2021site} and will benefit from future FEL upgrades, such as repetition rate, pulse duration, seeding, and polarization control.

The access to XFEL pulses of the intrinsic atomic units of time (\textit{i.e.}, attoseconds) will make it possible for the first time to study how the electron dynamics control the nuclear motion. In femtochemistry pump-probe techniques were developed to directly probe the nuclear motion, while the electron dynamics was inaccessible. With isolated X-ray pulses with a duration in the attoscond regime, one can perform pump-probe experiment where the pump laser is in the UV/VIS part of the spectrum and with a duration of a few femtoseconds, whereas the probe pulse is an X-ray attosecond pulse. The X-ray pulse will ionize the molecule, with the electrons carrying information about what happened during the pump phase. 
FELs generating circularly polarized attosecond X-ray pulses will open the door to locally sensing electron dynamics in chiral molecules on the attosecond-timescale. It can in particular be applied to study whether the electron dynamics in the L- and D-forms of chiral molecules is the same, or whether there are differences owing to the handedness. It can in particular be applied to study how the electronic structure and its dynamics in the L- and D-forms of chiral molecules differ owing to the structural handedness and how this stereochemically relates to the role of the Coulomb potential.



\onecolumngrid \newpage \twocolumngrid

\subsection{Nonlinear circular dichroism studies} 
\label{sec:NLCD}

\begin{center}
Michael Meyer and Tommaso Mazza
\end{center}

Circular dichroism (CD) in photoionization is defined as the difference in electron emission probability after the interaction with right and left circularly polarized light. In one-photon processes, a non-zero CD is given, if the target itself has non-zero chiral properties, since unpolarized atoms or non-chiral molecules will show a CD equal to zero. However, in two-color pump-probe experiments it was demonstrated that polarized target states can be prepared by the first color photon, so that a non-zero CD can be measured using the second color photon \cite{OKeeffe04,Aloise05, Meyer11}. These studies were performed in the linear intensity regime using synchrotron radiation in combination with optical lasers and the intermediate polarized state was prepared in general by resonant excitation to increase the efficiency of the two-photon process. 

Similarly, the CD from non-oriented targets is generally also zero in the non-linear regime, when the ionization is driven by several identical photons \cite{lambro72}. Non-zero CD is only observed by realizing a two-color multi-photon ionization \cite{kazan11}. Here, a circularly polarized photon of one color orients the systems via single- or multi-photon interaction and causes thereby a different response to the photon of different color for right-and left-handed circular polarization \cite{kaba96}.

The prototypical two-color non-linear photoemission (ionization) process is laser assisted photoemission or two-color above-threshold ionization (ATI) \cite{glover96}. Experimentally this process is observed by the appearance of so-called “sidebands”, which show up in the photoelectron spectrum on both sides of the main photoemission line and which can be interpreted as additional absorption or emission of one or more optical photons. This interaction scheme allows accessing detailed information on the electron dynamics of small quantum systems on the femtosecond time scale, exploiting the modification of the x-ray induced ionization, specifically of the kinetic energy and the angular momentum of the outgoing electron, within a strong optical dressing field \cite{meyer12,mazza15}. . Performing the investigations with well-defined, but changeable linear polarization for the photon sources has demonstrated to provide unique information of the photoionization process \cite{meyer08,mazza15}. The intensities of the two-color ATI lines, which are determined by the relative strengths of the partial electron yields at a given photon energy, change in a characteristic way as a function of the relative orientation between the two linear polarization vectors and show strong modulations as a function of the intensity of the optical laser field.

\begin{figure}
\includegraphics[width=1\linewidth]{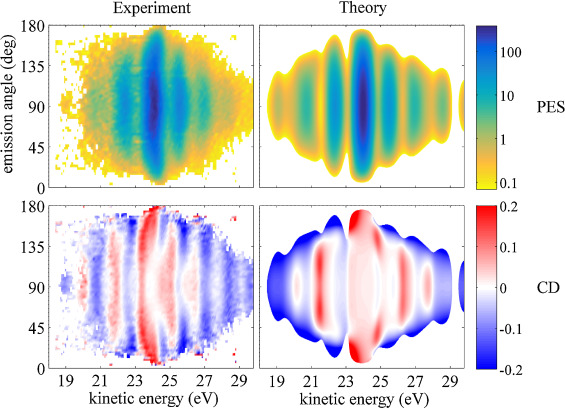}
\caption{Experimental (left) and theoretical (right) angle-resolved photoelectron spectra of atomic helium recorded at an XUV wavelength of 25.6 nm and a peak NIR intensity of 7.2 × 1012 W/cm2 (top) and the corresponding patterns for the angle-resolved CD (bottom). The theoretical spectra are calculated accounting for the spatial distribution of the NIR intensity, volume integrating the different values contributing to the two-color ionization process. The figure has been adapted from \cite{mazza15}}
\label{fig:fig1_MMTM}
\end{figure}

The availability of intense, circularly polarized short-wavelength radiation provided by free electron lasers such as FERMI \cite{Allaria2012}, LCLS \cite{lutman2016polarization} or, in the near future, the European XFEL \cite{LI2017103} has opened recently the possibility to extend these nonlinear studies also to the observation of CD in photoionization \cite{mazza15}. For the two-color ATI process, measurements of the CD enable much deeper insight into the photoionization dynamics, in particular by recording the photoelectron angular distributions (figure \ref{fig:fig1_MMTM}). The anisotropy parameters, which are characterizing the angular distribution change strongly for the different sidebands as a function of the laser intensity \cite{mazza16}. Oscillations between +1.5 and -1.5 were predicted e.g. for the $\beta_2$ parameter of the first sideband for laser intensities in the range of 1 to 10 x 1012 W/cm2. The experimental verification of these strong oscillations has been a major challenge, since effect of volume integration, and therefore the influence of a large range of intensities, has to be considered in the modelling of the observation \cite{mazza15}. 
In general, measurements of the CD in the electron angular distribution have been very important, since they were identified as a possible route for the realization of a so-called “complete” experiment, i.e., the determination of the photoionization amplitudes and their phases \cite{mazza16}. Moreover, CD in two-color ATI has been successfully employed as a metrology tool to determine the circular polarization of XUV light \cite{mazza2014determining, hartmann2016circular}, allowing to quantify the contribution of unpolarized radiation in a circularly polarized beam.

In a pump-probe approach, circularly polarized XUV FEL radiation is also used to resonantly excite the sample. For example, helium ions prepared in the He+(3p, m = +1) oriented state by a sequence of ionization and resonant excitation were probed by non-resonant multiphoton ionization by IR photons co-rotating or counter-rotating respect to the XUV \cite{ilchen2017emitter}. In this study, and in the experimental work that followed it \cite{wagner22}, a strong CD was observed in the nonlinear ionization of the oriented resonance. In addition, a strong dependence of the CD on the probing IR intensity was reported. A CD close to unity, i.e. a dominating ionization pathway for co-rotating pulses, was measured at low intensities. This strong CD was explained by destructive interferences between different ionization pathways in case of counter-propagating pulses, compared to co-rotating pulses where the ionization proceeds dominantly by only one pathway. This interpretation was corroborated by the angular distributions measured for the outgoing electron, which were in perfect agreement to the theoretical simulation of the process. At higher intensities, a strong change leading to the inversion of the sign is observed. This is interpreted  including strong field effects such as the polarization dependent AC stark shift as well as the role of resonant excitation processes \cite{Grum19}. Moreover, Freeman resonances \cite{freeman1987handbook} are shown to dominate the evolution of the intensity dependent CD for this multiphoton process and lead to strong differences in the electron angular distribution for different combinations of left- and right handed circular polarization \cite{wagner22}. The observed CD in these experiments performed at laser intensities up to 10x13 W/cm2, i.e. within the perturbative regime, is in general positive as the consequence of a higher ionization yield for co-rotating pulses. This is different from what was measured at higher intensities (>1014 W/cm2), where the electron emission in the regime of tunnel ionization \cite{eckart2018ultrafast} was dominating for counter-rotating pulses.

\begin{figure}
\includegraphics[width=1\linewidth]{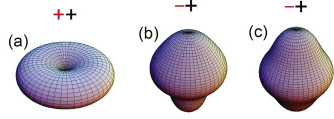}
\caption{PADs of the ATI of the inner 1s shell of neon for a XUV photon energy of 2 keV and an optical (NIR) photon energy of 1.5 eV. The helicity of the beams is indicated on top of each representation. (a) and (b) include the first-order non-dipole corrections, whereas panel (c) shows the PAD under the dipole approximation. The figure has been adapted from \cite{mazza15}}
\label{fig:fig2_MMTM}
\end{figure}

XUV radiation brings wavelength selectivity as a most relevant feature, which can and was exploited in the mentioned experiments to prepare a well-defined oriented system either in a bound (autoionization) or a free (continuum) state. In perspective, the extension of this scheme to even shorter wavelengths will open the path to new classes of experiments, namely the investigation of CD in the excitation of core resonances and the study of non-dipole effects in the photoemission. These perspectives are briefly addressed in the following.
An elliptical dichroism has been predicted for the two-photon ionization of the 1s electron in atomic Neon \cite{hofbrucker2018maximum}. The observable is the electron angular distribution, which also shows a strong photon energy dependence. In addition, using intense circularly polarized photons for the pump process, strongly oriented intermediate states of high angular momenta can be populated by multi-photon excitation. As a consequence, a strong CD is expected, but to date not observed, in the excitation of different multiplet states of the core hole, especially considering that the excitation of states with smaller angular momenta than the intermediate state is only possible for counter-rotating, but forbidden for co-rotating pulses. Finally, observing the angular distribution of a two-color ATI signal at high photon energies will reveal the to date unexplored sensitivity of this non-linear two-photon process to non-dipole effects (figure \ref{fig:fig2_MMTM}), which become more and more relevant for increasing momentum of the outgoing photoelectron. 

With different partial electron waves involved in the ionization process, their relative amplitude will also determine the influence of non-dipole contributions. The theoretical treatment of this particular aspect has already been developed \cite{Grum14}, but experimental verification is still lacking, due to the challenging requirement of the simultaneous need for high intensity and high energy resolution at short wavelengths. To this respect, double core hole ionization on atoms has been recently observed at the European XFEL using monochromatized beam with a resolution E/DE $>$ 2000 \cite{mazza23} demonstrating that the fast technological advances will make feasible the realization of these above discussed theoretical predictions.

\onecolumngrid \newpage \twocolumngrid

\subsection{Ultrafast dichroic phenomena explored with COLTRIMS reaction microscopes} 
\label{sec:COLTRIMS}

\begin{center}
Till Jahnke and Reinhard Dörner
\end{center}

Employing X-ray driven photoionization for the determination of structural properties of atoms or molecules has a longstanding history. Initially, these studies targeted mainly the electronic structure, for example, through photoelectron spectroscopy. It was later realized, that in case of molecules the emitted photoelectron contains information on the geometrical structure, as well. This information is encoded in the angular emission distribution of the photoelectron. The photoelectron wave is scattered by the molecular potential yielding distinct interference patterns that are directly connected to the shape of the molecule. This approach was first applied to molecules being absorbed on surfaces in the 70s of the last century (see e.g. \cite{Fadley1984PSS,Woodruff_Bradshaw_1994} for a early reviews). It took until the late 90s until single, isolated molecules in the gas phase were addressed in pioneering works \cite{Golovin1992, Shigemasa1995}. Since then, in particular (coincident) imaging methods have proven to be able to measure these electron interference patterns of molecules in the gas phase and in the following we will focus on so-called COLTRIMS reaction microscopes \cite{Doerner2000, Ullrich2003, Jahnke2004} as devices for such studies.

In reaction microscopy, a well-localized, typically internally cold supersonic gas jet is crossed at right angle with an ionizing beam of photons (or other projectiles). As a reaction occurs, the charged reaction-fragments (electrons and ions) are then guided by weak electric and magnetic fields towards two time- and position-sensitive detectors. By measuring the flight time and the position of impact of each fragment in coincidences, the initial momentum vector of the detected particle can be reconstructed in an offline analysis. As momenta are measured, all derived quantities as kinetic energies and (laboratory-frame) emission angles are obtained, as well, and the coincidence measurement allows to explore relative quantities of several fragments (e.g. relative emission angles) in addition.

\begin{figure}
\includegraphics[width=0.8\linewidth]{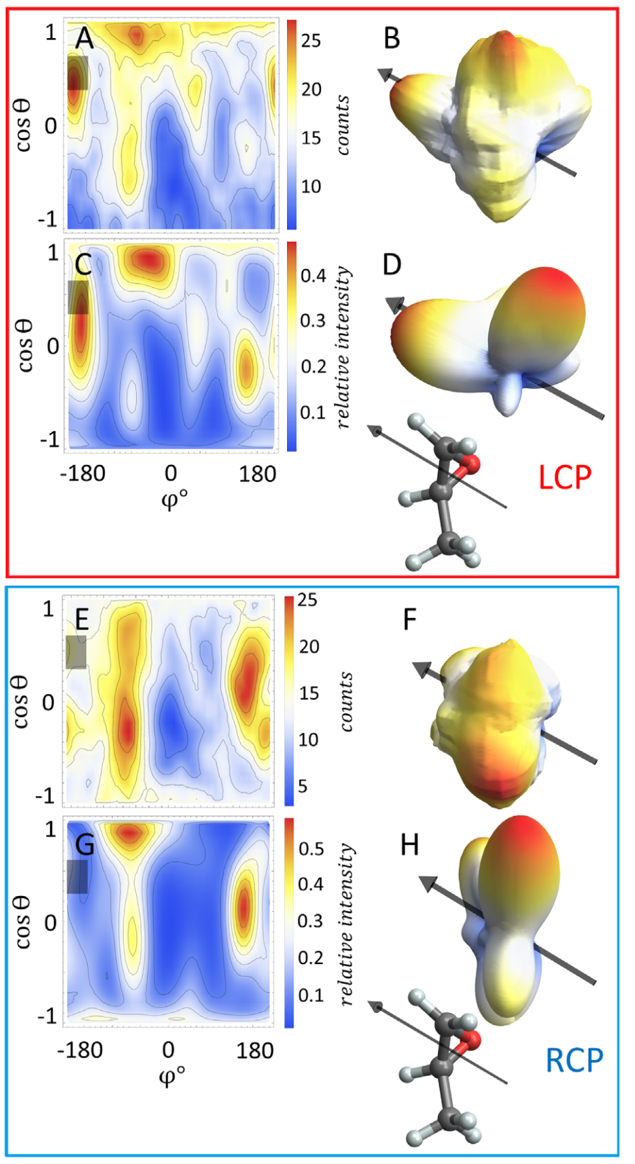}
\caption{Examples of measured (A,B,E,F) and computed (C,D,G,H) 3-dim. molecular-frame electron angular distributions using left circularly and right circularly polarized photons (top and bottom, respectively) for the ejection of the electron. The figure has been adapted and reprinted with permission from\cite{Fehre21}.}
\label{fig:MFPAD}
\end{figure}

If (soft) X-rays are employed for a photoelectron diffraction measurement, the photoelectron wave is typically launched at a well localized site inside the molecule and the molecule is ''illuminated from within'' \cite{Landers01}. As the diffraction occurs in the body-fixed frame of the molecule, the random orientation of molecules in the gas phase poses a first experimental challenge. Accordingly, the molecules can either be actively aligned using IR-laser pulses \cite{Stapelfeldt2003RMP}, or the molecular orientation at the instant of ionization is deduced from a coincident measurement of ionic fragments that are ejected. The latter approach works if this fragmentation is rapid  \cite{Zare72}, which is often the case, if it is triggered, e.g, due to secondary ionization processes such as Auger decay or Auger cascades following the inner-shell photoionization. The momentum vector of the electrons, that are measured in coincidence with the momentum vector of the ionic fragments, can then be transformed into the molecular-frame yielding the 3-dim. photoelectron diffraction pattern. Examples of such patterns are shown in Fig. \ref{fig:MFPAD}, which has been taken from \cite{Fehre21}. In the context of this roadmap article, the following point is particularly noteworthy: As the photoelectron diffraction is sensitive to the molecular structure it is also sensitive to the handedness of this structure, i.e., it is enantio-sensitive. In addition, if the photoelectron wave is launched by circularly polarized photons this enantio-sensitivity is typically enhanced, since the photon imprints an angle-dependent phase onto the electron wave during its birth. It is the combination of the scattering by the 3-dim. molecular potential and this angle-dependent phase, that causes finally the photoelectron circular dichroism (PECD) introduced already in Section \ref{sec:InstrumDiag} of this article. PECD is a light-helicity dependent forward/backward asymmetry of the photoelectron flux, that survives even an integration over molecular orientation. This property makes it directly accessible by means of angle-resolved electron spectroscopy. In most cases, however, (when ejecting the electron from an achiral orbital), the scattering of the electron wave by the molecular potential is the mechanism underlying PECD. Accordingly, the strength of the PECD-signal increases already drastically if the molecular orientation is only partially fixed in the laboratory-frame \cite{Tia17} and PECD in some cases becomes a 100\% effect if the chiral molecule is fully fixed in space \cite{Fehre21}. A corresponding example is shown in Fig. \ref{Fig_4D-PECD_MOx} in Section \ref{sec:Demekhin}. There, the heatmap on the left shows the measured PECD, the one one the right the results from theoretical predictions. The molecule is aligned with respect to the photon propagation direction as depicted by the sketches on the right. 

PECD and photoelectron diffraction measured in a COLTRIMS reaction microscope has grown into a much used powerful technique during the last decade and much of the underlying physics of these effects can be considered understood today. With the advent of circularly polarized XFEL radiation, the field will most obviously target the time domain in the near future and time-resolved studies of transient chiral structures are a first target of choice. Related time-resolved (pump-probe) studies using linearly polarized XFEL light are already underway. These cannot observe PECD as such for obvious reasons, but yet image the underlying 3-dim. molecular geometry  using photoelectron diffraction or Coulomb explosion imaging \cite{Kastirke2020,boll2022x}. As this roadmap article addresses future opportunities for polarization-controlled free-electron laser light, we want to speculate about two other possible, yet less studied chiral effects. These do not emerge from the chiral molecular structure directly, but may occur due to a (transient) chiral structure in the molecular (or even atomic) electron cloud. 

\begin{figure}[b]
\includegraphics[width=0.7\linewidth]{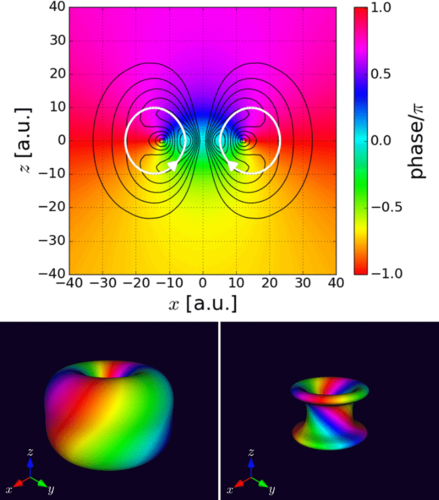}
\caption{Example for a chiral hydrogenic wave function $\chi$ in form of a cut along $y=0$ (top) and isosurfaces for $|\chi|=0.003$~a.u. and $|\chi|=0.005$~a.u., please refer to the original publication for details. The wave function's phase is indicated by the colors. The white circles in the top figure indicate the direction of the probability current in the plane $y=0$. The figure has been adapted and reprinted with permission from\cite{Ordonez2019}.}
\label{fig:ChiralAtom}
\end{figure}

As a first example we refer to Figure \ref{fig:ChiralAtom} which has been taken from \cite{Ordonez2019}. It shows a chiral electron cloud of an excited H atom. In this case of a centrosymmetric potential, this state is formed by the superposition of degenerate higher angular momentum states. In general there are two classes of such states:
those where the square of the wave function, i.e. the probability density, is a chiral object and secondly those where the square of the wave function is achiral, but the phase structure of the wave function makes it a chiral object. As the derivative of the phase of the wave function is related to the flux, this corresponds to a chiral nature of electron flux and thus a chiral  momentum distribution. A most simple example would be a state l=1,m=+1 which has a ring current and is prochiral. Real chiral distributions can be created by excitation of K-shell electrons by circularly polarized photons in a chiral molecule. Most probably, this can also be achieved in achiral molecules if the molecule is fixed in space under distinct angles with respect to the light propagation. Such transient chiral electron distributions are fascinating objects for time-resolved ionization experiments at  XFELs being capable of providing circularly polarized light. At low photon energies these have been studied already in pioneering work (where the process has been termed  ''photoelectron  excitation circular dichroism'') using high harmonic laser light sources \cite{Beaulieu2018}. We envision two tools to interrogate such chiral electronic wave functions. The first is direct photoionization of the orbital carrying the chiral flux. This is best done with high energy photons. When the photoelectron wavelength is short enough, the final continuum state of electron is well described within the Born approximation, which entails that one neglects the scattering of the photoelectron wave by the potential of the molecular ion after its emission. As shown in \cite{Waitz2017NatComm} under these circumstances the photoelectron angular distribution maps directly the initial state momentum distribution of the bound electron. Thus, one can hope to obtain a fingerprint of the chiral phase structure (and thus the chiral current) in angular emission distributions of high energy electrons. Alternatively, one can envision to not eject the chiral electron itself but a different (e.g. K-shell) electron instead. This photoelectron wave then probes the chiral electron cloud, i.e., similar to photoelectron diffraction this will probe the molecular potential created by the excited electron. Such scattering probes the potential which itself is determined by the chiral electron density. Thus, we have two tools at hand which probe different aspects of the chiral electron cloud: scattering of electrons from K-orbitals which senses chiral densities of the cloud, and ejecting the electron from the chiral orbital itself which probes the chiral flux rather than the density.

All of this promises a rich harvest for time-resolved circularly polarized X-ray photoemission studies of transient chiral objects in the molecular frame by coincidence techniques. The time scales of interest spans from the attosecond regime expected for changes in the chiral electron cloud to the femto- to picosecond regime on which the chiral molecular structure evolves.  

\onecolumngrid \newpage \twocolumngrid

\subsection{Ultrafast chirality at FELs: new enantio-sensitive observables, site specific PECD and charge-directed reactivity}
\label{sec:PECD2}

\begin{center}
Francesca Calegari, Olga Smirnova, Caterina Vozzi
\end{center}
Conceptually new \textit{local} approaches to inducing, probing and controlling chirality with light are known as the electric dipole revolution in chiral measurements \cite{Ordonez2018generalized,ayuso2022ultrafast}.  In contrast to traditional methods such as absorption circular dichroism (CD) and optical rotation,  these new methods do not rely on the interaction with the magnetic field component of the light wave but instead use exclusively electric dipole transitions to excite and probe dynamics in electronic, vibronic or rotational states of chiral molecules via detection of emitted photons or photoelectrons.

One of the most advanced in this family of methods is photoelectron circular dichroism (PECD).  Predicted in Refs. \cite{Ritchie,Powis2000JCP}, and first observed in Ref. \cite{bowering2001asymmetry}, PECD constitutes the emergence of enantio-sensitive photo-electron current upon one-photon ionization of chiral molecules by circularly polarized light (see e.g. reviews \cite{Janssen2014PCCP,Nahon2015JESRP}).
Opportunities for ultrafast chiral measurements utilizing PECD emerged after recent works demonstrating few-photon \cite{Lux2012Angewandte,Lehmann2013JCP,Bhargava2013,Fanood2014,Fanood2015,Lux2015,Lux2015JPB,Kastner2016CPC,Miles2017,Kastner2019,Ranecky2022} and strong-field \cite{Beaulieu2016NJP} analogues of PECD using femtosecond pulses, leading to time-resolved PECD \cite{Comby2016JPCL,Beaulieu2016,fabre21}, photo-excitation circular dichroism in bound molecular states (PXCD) \cite{Beaulieu2018}, two-color PECD \cite{Demekhin2018PRL,Demekhin2019PRA,Rozen2019PRX,ordonez2022disentangling}, and even detection of attosecond (1 as=0.001 fs) photoionization delays from chiral molecules \cite{beaulieu2017attosecond}.

FELs with polarization control add site specificity to these measurement protocols and thus enable local imaging and control of chiral molecules, which is highly desirable for applications in chiral analytics but is currently missing. For example, a large chiral molecule, such as a peptide, can contain smaller units, such as amino acids, of different chirality. Resolving the handedness of such units and detecting their spatial structure is an important challenge.    

\paragraph{Geometric magnetism and synthetic chiral light at FELs}\

For two decades PECD stood out as \emph{one-of-a-kind} effect with record enantio-sensitivity in the range of tens of percents, representing a "golden standard" for chiral detection.
Recently it has been shown that PECD is only a "tip of the iceberg":  new classes of observables with similarly high enantio-sensitivity, united by the concept of geometric fields \cite{ordonez2023} have been predicted. What's more, these observables are unique messengers of chiral currents in electronic,  vibrational, rotational or spin degree of freedom, because they only exist if an arrow of time is introduced into the system. One example of such a new observable is enantio-sensitive molecular orientation after photoionization by circularly polarized light (PI-MOCD, molecular orientation circular dichroism by photoionization)\cite{ordonez2023}.

Physically, the chiral molecular geometry together with broken time-reversal symmetry via excitation of currents of any kind leads to the emergence of the Berry curvature -- a geometric field, much like in the quantum Hall effect. As a result, randomly oriented molecules may acquire new properties, dictated by the excitation, such as e.g. ensemble-averaged permanent dipoles proportional to the Berry curvature and oriented in opposite directions in left and right enantiomers.  
This effect opens the opportunity to control the direction and degree of molecular orientation by controlling electronic or vibronic molecular excitations. 

FEL pulses offer an exciting opportunity to induce PI-MOCD in randomly oriented molecules via core excitation with few-femtosecond or attosecond X-ray pulses. In contrast to excitations in valence shells, localised site-specific core excitation could initiate currents from different locations inside the molecule and may result in more accurate and versatile control of the Berry curvature.  This excitation can then be used by a delayed probe pulse with a much lower carrier frequency to achieve enantio-sensitive reactivity directed by the excited charge dynamics.

As a result of the excitation by an ultrashort FEL pump pulse, the left enantiomer of a chiral molecule, when photoionized by a time-delayed circularly polarized probe pulse, orients itself in such a way that the  Berry curvature generated by the pump-excited electronic current points in the positive direction of $\mathbf{k}$, $\mathbf{k}$ being the propagation vector of the ionizing probe pulse. In contrast to the left enantiomer, the right enantiomer orients itself in the exactly opposite direction. Fragmentation of an oriented molecule will lead to enantio-sensitive fragment distributions, such that chemically identical fragments originating from opposite enantiomers will fly in opposite directions of the light propagation axis.  
Thus, we can control \emph{vectorial} properties of molecules with an efficiency comparable to the "golden standard" of the PECD.

The concept of geometric fields also predicts new geometry-driven \emph{scalar} observables, such as photoionization or photoexcitation yields, which are proportional to the Berry phase. None of these geometry-driven observables has been observed so far, but they promise spectacular applications, such as efficient and robust control of enantio-sensitive chemical reactions. The efficiency stems from the electric-dipole nature of these new observables, while robustness may come from their purely geometric origin. For FELs, this opportunity translates into enabling local enantio-sensitive imaging. For example, the challenge of enantio-discriminating small chiral units embedded in a large molecule, such as amino acids attached to peptides, could be addressed by efficient selective excitation of, say, left units.

Looking broadly, the polarization shaping at FELs could be extended towards creating new efficient schemes for chiral low-order non-linear processes such as chiral sum-frequency generation\cite{vogwell2023ultrafast}, non-linear optical rotation\cite{ayuso2021ultrafast} or chiral CARS in electric dipole approximation. Non-collinear or tightly focused FEL beams could be used to create synthetic chiral light \cite{Ayuso2019NatPhot} in the X-ray range. In contrast to standard chiral light, which uses the (non-local) spatial helix of circularly polarized light, synthetic chiral light encodes chirality in the temporal dimension and may allow higher enantio-sensitivity at photoexcitation and X-CARS.

\paragraph{Chiral femtochemistry probed by core-level photoemission}\ 

Studies of ultrafast structural dynamics in chiral molecules can provide insights into fundamental questions in chiral chemical and physical processes, from the role of chirality in chemical reactions to its influence on biological processes. 
The development of ultrafast X-ray FELs enables such studies and opens avenues for understanding the chiral activity of photo-excited states.

In this respect, time-resolved X-ray photoelectron spectroscopy (TR-XPS) is a powerful approach for the investigation of photochemistry in molecules providing information about the local charge distribution with atomic specificity in the electronic ground states. Furthermore, the localization of core orbitals makes this approach site-selective. Since the ionization potential (IP) gives the difference between the neutral state and the core-ionized state for a specific atom in the molecule, the chemical shift of the IP gives information on the charged environment near the specific probed atom. In contrast to standard XPS, TR-XPS enables access to non-equilibrium charge distribution and ultrafast relaxation pathways with chemical selectivity. A recent pioneering experiment performed at Flash2 demonstrated the capability to capture the UV-excited dynamics of 2-thiouracil and deduced charge distribution changes in excited molecules in the framework of a potential model \cite{mayer2022following}. The photoexcitation can trigger a change in chemical shifts, which can be used to distinguish between the contributions of the different atoms of the same species in the molecule, thus enhancing the site specificity of the technique. This approach is indeed relevant for the study of chiral dynamics since the most sensitive technique for the study of these dynamics is time-resolved photoelectron circular dichroism (TR-PECD).


The availability of circularly polarized femtosecond pulses in the soft X-ray range at FELs enabled the extension of TR-PECD to core-level photoemission for chemical-specific and site-selective probing of the chiral electronic structure and its relaxation dynamics. 
Recently, Ilchen et al. reported a chemical-specific investigation of the dissociating trifluoromethyloxirane chiral cation (C$_3$H$_3$F$_3$O$^+$) \cite{ilchen2021site} performed with an X-ray pump and X-ray probe at the LCLS-XFEL at SLAC. They measured PECD at the binding energy corresponding to the fluorine K-edge during the molecular dissociation dynamics, however, the contribution of the three fluorine atoms in this molecule could not be distinguished.

\begin{figure}
\includegraphics[width=0.8\linewidth]{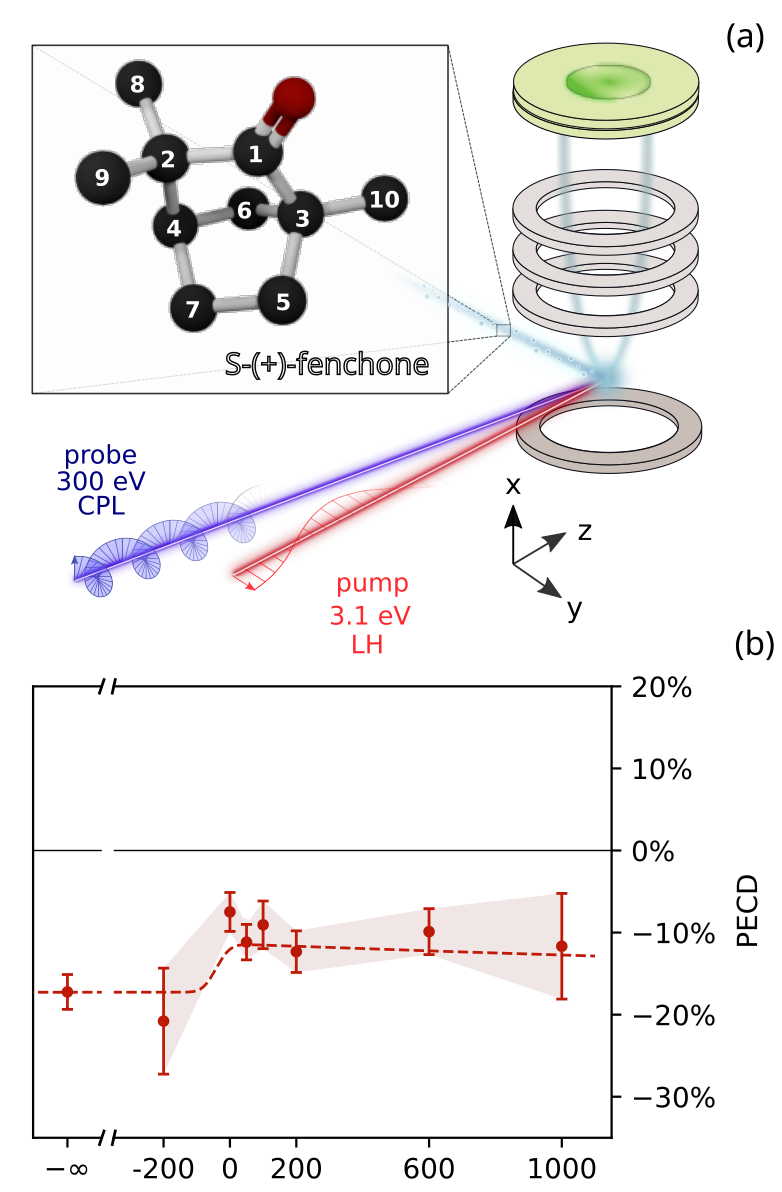}
\caption{(a) Experimental setup for TR-PECD. The circularly polarized XUV probe and the visible pump with linear horizontal polarization are focused quasi-collinearly in the molecular beam of S-(+)-fenchone enantiomers. The inset shows a 3D representation of the S-(+)-fenchone, where only carbon (black) and oxygen (red) atoms are shown, while hydrogen atoms are hidden for the sake of clarity. (b) Measured TR-PECD at 292.53 eV, shown with violet dots along with the error bars. The figure has been adapted from \cite{PhysRevX.13.011044}.}
\label{fig:MFPAD2}
\end{figure}

Another experiment performed at FERMI investigated structural dynamics in fenchone (C$_{10}$H$_{16}$O) induced by a UV pump and probed by core-level TR-PECD at the carbon K-edge\cite{PhysRevX.13.011044}. Core-level probing was used to study the electronic scattering potential of the photo-excited molecules. The delocalized excitation to the 3s Rydberg state has been shown to affect the PECD of the carbon 1s photoelectrons. Thanks to a substantial chemical shift in the carbonyl group and an additional chemical shift caused by the excitation, the experiment was able to distinguish the chiral signal from  C(1s)=O and its two neighbouring carbons from the rest of the carbon atoms by detecting the photoelectron dichroism at different pump-probe delays and photoelectron kinetic energies. A sketch of the experimental setup is shown in fig.\ref{fig:MFPAD2} (a) and the temporal evolution of the measured PECD at 292 eV is shown in fig. \ref{fig:MFPAD2}(b). 

The combination of the experimental results with theoretical calculations was important for understanding the role of the different carbon atoms in the ground state relative to the contribution to the total PECD signal coming from the excited molecule.

These results represent the first step towards chemical, site-specific, enantio-sensitive observations of the electronic and structural dynamics of photoexcited chiral molecules. Such studies will greatly benefit from the availability of high repetition rate x-ray FELs with circularly polarized ultrafast pulses in the future.

\paragraph{Enantioselective charge-directed reactivity}\

Another fascinating route in the ultrafast control of the chiroptical response of molecules is offered by charge migration \cite{cederbaum1999,kuleff2007, kuleff2009} and charge-directed reactivity \cite{weinkauf1997nonstationary,remacle1999, remacle2006}. Charge migration occurs after sudden ionization or excitation of a molecule by a broadband laser pulse and may involve processes driven by electron-electron correlations \cite{cederbaum1999,kuleff2007, kuleff2009}. The superposition of electronic states created by such excitation leads to non-stationary charge density distribution that typically oscillates along the molecular backbone on a time scale from a few tens of attoseconds to a few femtoseconds before the nuclei have had a chance to move. In turn, charge-directed reactivity implies chemical change, i.e. the dynamics of the nuclei following as a result of such attosecond electron dynamics. Charge migration has been theoretically predicted \cite{cederbaum1999, remacle1999, remacle2006, kuleff2007, kuleff2009} and subsequently experimentally demonstrated to occur in aromatic aminoacids \cite{Calegari2014, Lara-Astiaso2018} and nucleobases \cite{Mansson2021} after prompt ionization by an XUV attosecond pulse. Recently, charge migration has been observed in inner-valence ionized molecules exploiting a two-color attosecond time-resolved approach at the LCLS free electron laser \cite{barillot2021}. The role of the nuclear dynamics in the dephasing and the revival of these coherent electronic oscillations has been studied by soft x-ray attosecond transient absorption spectroscopy of strong-field excited molecules \cite{matselyukh2022}. \\
\indent In this context, there has been an increasing interest in understanding the impact of coherent electron dynamics on the chiral response of molecules. Local chirality is a property of both the nuclear and electronic structure. Where the nuclei create a chiral structure, the electron density distribution is expected to follow suit. However, one may ask what happens to the chiroptical response, i.e. the interaction with chiral light, when this charge density distribution is no longer static. Is the circular dichroism (CD) affected by the electronic current originated along the molecular backbone? Is it possible to switch the sign of the CD in the same enantiomer by exploiting its electron dynamics? Preliminary work has been conducted both theoretically and experimentally to address these fundamental questions. 

Enantio-sensitive charge migration and enantio-sensitive charge-directed reactivity are the next challenging milestones, which require  efficient excitation of attosecond electron dynamics in chiral molecules as well as enantio-sensitive detection of its impact on the dynamics of the nuclei. 
The recently predicted PI-MOCD \cite{ordonez2023} resulting upon excitation of non-equilibrium charge distribution in chiral molecules is an example of such protocol. The first step - charge migration in a neutral chiral molecule - has been detected in a very recent  experiment \cite{wanie2023}  preparing a coherent superposition of Rydberg states via broadband UV-light excitation of the chiral molecule methyl-lactate and observing its periodic evolution on a few-femtosecond time scale by TR-PECD. 
Chiral currents observed in this experiment can lead to PI-MOCD:  
molecules whose orientation is such that the current behaves near in-plane and in-phase with the circularly polarized electric field of the ionizing probe pulse will be preferentially ionized, leading to an oriented ensemble of ions, with the left and right enantiomers oriented in opposite directions.  
Detecting the direction of the fragments ejection (forward or backward with respect to the  propagation of the circularly polarized ionizing pulse) provides an enantiosensitive observable. Even more strikingly, by adjusting the delay between the pump pulse preparing the electronic coherences and the probe pulse photoionizing the molecule, one can control the direction of the fragments ejection without changing enantiomer, therefore opening a route towards the implementation of the enantio-sensitive charge directed reactivity.  \

\indent The use of attosecond X-ray Free Electron Lasers offers new avenues for exploring these phenomena. In particular, employing a two-color scheme, charge migration could be selectively activated via single or double core-hole ionization and the resulting chiral currents could be probed with attosecond time resolution, while the Auger-Meitner decay occurs, and with site and stereo selectivity at the k-edge of the molecular chiral center. Circularly polarized attosecond FEL pulses are key for this investigation, since probing or pumping with circularly polarized light would offer the possibility to explore the attosecond TR- PECD or PXCD response, respectively. In the PXCD geometry, the selective chiral excitation of the electronic currents would allow the control of the chemical reactivity of the molecule to be achieved at the electron time-scale.
\clearpage
\onecolumngrid \newpage \twocolumngrid

\subsection{Interplay between photonic spin and orbital angular momentum as a tool for new spectroscopies and for tailoring matter properties at the nanoscale} 
\label{sec:OAM}

\begin{center}
Giovanni De Ninno, Jonas W\"atzel, and Jamal Berakdar
\end{center}

\begin{figure}[h]
    \centering
    \includegraphics[width=0.9\linewidth]{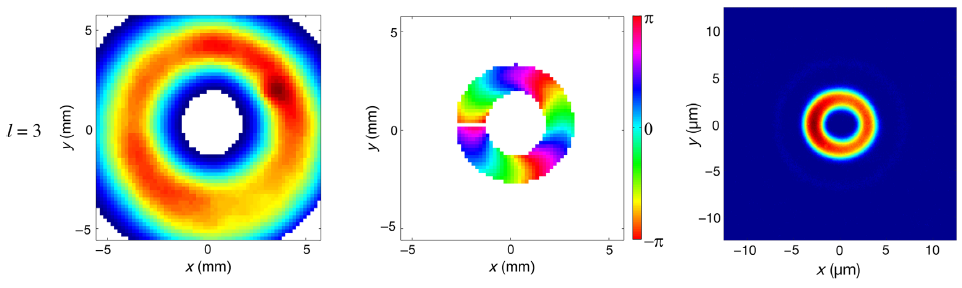}
    \caption{Measured far-field intensity (left column) and phase (middle column), and the intensity distribution in the focal plane (right column) for ZPs with charge l=3. The figure has been adapted from \cite{Ribi2017}.}
    \label{fig:farfield}
\end{figure}

Photons have a fixed spin (SAM) and an unbounded orbital (OAM) angular momentum. While the former is related to light polarization, the latter is manifested in a particular spatial structure of the orbital part of the electric field. The expectation value of the optical OAM with respect to the propagation axis of the light wave is $\hbar\ell$, where $\ell$ is an integer and indicates the number of windings in a wavelength \cite{Allen1992}. The corresponding non-uniform spatial texture of the photonic vector potential can act as a source for additional scattering during light-matter interaction. This may lead to new phenomena such as the transfer of optical OAM to charges as well as the possibility of spatially-resolved control of the light-matter interaction below the optical diffraction limit. The interaction with OAM results in electronic transitions beyond the dipolar propensity rules, as they occur during photoionization. The structured-light induced generation of unidirectional charge current loops with an associated orbital magnetic moment  has been demonstrated \cite{Ninno2020, Waetzel2022, Waetzel2016, Waetzel2020}, as well as the production of skyrmionic defects \cite{Fujiata2017} with potential applications in nanoscale magnetic memory devices. A particularly useful setup is the combination of XUV with IR laser pulses simultaneously acting on matter. The benefit is that by varying the beams’ focusing, the SAM, and the OAM transfer to matter can be spatio-temporally controlled \cite{Ninno2020, Waetzel2022}.  Generally, technical advances enabled new ways for material characterization techniques such as OAM dichroism \cite{Ninno2020, Waetzel2020}, OAM ptychography \cite{PrivComm1} and stimulated emission depletion-like microscopy \cite{Gariepy2014}. Technically, ultra-short, high-intensity XUV or x-ray sources are needed that are  capable of producing vortex beams. In contrast to optical vortex generation in the visible and infrared (IR) spectral regions, where such beams have been successfully used in a wide array of applications ranging from particle manipulation \cite{He1995, Kuga1997} and detection of spinning objects \cite{Lavery2013, phillips2014} to optical communications \cite{wang2012} and super-resolution imaging \cite{jesacher2005}, the generation of intense XUV or x-ray vortices remains a challenging task. Recently, two groups reported the production of XUV beams with tunable topological charge using high-harmonic generation in a wave-mixing setup \cite{Kong2017, gauthier2017}. In Ref. \cite{denoeud2017interaction}, significant experimental progress was made towards plasma-based vortex generation, which shows promise for producing intense harmonics of the driving laser with adjustable OAM. However, the current performance of such tabletop systems in terms of the intensity of the generated XUV light is orders of magnitude lower compared to the new generation of accelerator-based sources such as FELs, and may not be sufficient to perform some of the proposed experiments. Therefore, several schemes have been studied for producing optical vortices at FELs \cite{Hemsing2012, Ribi2014}. 
At the FERMI FEL, two methods are currently used for generating intense XUV beams carrying OAM \cite{Ribi2017}. The first one takes advantage of the vortex nature of harmonics emitted in a helical undulator \cite{Katoh2017}, allowing to obtain intense, femtosecond, coherent XUV vortices.  The second technique relies on the use of aspiral zone plate, which is placed directly into the FEL beam path. 
The setup produces a focused, micron-sized, high-intensity optical vortex without requiring extensive modifications of the FEL beamline, see Fig. \ref{fig:farfield}

\begin{figure}[h]
    \centering
    \includegraphics[width=0.9\linewidth]{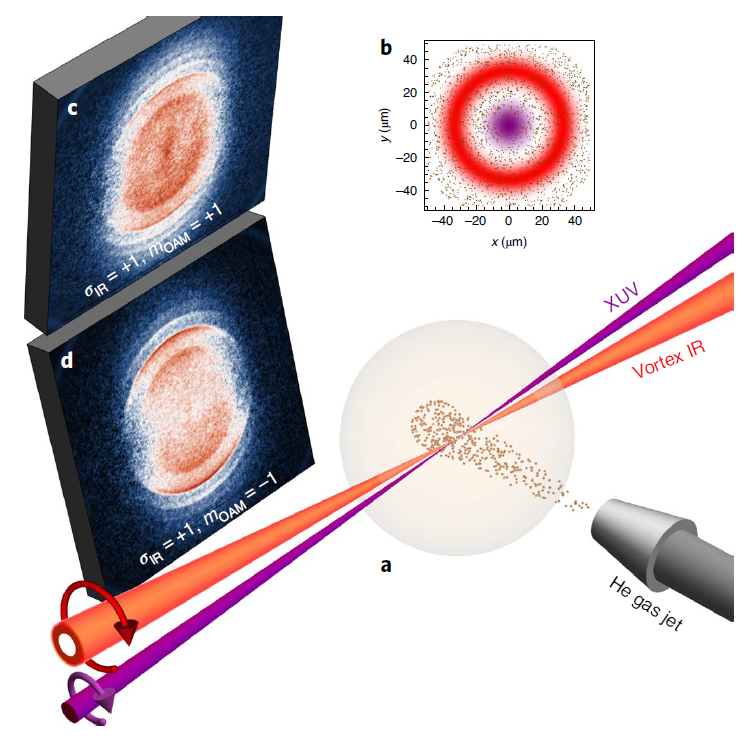}
    \caption{Setup demonstrating the OAM transfer to a photoelectron. a) Two co-linearly propagating lasers (XUV and IR vortex field carrying OAM) interact with a Helium gas cell. b) A cross sectional view on the intensity distributions: violet dot (red circle) is the XUV (IR) laser intensity profile. c) Measured Photo-ionization cross section (DCS) for the laser setup with an IR field carrying OAM of +1$\hbar$ d) Same as c) for an IR field with OAM of -1$\hbar$. The different rings in the cross section belong to different “side-bands” caused by IR multiphoton processes. All measurements and parameters of the XUV laser field and the polarization state of the IR field are kept the same. The figure has been adapted from \cite{Ninno2020}.}
    \label{fig:OAM1}
\end{figure}

The distinctive way in which the photon spin dictates the electron motion upon light–matter interaction is the basis for numerous well-established spectroscopies. By contrast, imprinting OAM on a matter wave, specifically on a propagating electron, is generally considered very challenging and the anticipated effect undetectable. In ref. \cite{Kaneyasu2017}, the authors provided evidence of OAM-dependent absorption of light by a bound electron. Recently, two experiments have been carried out at the Low Density Matter beam line \cite{Ninno2020, Waetzel2022} of the FERMI FEL, on a gas-phase sample. In these experiments, the interplay between an IR laser pulses with OAM and a circularly polarized XUV FEL pulse allowed to demonstrate that: a) OAM can be imprinted coherently onto a propagating electron wave \cite{Ninno2020}; b) an optical vortex can induce orbital magnetization at the nanoscale \cite{Waetzel2022}. The obtained results are summarized below.   
In \cite{Ninno2020},  using extreme XUV radiation from the FERMI FEL, electrons in a He gas cell are elevated to just above the single-ionization threshold by one photon. During this process an intense infrared laser field is present (Fig. \ref{fig:OAM1}). 

The FEL beam has a fixed (right-handed) circular polarization and the IR field has an orbital phase corresponding to a well-defined OAM with respect to the propagation direction. The question of concern is whether it is possible to observe a vortex-dependent OAM dichroic signal (meaning a dependence of the ionization probability on the sign of the OAM), even though the atoms are randomly distributed across the laser spot of the OAM-carrying  beam. This would be surprising: the atomic wave function is extremely localized on the scale of the OAM beam’s waist (Fig. \ref{fig:OAM1}b). Off-axis atoms experience the infrared laser field as an ordinary Gaussian beam and atoms close to the optical axis, where the OAM is well defined, experience a vanishingly small field. Besides, the fraction of near-axis atoms is small. Although the transfer of optical OAM to photoelectrons seems unlikely, if it happens, it should involve a new type of non-dipole transition, related to the OAM-carrying vector potential of the IR field.
To clarify experimentally the above issue, atoms are irradiated with XUV beam copropagating with an infrared laser with adjustable right- or left-handed circular polarization, and variable OAM  l= $\pm$ 1 .The two beams are focused and intersect a gas jet of He atoms in a vacuum chamber equipped with a velocity map imaging (VMI) spectrometer \cite{Waetzel2022}. The velocity distribution of electrons produced in the interaction volume is projected onto a two-dimensional (2D) imaging detector parallel to the optical axis of both beams. For sufficiently high infrared intensities, the resulting photoelectron spectra (Fig. \ref{fig:OAM1}c,d) show a main band, corresponding to direct (XUV-induced) photoemission, together with a series of weaker rings, called sidebands, whose radius increment (decrement) represents the energy acquired (lost) by the photo-emitted electrons upon absorption (emission) of one or more infrared photons. All bands depend on the polar angle at which photoelectrons are emitted. They reflect the interplay between XUV and optical fields during the photoionization process, providing information on the transfer of light properties, for example, angular momentum, to matter. When the two beams are circularly polarized and l = 0, there is a few-percent difference in the angular electron distributions for co-rotating and counter-rotating XUV and infrared fields, resulting in a dichroic contrast in the range of 5–10\%  \cite{mazza2014determining}.
Now what happens if l= $\pm$ 1. The two VMI spectra shown in Fig. \ref{fig:OAM1}c, d are obtained for fixed (right and left-handed) infrared circular polarization and alternate OAM (+$\hbar$ and -$\hbar$, respectively), with all other experimental conditions kept unchanged. The images demonstrate that different topological charges l result in qualitatively different photoelectron spectra, attesting  a clear OAM-dependent dichroic effect. To clarify the underlying mechanism for the experimental evidence of the OAM-dependent dichroism, a first-principle theoretical model has been developed \cite{Ninno2020,Waetzel2020,Waetzel2022} that accounts for all nonlinear effects in light–matter interaction in a spatially structured vector potential. As in the experiment, the challenge is to account for the two vastly different length scales of the bound electrons and the spatial structuring of the vector potential. In addition to the full numerical solution of the time-dependent Schrödinger equation on a space-time-grid, perturbation theory \cite{dahlstrom2013theory, Bray2020} accounts for the coherent interaction of both lasers with the atoms, endorsing the following picture. The initially bound wavefunction is excited with one single XUV photon to just above the threshold inflating its extension. At this event in time and space, the  electron wavepacket  starts scanning over the structured vector potential of the IR laser, acquiring information on its orbital phase. As shown in \cite{Ninno2020}, the well-defined OAM of the IR laser implies that, in the azimuthal direction the phase gradient of the vector potential of the IR laser is constant. Hence, any atom at a certain radial distance from the optical axis experience the same phase gradient of the IR laser, regardless of the atoms’ orientation, which explains the finite OAM dichroism, even after performing a thermal average over the ensemble of the atoms in the gas cell. 

\begin{figure}[t]
    \centering
    \includegraphics[width=0.9\linewidth]{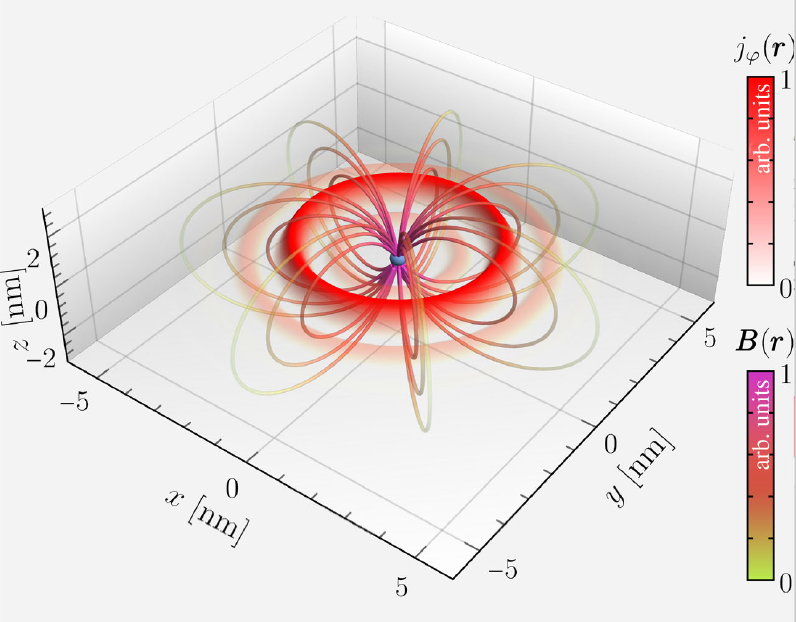}
    \caption{Irradiating atoms by a co-propagating  circularly polarized XUV laser pulse  and IR vortex field carrying OAM induces in a He atom a circular unidirectional charge Rydberg current density j\textsubscript{$\varphi$}(r). The current leads to a nano-scale localized, persistent magnetic field (B) around each atom. The figure has been adapted from \cite{Waetzel2022}.}
    \label{fig:OAM2}
\end{figure}

Having understood the nature of the OAM transfer to a random ensemble of atoms, one may think of orbitally magnetizing the distribution of the atoms. Thereby, the laser beams act as the bias field which orients each atom individually by driving a unidirectional charge current upon the transfer of SAM and OAM from the laser fields. The orbital magnetic moment associated with the charge current is robust since the excitations occur predominantly via a single-photon resonant excitation to Rydberg states which subsequently decays in absence of fields on the microsecond time scale. Interatomic interaction is negligible due to the low atom density. Obviously, such a method, if successful, can generate femtoseconds magnetic fields within micrometers. Even though the field strength averaged over a macroscopic scale might be very small, the local field can reach values useful for triggering local magnetization dynamics in nearby magnetically ordered material. To validate this scenario, experiments have been performed. For that, use has been made of a gas of He atoms. The spatio-temporal control of the excited magnetic moment was obtained by exciting atom’s ground state to a Rydberg state with the XUV circularly polarized photon beam generated by  FEL. The spatial extent of the generated Rydberg orbital is set by the XUV frequency. A second copropagating IR vortex pulse assists in generating the unidirectional current. Tuning the intensity ratio between the XUV and the IR pulses controls the cross-sectional area where the magnetic moments reside, allowing in this way for a nanoscale tuning of the generated magnetic fields with a resolution well below the diffraction limit of the IR laser. The magnetic field associated with the quantum mechanical current density  \textbf{J}\textsubscript{$\varphi$} is obtained from classical electro-dynamics, see Fig.\ref{fig:OAM2} \textbf{J}\textsubscript{$\varphi$} itself is deduced from full ab-initio quantum evolution of the initial state wave functions in the laser fields.

The proposed scheme is not restricted to He. Rydberg states are a generic feature of electronic compounds, and their photo-excitation does not require a symmetry breaking in the sample. For example, the approach can be applied to adsorbates such as rare gases that are physisorbed at magnetically active surfaces \cite{jacobi1982photoemission}, allowing one to study the spatio-temporal material’s magnetic response to the fields of the photo-generated magnetic moments. 
A further future development is the spatio-temporal modulation of the spin part (polarization) of orbitally structured  laser electric fields. This would pave the way for the generation of new types of topological optical fields in the XUV regime with the prospect of new insights and applications, particularly to 
topological and vortex materials, as shown for superconductors \cite{Niedzielski2022}.
Femtosecond polarization shaping of spatially homogeneous free-electron laser pulses has been demonstrated recently \cite{PhysRevLett.131.045001}. 

\onecolumngrid \newpage \twocolumngrid

\subsection{Perspectives for stereochemistry in complex molecules} 
\label{sec:ESI}

\begin{center}
Sadia Bari and Lucas Schwob
\end{center}

Amino acids, which are the building blocks of proteins, are all left-handed, while the sugars in DNA and RNA are right-handed (see also section \ref{sec:PECD}). Yet the origin of this homochirality of life still remains unsolved. Circular dichroism (CD) studies exploit the chiral properties of biomolecules to reveal structural information \cite{Miles_2006,Kypr_2009} The availability of commercial CD instruments covering the energy range from the far UV (190~nm) to the NIR (1600~nm) makes it easy to address different absorption bands and thus different structural aspects and molecular systems. Here we focus on the electronic CD, where for example, in the UV the absorption by DNA bases reveals their relative positions and thus the different conformational forms of DNA \cite{Kypr_2009}. CD of proteins in the far-UV analyzes the fraction content of the secondary structures due to a strong $\pi-\pi^{\ast}$ transition around 190~nm and a weaker but broader $n-\pi^{\ast}$  transition between 210-220~nm in the peptide bond \cite{Miles_2006}. In the near UV the CD is dependent on the aromatic side chains and disulphide bonds and can give information about the tertiary structure of a protein. Synchrotrons allow the use of photon energies in the VUV with orders of magnitude higher photon fluxes for CD measurements than in commercial CD instruments, which compensates for the stronger absorption in the VUV region due to e.g. air, buffer solutions and salts and generally improves the signal-to-noise ratio of the spectra significantly. The lower wavelengths down to 120 nm lead to the observation of more CD bands in biomolecules and therefore more conformational information. First measurements in the soft X-ray regime on amino-acids films show the feasibility of observation of X-ray natural absorption CD and pave the way together with theoretical calculations for a new local probe by the element and orbital specific transitions \cite{Tanaka_2005,Kimberg_2007}. When combined with "near edge X-ray absorption fine structure" (NEXAFS) spectroscopy, a powerful tool for studying of the molecular electronics and geometric structure, it could provide additional information due to the fact that particular atoms or functional groups which have very similar X-ray absorption spectra may have different chiral properties. \cite{Kimberg_2007}

Most biomolecular CD studies are done in the condensed phase and can be performed on molecular systems of any size and under various solvent and temperature conditions without the need for crystallization. In contrast to X-ray crystallography, CD allows obtaining additional conformational information, such as protein denaturation \cite{greenfield2006using, sreerama2000estimation}.
However, studies on complex isolated biomolecules in the gas phase are still lacking to decipher the purely intramolecular effects from the influences of the environment. These studies would allow careful chemical control, a bottom-up approach by gradually increasing the complexity of the systems, and easier comparison with theoretical models because intermolecular interactions can be excluded. Furthermore, in the gas phase, i.e. in vacuum, one would avoid the strong photoabsorptions into the environment mentioned above.
Due to the lower target densities in the gas phase, CD measurements are performed using photoionization and measuring differences in the photoelectron emission rather than differences in absorption yields. For example, resonance-enhanced multiphoton ionization (REMPI) used in combination with mass spectrometry and table top UV lasers have detected a CD asymmetry up to almost 30~\% for small organic molecules \cite{Boesl_2013}. Strong effects were as well observed for small organic molecules by photoelectron circular dichroism (PECD) using a velocity map imaging spectrometer in the VUV \cite{Tia2014} and a hemispherical analyzer in the soft X-ray regime \cite{Ulrich_2008}. That means that CD studies in the gas phase can reach asymmetries orders of magnitude higher than in conventional CD such that the applied techniques should be adapted for more complex biomolecules.

Studying in the gas phase larger, thermally fragile and non-volatile, biologically relevant molecular systems requires the use of soft-ionization techniques. Electrospray ionization (ESI), developed in the 80s by Nobel laureate John B. Fenn \cite{Yamashita_1984,Fenn_1989}, is such a technique. It allows transferring the analyte molecule from the liquid phase to the gas phase by the production of a spray of charged microdroplets under a strong electric field. The desolvation of the droplets eventually produces a beam of the molecular ions of positive or negative charge depending on the solvent used and direction of the electric field. Working with ions enables the use of advanced ion-manipulation techniques employed in mass spectrometry (MS). These techniques are based on static and radio-frequency electric fields and enable ion beam guiding and focusing, mass and conformer selection of analyte ions, accumulation and trapping. ESI-MS has become popular, noteworthy in the analytical field, for its high sensitivity and low sample consumption (typical analyte solution concentration of nmol/L to umol/L).

As an analytic tool, mass spectrometry as such is blind to chirality. Being widely used for the analysis of complex biological samples and pharmaceutical drugs, much research effort has been made in the past 20 years to couple MS to chiral-selective methods \cite{Awad_2013}. MS has primarily been combined with separation tools such as liquid or gaseous chromatographic techniques in order to differentiate enantiomers of chiral molecules prior to MS analysis. However, incompatibilities with the process of electrospray ionization are limiting the use of hyphenated MS techniques. In parallel, chiral analysis solely based on ESI-MS has been developed under the principle of the chiral recognition mechanisms. The methods rely on intricate ion-molecule reactions as well as on formation and/or dissociation processes of molecular complexes between a guest enantiomer and a host chiral selector, such as crown ethers \cite{Wu_2012}. 

As discussed thoroughly earlier, polarized light is a most straightforward and universal tool to probe chirality. In the past decades, ESI-MS has been successfully/routinely coupled with a variety of light sources, from tabletop lasers to synchrotron and free-electron lasers, to perform action-spectroscopy on gas-phase biomolecular ions in the IR, UV, VUV and soft X-ray energy ranges. To compensate for the low ion density produced by ESI sources, the molecular ions are exposed to light while stored in an ion trap. In this configuration, the overlap between the target ions and the photon beam is facilitated and  the duration of the light exposure can be controlled to achieve a good signal-to-noise ratio. However, the detection of photoelectrons is impossible due to radiofrequency field in the ion trap and their close geometry.

Circular dichroism spectroscopy of mass-selected, trapped, biomolecular anions using ESI-MS and circularly polarized UV light has been first reported in 2020 by Daly et al. \cite{Daly_2020}. The authors investigated a series of negatively charged right- and left-handed DNA G-quadruplexes and G-duplexes ($>$ 5000 Da) by means of electron photodetachment (EPD), which is a single-photon process thanks to the low electron detachment threshold in polyanions. The CD spectra is thus obtained by recording, as a function of the laser wavelength, the yield of $ [M-nH]^{(n-1)-}$ from the reaction $[M-nH]^{n-}+h\nu \rightarrow [M-nH]^{(n-1)-}+e^-$. Overall, by comparing the solution and gas-phase CD spectra in the 200 to 300~nm wavelength range, they observed that the DNA polyanions have similar CD spectra in terms of sign and position of maximum and minimum intensity. Most importantly, their measured gas-phase CD spectra showed up to five-fold more intensity in the asymmetry factors than those measured in the solution phase. The authors concluded that the stacking pattern of the nucleobases is conserved in the gas phase and that the excited states responsible for CD in the UV range also trigger the photodetachment process.
\begin{figure}[h]
    \centering
    \includegraphics[width=0.9\linewidth]{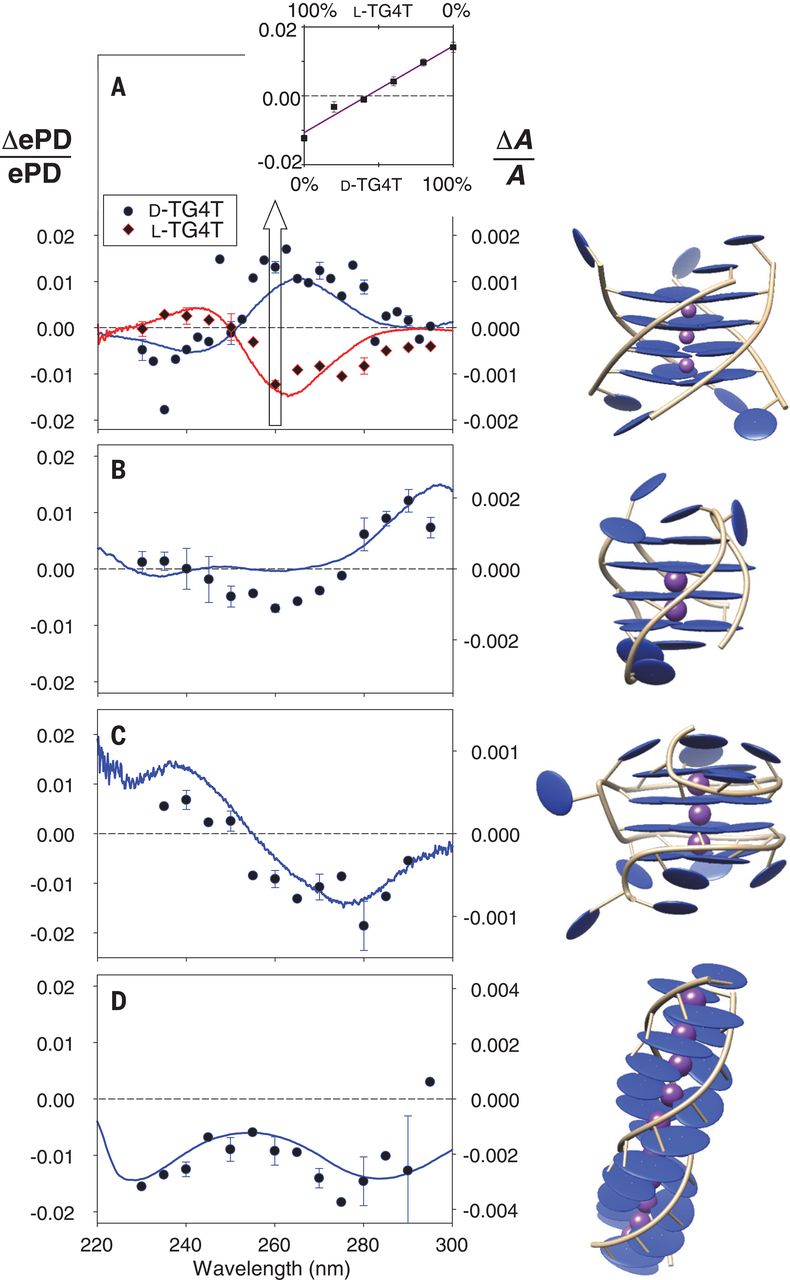}
    \caption{Gas-phase circular dichroim spectra ($\Delta$ePD/ePD) compared to solution-phase spectra ($\Delta$A/A). Symbols and lines denote gas-phase and solution-phase CD data, respectively. \textbf{A}: right-handed $\scriptstyle D-$ and left-handed $\scriptstyle L-$TG4T $([(TGGGGT)_4 \cdot (NH_4^+)_3]^{5-})$ with inset showing CD at 260 nm for different ratio of $\scriptstyle D-$TG4T and $\scriptstyle L-$TG4T. \textbf{B} to \textbf{D}: CD spectra for the antiparallel G-quadruplex $5YEY^{5-}$ (B), left-handed G-quadruplex $ZG4^{6-}$ (C) and G-duplex $GAgG^{5-}$ (D). The figure has been reprinted with permission from \cite{Daly_2020}.}
    \label{fig:FBS}
\end{figure}

While the latter study is based only on ion yields as an observable for the circular dichroism, it appears evident to the reader that measurement of the photoelectrons, such as the backward/forward direction, would bring unequivocal information on the CD, even in case when CD cannot be distinguished from the ion yields. As detailed thoroughly in the previous sections, PECD is mostly recorded in VMI-type instruments in crossed-beams configuration. Such experiments are extremely challenging for ions produced by an ESI source due to low target density available and complications inherent to the use of charged target molecules instead of neutral. UV PECD of electrosprayed molecular anions has been successfully reported for the first time by Krüger and Wietzel in 2021 \cite{Krueger_2021}. Their proof-of-principle experiments were performed on deprotonated left- and right-handed glutamic acid and 3,4-dihydroxyphenylalanine (DOPA), using a circularly polarized UV laser at 355~nm. Like in the work by Daly et al. \cite{Daly_2020}, electron photodetachment of anions requires a single UV photon, making it a more simple technique than REMPI-based CD, both in terms of technical requirement and theoretical description. Strong CD effects of up to 4.5~\% were measured in the forward–backward asymmetry in the photoelectron angular distribution for both molecules, although no CD effect could be identified from the ion yields only. This highlights once again the power of PECD spectroscopy to resolve chiral compounds.

Overall, circularly polarized light-based chiral analysis of gas-phase complex biomolecules produced by soft ionization source is an emerging field based on well known experimental tools. In pioneering experiments using electron photodetachment of anions, Daly et al. \cite{Daly_2020} and  Krüger and Wietzel \cite{Krueger_2021} paved the way for CD spectroscopy of electrosprayed molecules using UV lasers. This work has now to be extended to more systems and also to cations, for which a single UV photon is not sufficient to induce photoionization. As was shown for the extensively studied fenchone molecule, extending PECD in the VUV energy range \cite{Nahon_2016} and even in the soft X-ray regime \cite{Ulrich_2008,Pohl2022} allows reaching very high (up to 20~\%) asymmetry factors. Thus, by interrogating chirality of large gas-phase biomolecules in such an energy range, one would benefit from stronger CD effects in addition to overcoming the limit of UV-induced electron photodetachment. Especially due to the low ion density available with ESI sources, highly brillant X-ray sources are needed. XFELs would therefore be of prime importance to extend the capabilities of gas-phase CD spectroscopy. Finally, biomolecular chirality studies using polarized soft X-rays would profit from the high absorption site selectivity of soft X-ray photons, which in the form of ultrashort pulses would enable dynamical investigation of structural and chiral properties.

\onecolumngrid \newpage \twocolumngrid

\section{Theoretical Perspectives for Polarization-Controlled FELs}

This section is devoted to selected theoretical aspects of the topic. 
We will not attempt to provide a comprehensive summary of all the
tools that are currently available, the ideas behind them, or a discussion of
their strengths and (often severe) limitations. Instead, we will first illustrate the current state-of-the art by reporting a few selected benchmark studies within the scope of the topical orientation of this roadmap. 

We will start with the popular 
single-active-electron (SAE) model to solve the time-dependent Schr\"odinger equation (TDSE).  This model is, of course, the ideal tool for truly one-electron targets such as atomic hydrogen or the He$^+$ ion. It can also be expected to be successful for quasi-one-electron systems such as alkali-like atoms or ions with an appropriate core potential, as long as all the processes of interest only involve the valence electron. One example of the latter are recent studies involving optical lasers~\cite{PhysRevLett.126.023201,PhysRevA.104.053103,PhysRevA.106.023113}.
Nevertheless, the SAE approach has been used extensively for helium (e.g.,~\cite{PhysRevA.102.062809}) 
and even more complex targets such as the heavier noble gases neon~\cite{prince2016coherent} or argon~\cite{PhysRevA.107.022801}.
While none of these studies employed undulator-based polarized XFEL radiation, increasing the XUV intensity
using the latter may open up additional possibilities for studying non\-linear phenomena also in asymmetric processes. 

After presenting one undulator-based circular-dichroism study, we will discuss selected already available approaches that go beyond the SAE model, mostly for more complex atoms and relatively small molecules.  Additional subsections are then devoted to observations that have become possible by polarization-based XFEL radiation, but whose theoretical description will require the development of new theoretical and computational tools due to the complexity of the targets (large molecules) involved. 

\subsection{The Single-Active-Electron Approximation}
\label{sec:BartschatDouguet}

\begin{center}
Klaus Bartschat and Nicolas Douguet
\end{center}
At the fully quantum-mechanical level, the simplest approach that starts
with the time-dependent Schr\"odinger equation (TDSE) is the so-called
``single-active-electron (SAE) approximation''.  As the name indicates, only
one electron in the target is considered to be affected by the laser field.
Even without the field, the interaction of this electron with the nucleus 
and any other electron is only treated approximately.  Except for atomic hydrogen,
He$^+$, and other true one-electron systems,
therefore, the active electron is assumed to move in some average, 
usually spherically symmetric, field  
that is made up by both the nucleus and the other
electrons.  The pure (point-like nucleus) non\-relativistic Coulomb problem 
for this setup  can, indeed, be solved numerically with very high accuracy, 
although the choice of laser parameters can still make the problem challenging. 
Furthermore, currently available experimental 
techniques involving XFELs are not sufficiently sensitive to require accounting
for very small 
effects that, for example, should be considered in highly sophisticated
atomic-structure calculations.

Moving on to helium as a two-electron target, it turns out that the SAE approach
{\it might} (see next subsection for more sophisticated approaches) still 
be applied with some confidence to support many ongoing experimental investigations.
This is due to the fact that the non\-active 1s electron is very tightly bound
and often only acts as a spectator.  Even though there is a significant
difference between the Hartree-Fock 1s orbital for the ground state and the 
1s orbital of He$^+$, which is a very good approximation for all the 
excited 1s$n\ell$ bound states of helium, does not seem to seriously affect
the quality of SAE predictions. The most important aspect here is
to represent the effect of the spectator electron by a suitably chosen 
potential.  In reality, this potential would have a non\-local component due
to the exchange interaction, but in practice local terms to approximate
exchange as well as core polarization and possibly even absorption effects
have been quite successful.

The very same approach can be employed in particular to describe studies involving
the valence electron of alkali atoms.  In fact, the well-known
spectra of these atoms allow the design of very appropriate potentials to be
used in the SAE approach.  Not surprisingly, therefore, the theoretical predictions
are often in excellent agreement with experiment.  In fact, if there is disagreement,
it may well be worth to check the experimental data even more thoroughly than it is
done anyway.

Coming back to helium, it is very important to note that there are now 
highly sophisticated codes around (see further below) 
that can fully treat the interaction of
both electrons with each other as well as the laser field.  Hence, it is possible
to determine with high confidence the range of applicability of the SAE approaches.
In addition, true correlation effects involving, for example, auto\-ionizing
resonances, can be studied with these methods.  This has, indeed, been used
to image the time evolution of the famous Fano resonances~\cite{PhysRev.124.1866, wickenhauser2005time, kaldun2016observing}. 
Another interesting two-electron system is H$^-$, where correlation effects 
in the ground state are stronger than in helium, while no excited states exist at
all. However, this target does not seem to be particularly suitable for FEL studies
and hence will not be discussed further.

Moving on to even larger atomic targets, both SAE and
fully-correlated two-electron approaches are undoubtedly reaching their limits
when it comes to the experimentally often favored heavy noble-gas targets 
Ne, Ar, Kr, and Xe.  To begin with, 
the active electron is often the outer \hbox{p-electron}, although the slightly tighter
bound \hbox{s-electron} in the same major shell can affect the outcome as well -- even by
just making the signal less clear and hence more difficult to interpret.
The first challenge, of course, is to find suitable SAE potentials. There are
well-known problems already when it comes to reproducing the standard
photo\-ionization cross section.  Such problems, of course, would
cast doubt on any SAE results in certain energy regions.

Next,
the calculations need to be performed for individual magnetic sublevels.
For linearly polarized laser light, the results for $ m = \pm 1$ are 
related in a simple way.  For example, the ionization probabilities starting
from either one are the same due to the cylindrical symmetry of the scheme.
However, they are different for the initial $ m = 0$ sublevel, 
and hence the results have to be combined appropriately.  
When circularly or elliptically polarized beams are used, 
the situation is even more complicated.  In this case, a thorough general
description of the observables in terms of the relevant amplitudes needs to 
be developed as well.

Another challenge is the structure description of these targets.  Even neon as
the lightest of these species is not well described in pure $LS$-coupling.  
As a result, instead of one $\rm 2p^5 4s$ state with a total electronic angular
momentum $J=1$ (namely $\rm (2p^5 4s)^1P_1$ in $LS$-notation) being available 
as a stepping stone in an $\omega + 2\omega$ setup~\cite{prince2016coherent}, a second one
(usually labeled $\rm (2p^5 4s)^3P_1$ in $LS$-notation) could also be accessed. 
While it is possible to select one of the two by tuning the photon energy, this 
shows that the complexity increases substantially when the simplest and
strictest selection rules are no longer applicable.  Similar, even 
more complex, considerations need to be given to the 
$n_1{\rm p}^5 n_2{\rm d}$ configuration.  And, finally, 
the fact that the target s- and d-orbitals are
actually term-dependent further complicates the situation.

By carefully comparing predictions from perturbative calculations with
a sophisticated target description and non\-perturbative SAE calculations
with a much simpler structure model, the similarities and remaining 
differences have been analyzed in a number of cases. Similarly, predictions
from SAE models have occasionally been compared with those from
two-electron codes and, for heavy noble gases, even 
all-electron approaches such as R-matrix (close-coupling) with time dependence (RMT)~\cite{RMT_cpc}.  
Unfortunately, it is generally a major challenge to define a 
starting point that is sufficiently close to allow testing of a single
aspect of these models. Even though this often cannot be done to full
satisfaction, agreement between results obtained with vastly different
methods can provide confidence in these predictions and subsequently
be used as a basis for the interpretation of the experimental studies
that these calculations are supposed to support. 

One example of an undulator-based circular-dichroism study on a simple target is the experiment reported
by Ilchen {\it et al.}~\cite{ilchen2017circular}. 
A circularly polarized XUV pulse was applied to neutral helium in the $(\rm 1s^2)^1S$ state to ionize
the target and, in the same step, the remaining electron in the He$\rm ^+(1s)$ ground state was
excited to the $m = +1 $ magnetic sub\-level of
the $\rm 3p$ state. This excited and oriented state was then ionized by a co-rotating or
counter-rotating near-infrared (NIR) field. 

\begin{figure}[t]
\includegraphics[width=0.9\columnwidth]{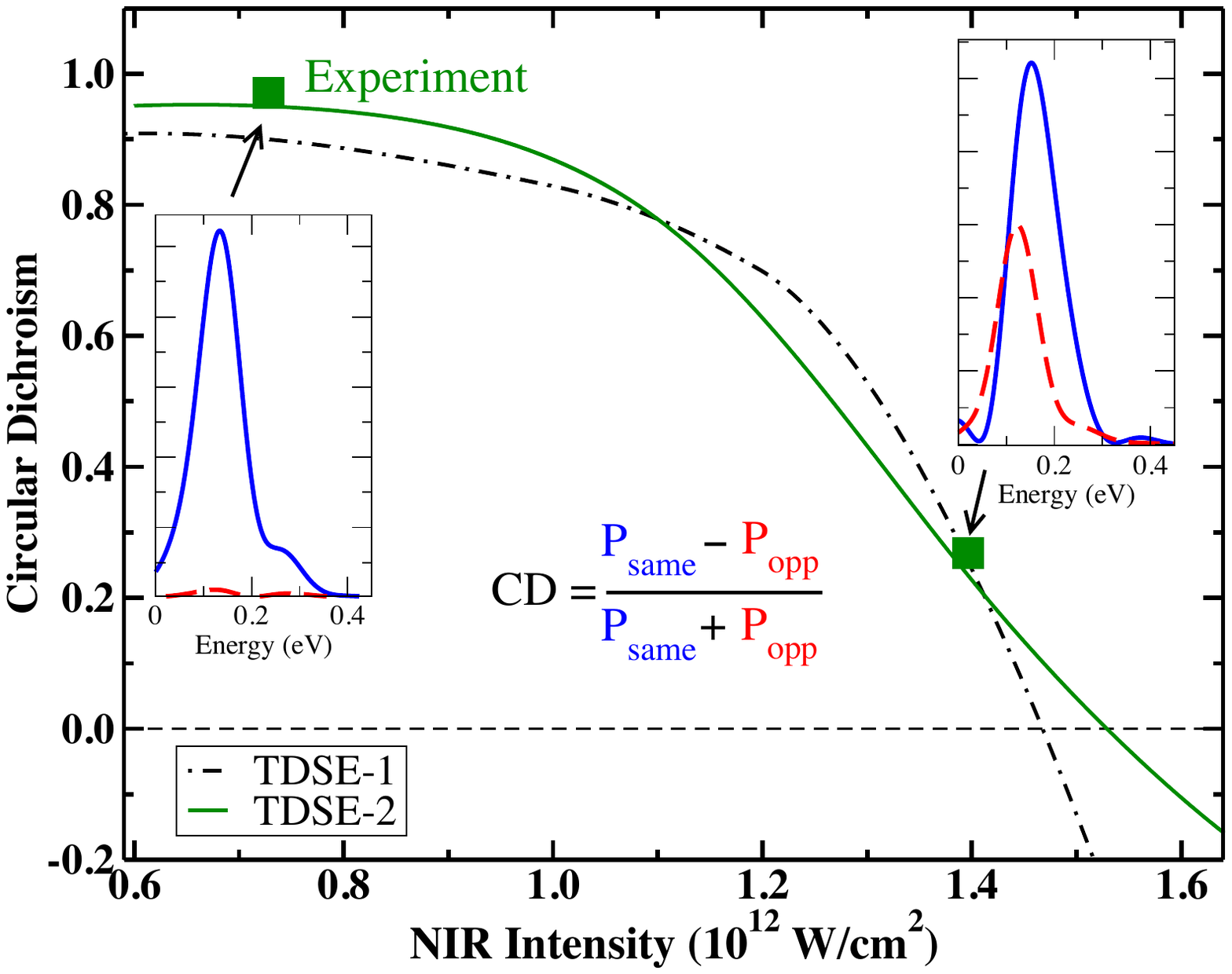}
\caption{Circular dichroism in the low-kinetic-energy peak
as function of the NIR peak
intensity for an XUV peak intensity of $1.0 \times 10^{13}$ W/cm$^2$.
The two experimental points are compared with predictions from two independent \hbox{TDSE} calculations
 (\hbox{TDSE-1}~\cite{Kazansky_2007}, \hbox{TDSE-2}~\cite{PhysRevA.93.033402}).
The insets show the low-energy spectra from the TDSE-2 calculation for the two experimental cases.The figure has been adapted from \cite{ilchen2017circular}.]}
\label{fig:Ilchen2017}
\end{figure}

Figure~\ref{fig:Ilchen2017} shows the angle-integrated circular dichroism (CD) of the low-kinetic-energy peak
at about 0.12~eV as a function of the NIR intensity.
With increasing NIR intensity, the CD rapidly decreases and is predicted to even change sign at
$I_{\rm NIR} \gtrsim 1.5\times 10^{12}$ W/cm$^2$, i.e.,
ionization by counter-rotating fields becomes more effective. This can be clearly seen in the very large differences in the predicted signals for the co- and counter-rotating cases when the intensity is changed.
Given that this is an ideal problem for a single-active-electron method (one can actually start with
only the electron in the excited ionic state), it is not surprising that there was excellent agreement 
between experiment and theory, albeit the comparison at the time was limited to just two points.

The explanation for the strong dependence on the NIR intensity proposed by Ilchen {\it et al.}~\cite{ilchen2017circular} was a different, helicity-dependent shift of the He$^+({\rm 3p},m=\pm 1)$ states due to the presence of the
circularly polarized NIR field. This explanation was, indeed, confirmed later by 
Grum-Grzhimailo {\it et al.}~\cite{Grum19}.  Hence, even this apparently simple system, which can be used for a benchmark comparison between experiment and theory, carries surprises that justified further studies. Indeed, Wagner {\it et al.}~\cite{wagner22} recently performed a follow-up experiment, in which a delay between the XUV and the NIR was introduced to 
prepare the initial He$^+$ state without the NIR already being present.  The results are currently being written up for publication. 

Photoionization of the simplest molecular ion, H$_2^+$, can be treated exactly, within the fixed-nuclei approximation, by using an SAE approach with elliptical coordinates. The most general methods usually employ compact \hbox{B-splines} or Finite-Element Discrete Variable Representations (FEDVRs). In H$_2^+$, it is also possible to treat the nuclear motion explicitly, which provides an interesting case study. Molecular hydrogen is another interesting case, as it can be treated accurately with a two-active-electrons method and thus represents a benchmark for theory for electron correlation in molecules.

Affine to the SAE method, where one electron is described in the field of a fixed potential, are those many-body methods for atomic and molecular ionization where both the initial bound state of the system and the final states in the continuum are restricted to a single-reference.

The \textsc{xatom} code, an ``integrated toolkit for x-ray and atomic physics'' developed at CFEL by Sang-Kil Son and Robin Santra, is based on the Hartree-Fock-Slater method. The code is ideally suited to describe the generation of multiply-charged ions generated by the interaction of intense XFEL radiation with atoms by means of sequential ionization mechanisms~\cite{Young2010,Son2011,Rudek2012,Fukuzawa2013,Murphy2014}.
\textsc{xatom} uses the static-exchange approximation to compute bound-state energies, ionization cross sections, Auger-decay rates, and fluorescence rates of all charged states involved. Even though \textsc{xatom} neglects any single or multiple excitations from the ionic reference states, the mere process of creating vacancies in the target, as well as in the product of its subsequent Auger and radiative decays, gives rise to millions of configurations and a manyfold of more ionization paths. \textsc{xatom} employs a Monte-Carlo method to sample the most relevant reaction paths in the rate equations for the system.
The \textsc{xmolecule} code~\cite{Hao2019} is an extension to molecules of the Hartree-Fock-Slater methodology used in \textsc{xatom}. It includes a classical treatment of nuclear motion, which is solved on-the-fly alongside the electron dynamics, in association with a Monte Carlo approach. \textsc{xmolecule} is capable of determining the yields of fragments that emerge from the interaction of a molecule with an intense XFEL pulse, and which can undergo  multiple photoionization and non-radiative decay steps~\cite{Rudenko2017,Inhester2016,Rudenko2017,Kumagai2018,Gianturco1994}.

The \textsc{e-Polyscat} code~\cite{Natalense1999,Gong2022}, developed by Robert Lucchese and co-workers, can describe the single-ionization electronic continuum of non-linear molecules, within the fixed-nuclei and the static exchange approximation. The continuum functions are based on a single-center expansion. \textsc{e-Polyscat} has been applied to the calculation of electron-molecule scattering cross sections~\cite{Gianturco1994}, and of multiphoton ionization observables of non-linear molecules.

The \textsc{tiresia} B-spline density-function method is based on 
single-center~\cite{Venuti1998,Stener1998,Brosolo1992,Brosolo1994,Fronzoni1998,Venuti1999,Decleva1999,Stener2002}
and poly-center expansions~\cite{Toffoli2002,Decleva2022,Stener04,Toffoli2006c,Stener2007,Moitra2020}.
It has been used to predict the ionization yield in molecular strong-field ionization~\cite{Petretti2010,Awasthi2008,Farrell2011} and high-harmonic generation~\cite{Neufeld2019}. It has been applied to the calculation of one-photon vibrationally-resolved core photoionization of molecules, which was instrumental to conclusively characterize the interference effects associated to two-center emissions and to intramolecular photoelectron scattering~\cite{Canton2011,Plesiat2012b,Argenti2012,Kukk2013,Ayuso2014,Ueda2013,Patanen2014}. It has been applied to the study of charge-migration processes in biologically relevant molecules~\cite{Calegari2014,Lara-Astiaso2018,Lara-Astiaso2016,Ayuso2017,Lara-Astiaso2017}, where it has been key to interpret molecular fragmentation in attosecond pump-probe experiments.
 
\textsc{Octopus}~\cite{Octopus2020,Andrade2015} is a general-purpose scientific program based on density-function theory (DFT) and time-dependent density-functional theory (TDDFT)~\cite{MarquesGross2004} and on a grid representation of the wave function. \textsc{Octopus} has been applied to attosecond photoelectron spectroscopy~\cite{DeGiovannini2012} and transient-absorption spectroscopy~\cite{de2013simulating,Crawford-Uranga2014a,Crawford-Uranga2014,Sato2018}. TDDFT has also been applied with a focus on tunnel ionization and on the  bound-state dynamics in the residual ion~\cite{Li2020,Wahyutama2022}.
Single-active-electron methods reproduce the main features of strong-field ionization processes~\cite{Awasthi2008} and of those photoelectron spectroscopies were electronic correlation does not play a central role, such as in the case of the photoemission of high-energy electrons (several hundreds eV) from core orbitals~\cite{Kukk2013}. They are also useful for complex processes that lead to the complete fragmentation of a target~\cite{Ayuso2014}. When  multiply excited states or the entanglement between fragments play a fundamental role, however, more complex approaches are required.

\onecolumngrid \newpage \twocolumngrid

\subsection{Going beyond the SAE}
\begin{center}
Luca Argenti, Klaus Bartschat, Nicolas Douguet, and Kathryn Hamilton 
\end{center}

 \bigskip

As described above, the SAE approximation works well at energies low enough that the residual parent ion cannot be excited. It can also work well for the ionization of relatively uncorrelated core electrons, in atoms and molecules, when the released photoelectron is highly energetic. Indeed, in this case the interaction between different channels is negligible. Hence, the system can be described by a set of non-interacting single-active-electron effective potentials, one for each hole created in the single-configuration core of the target atom or molecule. 

The SAE, however, cannot reproduce all the dynamical states of polyelectronic systems. In particular, ionization in energy regions close to the opening of an excited threshold entails the formation of multiply-excited states and ionization channels with slow electrons, which are strongly coupled with each other, giving rise to Auger decay and inelastic scattering. These conditions are common in the ionization of atoms and molecules from the valence or inner-valence shells with VUV or XUV radiation. It is relevant also for near-edge core ionization. In the case of core ionization, furthermore, the parent ion itself is in a metastable state. The interplay between the photoelectron and the Auger electron is entirely neglected not only by the SAE method, but also by computer codes restricted to single ionization.

A first step towards an accurate representation of correlated ionization dynamics near threshold can be achieved with the close-coupling (CC) approach. This approximation assumes that the parent ion of a target system explores only a finite number of localized bound states, while the photo\-electron is unconstrained. The lowest-order approximation to this approach is the so-called ``CI-singles'' model, in which the configuration space comprises only the Hartree-Fock ground state of the target, as well as all single excitations from it. In this approach, the ionic states are assumed to be unrelaxed single determinants. Most of the correlation in the parent ion is ignored, which is a major limitation. However, in contrast to SAE approaches, the approach already  accounts for inter-channel coupling.

A better approach is to describe the parent ions by correlated instead of mono\-determinantal functions. In this case, the close-coupling framework allows one to represent with high accuracy the asymptotic portion of the wavefunction at any energy below the second ionization threshold. The inclusion of a few additional localized configurations for the neutral system not included in the CC expansion can greatly improve the quality of the wavefunction at short ranges as well. This approach to the definition of a system configuration space is used by most current state-of-the-art codes for atomic and molecular ionization.
These codes are able to reproduce with variable accuracy the valence multiphoton ionization cross sections of atoms and even small molecules in the fixed-nuclei approximation. 

The current limitations of these codes are threefold. First, the short-range correlation may require a very large number of configurations to reach convergence. Second, the calculation of matrix elements between the close-coupling space with strongly-correlated ions can be extremely complex. Many of the currently existing programs, therefore, restrict the configuration space of the ions, or neglect some exchange terms in the hamiltonian. Third, the size of the target ion is limited, either by the difficulty of carrying out many-body calculations in large systems, or because the numerical method forces the parent ion to be confined to a comparatively small sphere.
None of these codes is currently able to describe correlated double-ionization processes in poly\-electronic systems.

In the case of genuinely two-body systems, such as helium or the H$_2$ molecule, it is possible to include in the CC calculation the full-CI space of localized configurations built from an effectively complete set of orbitals in a sphere of the size of tens of Bohr radii. In this case, therefore, as far as single-ionization states are concerned, the representation of the system in the fixed-nuclei approximation can be virtually exact. 

As soon as the number of electrons liberated to the continuum or the number of active electrons in the ion increases, however, the complexity of the problem explodes exponentially. For example, even two-electron systems push the current limit of computation when they are explored above the double-ionization threshold. In this case, the state of neither of the two electrons can be restricted to a fixed small set of orbitals. Furthermore, the two electrons keep interacting appreciably up to large distances. For this reason, a large number of states, of the order of several thousands for each electron, must be considered.  This results in configuration spaces with sizes of the order of millions. Consequently, a complete spectral resolution of the hamiltonian is out of the question. 



In molecules, the nuclear and rotational motions should, in principle, be taken into account explicitly.  However, a first-principles approach combining the nuclear and electronic dynamics in short and intense laser pulses is at the moment computationally prohibitive. Nevertheless, impressive efforts have recently been pursued to combine, with some necessary approximations (e.g., the Born-Oppenheimer approximation and semi-classical nuclei propagation), the nuclear dynamics within the density-functional formalism.

Fortunately, many ultra\-fast processes in molecules subjected to short pulses can be modeled, to a good approximation, by assuming fixed nuclei. This approximation is justified considering the slow motion of the nuclei when compared to the ultra\-fast electronic motion. As a consequence, it is often possible to consider the effect of the laser at different relevant molecular geometries and invoke the Franck-Condon approximation when necessary. While rotational averaging can readily be performed for weak-field processes in randomly oriented molecules, non-perturbative processes require repeating calculations for  many molecular orientations, which can be computationally demanding. 



Below we list a number of methods and computer codes that are currently available.
A first group of methods, known as time-dependent (TD) multi-configuration (MC) self-consistent-field (SCF) approaches, bypass the demanding basis requirement of close-coupling approaches by optimizing a limited number of orbitals during the simulation. TDMCSCF methods achieve an extremely compact description for the wave function, paying the cost of propagating non-linear equations. These methods include multi-configuration TD Hartree~\cite{Lode2020} and Hartree-Fock~\cite{Haxton2011,Sawada2016,Liao2017,Haxton2014,Haxton2015,Haxton2015b},
TD restricted-active-space (RAS) self-consistent field (SCF)~\cite{Miyagi2013,Omiste2021,Miyagi2014a,Miyagi2014,Miyagi2017},
TD generalized-active-space (GSA) SCF~\cite{Bauch2014,Chattopadhyay2015a},
TD complete-active-space (CAS) SCF~\cite{Sato2013a,Sato2016,Sato2018a},
TDMCSCF~\cite{LiSato2021}, and TD coupled-cluster  methods~\cite{SatoCC2018,Pathak2020,Pathak2020a,Pathak2021,Sato2023}.
These methods are particularly suited to reproduce optical observables, such as high-harmonic generation. Since the parent ions can continue evolving close to the asymptotic limit, even in the absence of external fields, channel-resolved photo-electron distributions may be more difficult to converge.

The methods mentioned below are based on the close-coupling (CC) method. While more computationally expensive than the SAE or TDSCF methods, CC guarantees that the wavefunction of the photo fragments have the correct asymptotic behavior and that the time evolution is rigorously linear, which allows for stable and easily parallelizable solvers. The CC methods differ in how they represent electronic correlation in the ions, in the approximations used to compute the representatives of the relevant operators in the CC basis, and in the constraints they set on the size of the target system, and on its maximum excitation energy. 

The most elementary realization of the CC approach is the so-called configuration-interaction singles (CIS)~\cite{Toffoli2016,Hoerner2020,Carlstrom2022,Carlstrom2022a}, which restricts the configuration space to single excitations from a reference mono-determinantal ground state. While simple to implement, the CIS method can reproduce only a limited set of ionic thresholds and, lacking double excitations, it underestimates correlation.

In more accurate CC implementations, the parent ions are expressed in a multi-configuration basis, and the CC expansion may include a nourished group of localized configurations for the neutral system to reproduce the dynamic correlation correlations in bound and continuum states that is not captured by the conventional CC expansion alone. Different programs implement different strategies to mix Gaussian-type orbitals (GTO), which are best suited to represent states localized near the nuclei, with numerical bases, which are required to reproduce photoelectron states at intermediate and large distances from the molecule. Numerical functions include mono\-centric~\cite{Bachau2001} and poly\-centric~\cite{Toffoli2002} B-splines, or mono\-centric~\cite{Yip2008} and poly\-centric finite-element discrete variable representation (FEDVR) functions~\cite{Greenman2017}.

The R-matrix method for atomic, molecular, and optical processes~\cite{Burke2011} is a general approach to solve the close-coupling equations for electron collisions with atoms, ions, and molecules. This includes the final state after photo\-ionization or photo\-detachment.  Since the method is also applicable to the calculation of bound states, it can be used to generate the necessary matrix elements to describe photo-induced processes.
Originally implemented for electron scattering and stationary regimes, the atomic packages \hbox{\textsc{rmatrx-i}~\cite{BEN1995,RMATRXI-Badnell}} and \hbox{\textsc{rmatrx-ii}~\cite{RMATRXIIgit}}, as well as
\hbox{\textsc{ukrmol}~\cite{Gorfinkiel2005,Carr2012,brigg2014r}} have recently been extended to time-dependent photoionization processes, resulting in the general R-matrix with time-dependence \textsc{rmt} code~\cite{Brown2020}, which can treat atoms and molecules in arbitrarily-polarized short-pulse intense laser fields.  Specifically for molecules,
\hbox{\textsc{ukrmol+}~\cite{Harvey2014,Masin2020,brambila2015role,Bruner2016, brambila2017role,Benda2020}}
employs hybrid \hbox{Gaussian} \hbox{B-spline} bases and can deal with arbitrary hybrid integrals. 

A promising alternative to the standard implementation of the R-matrix approach is the B-spline \hbox{R-matrix} (BSR) method developed by Zatsarinny \hbox{\cite{BSR_CPC,BSR_review,BSRgit}}. In addition to the use of B-splines, which has since been adapted in the most recent versions of \hbox{\textsc{rmatrx-i}}, \hbox{\textsc{rmatrx-ii}}, and \hbox{\textsc{ukrmol+}}, the use of non-orthogonal sets of term-dependent one-electron orbitals can reduce the size of the CI expansion considerably while maintaing and often even improving the accuracy of the target description. Efforts to interface the output from the atomic \hbox{\textsc{bsr}} package with \hbox{\textsc{rmt}} are currently in progress.

%

The Complex-Kohn approach has an economical representation of the continuum functions, since it identifies, beyond a certain distance, the channel functions with a linear combination of their asymptotic regular and irregular components. The implementation of the Complex-Kohn method developed by McCurdy, Orel, Rescigno, Schneider, and others~\cite{Schneider1988,McCurdy1989,Schneider1990,Rescigno1993,Williams12,Lin2014,McCurdy2017}, which will refere to as \textsc{ck-mesa}, is based on the \textsc{mesa} program, and has the unique capability of reproducing dynamic correlation through an optical potential generated from a large configuration space of localized functions. This approach allows to compute accurate scattering and photo\-ionization cross sections. While \textsc{ck-mesa} was restricted to one-photon transition processes, it was recently extended to account for continuum-continuum radiative transitions and two-photon transitions processes~\cite{Douguet2018}. 
A variant called \textsc{attomesa}, in which the analytic asymptotic continuum functions are replaced by FEDVR, currently under development in the group of Nicolas Douguet, promises to be able to reproduce arbitrary non-perturbative processes, such as strong-field ionization and high-harmonic generation.

The Recursive indeXing code \hbox{(\textsc{t-recx})} is a code for integrating \emph{ab initio} the time-dependent Schr\"odinger equation of atoms and molecules~\cite{Majety2015c,Scrinzi2022,Majety2015b,Majety2017}. It is able to carry out close-coupling in a hybrid basis for systems in intense laser fields. It can reproduce both single and double ionization processes, non linear models (in which the Hamiltonian depends in a simple way on the wave function being propagated), and source terms.

The \textsc{xchem} code implements a close-coupling approach that leverages the multi\-configuration methods implemented in state-of-the-art quantum chemistry (QC) programs, such as \textsc{molcas}, to evaluate close-coupling matrix elements in the molecular region. Since QC programs contemplate only Gaussian orbitals, \textsc{xchem} defines a large set of single-center tempered Gaussian functions to fill the space of possible configurations for the photoelectron within a fixed radius from the molecular center of mass. Beyond this radius, the ionic orbitals are assumed to be negligible, and the polycentric character of any virtual polycentric orbital is assumed to be well represented by a single-center expansion. In this way, \textsc{xchem} reduces the calculations of all the CC matrix elements that involve numerical orbitals to simple monocentric integrals~\cite{MarantePRA2017,MaranteJCTC2017,KlinkerJPCL2018,Klinker2018,MarggiPoullain2019,Borras2021,Borras2023}. \textsc{xchem}, which can account for a short-range correlation space and can compute ions at the RAS-SCF level, is well suited to accurately reproduce the photoionization of small molecules, both in stationary regime and resolved in time.

Recently, Marco Ruberti has developed a new code for time-resolved molecular ionization that is based on the algebraic diagrammatic construction (ADC)~\cite{RubertiDecleva2018,Ruberti2019a,Ruberti2019,RubertiHHG2018,RubertiFD2021,Ruberti2022}. The code makes use of the polycentric B-spline basis developed for the \textsc{tiresia}.

The Multi-Channel Schwinger Configuration Interaction Method (\textsc{mcsci}), developed by Robert Lucchese, implements the graphical unitary group approach \cite{Bandarage1989, Stratmann1995, Stratmann1995, Lucchese2007b} for linear molecules, taking molecular symmetry fully into account. The method has been successfully applied to time-resolved photo\-electron spectroscopy \cite{Lucchese2007b, Sato2013a, Jin2010}.

The \textsc{astra} (AttoSecond TRAnsitions)~\cite{Randazzo2023} is a recently-developed wave-function close-coupling approach based on the formalism of transition-density-matrix close-coupling (TDMCC). \textsc{astra} implements the exact expressions of the matrix elements between close-coupling states in a hybrid basis, which allows it to tackle larger molecules, and it uses the efficient formalism of spin-strings~\cite{Olsen2000a} to compute transition-density-matrix elements between ionic states of arbitrary symmetry and multiplicity. \textsc{astra} relies, for the QC part, on \textsc{lucia}, a large-scale CI code developed by Jeppe Olsen, and for the hybrid-integral part on the GBTOlib library, developed by Zdenek Masin. Currently, the code is capable of reproducing the photoionization cross section and molecular-frame photoelectron angular distributions in molecules such as N$_2$, CO, and formaldehyde, at the same level of accuracy as the other state-of-the-art molecular-ionization codes.

\onecolumngrid \newpage \twocolumngrid

\subsection{Novel observables enabled by polarization control at XFELs}
\label{sec:RouxelMukamel}

\begin{center}
J\'er\'emy R. Rouxel and Shaul Mukamel
\end{center}

The appeal of polarization control at FELs to monitor molecular chirality has already been introduced in section \ref{sec:InstrumDiag}. Sections \ref{sec:PECD} to \ref{sec:PECD2} demonstrate how PECD can be a sensitive probe of molecular chirality with unsually large asymmetry ratios.
Here, we survey other chirality-sensitive techniques\cite{rouxel2022molecular} that make use of FEL polarization control, and in particular of circularly polarized light (CPL).

The precise control of FEL polarization is now enabling the design of novel windows for molecular chirality in the time-domain in a near future.
The description and interpretation of ultrafast chiral spectroscopic observables thus combines the challenges from both chirality and ultrafast dynamics. 
X-ray chiral signals naturally provide two structural information: the mirror inversion breaking probed by chiral signals and a window onto local atomic structures through the element sensitivity of X-rays at resonance.
The ultrashort time-resolution offered by FELs will allow to selectively target local ultrafast dynamics involving parity breaking, making and changes.
Examples include photodissociation at a chiral center \cite{baykusheva2019real, svoboda2022femtosecond}, low-energy vibrational dynamics \cite{mincigrucci2023element}, ligands twisting in chiral metallic complexes\cite{ashley2018ray}.
In the frequency-domain, synchrotron-based X-ray Circular Dichroism (XCD) experiments probing local chirality in crystals and molecules have been achieved in the last decades \cite{alagna1998x, goulon2000x, turchini2004core}. Experiments at FELs using circularly-polarized light are still few \cite{mincigrucci2023element}.

Many chiral-sensitive signals rely on CPL differential measurements to generate a dichroic response that cancels out in the electric-dipole (ED) response.
These signals include  X-ray CD\cite{Tanaka_2005,Kypr_2009,Mukamel_formamide,Mukamel_2017} or X-ray Raman Optical Activity (ROA)\cite{rouxel2019x}, and can be well-described by using the multipolar expansion truncated at the magnetic dipole (MD)/electric quadrupole (EQ) order\cite{turchini2004core}.
In this case, the interaction Hamiltonian coupling the X-ray pulses and the matter is:
\begin{equation}
H_\text{int}(t) = -\bm \mu\cdot \bm E(t) - \bm m\cdot \bm B(t) - \bm q\cdot \nabla \bm E(t)
\end{equation}
\noindent where $\bm \mu$, $\bm m$ and $\bm q$ are electric dipole, magnetic dipole and electric quadrupole operators respectively.
Signals relying on higher multipoles are intrinsically weak since they are a small correction over an achiral background. 
PECD\cite{Pohl2022} and nonlinear even-order techniques offer large asymmetry ratios since they chiral sensitive at the electric dipole order.
On the other hand, XCD relies on differential absorption which is a simple process to describe and implement experimentally.

In the EUV or X-ray regime, signals relying on the multipolar expansion can also greatly profit from quadrupolar interactions since X-rays have large wavevectors.
Static signals XCD cannot take advantage of this because the ED-EQ two-point response tensor $\langle \bm \mu \bm q\rangle$ vanishes upon rotational averaging.
For time-dependent signals, the situation is different. 
They involve higher rank matter response tensors and one must include quadrupolar couplings\cite{crawford2007current}.
For example, the perturbative description of a standard pump-probe experiment involves a four-point correlation function $\langle \bm\mu\bm\mu\bm\mu\bm\mu\rangle$ of transition electric dipoles $\mu$. 
Time-resolved XCD (tr-XCD) measures the circular dichroism of a delayed X-ray probe beam following an actinic pulse that triggers a chiral nuclear dynamics. It has contributions from 8 response functions ($\langle \bm\mu\bm\mu\bm\mu\bm m\rangle$, $\langle \bm\mu\bm\mu\bm\mu\bm q\rangle$ and their permutations of the  $\bm m$ and $\bm q$ interactions).
Here, both the magnetic dipole and electric quadrupole coupling with the ultrafast X-ray probe will contribute to the signal.
On computational aspects, the numerical calculation of multipoles must be done with caution. When using a truncated basis, the multipoles matrix elements can become dependent on the origin of coordinates and special care must be made to recover accurate observables by summing over states, using a gauge-including atomic orbitals\cite{juselius2004calculation, jiemchooroj2007near}. 
Finally, X-ray molecular chirality in the multi-scattering regime of EXAFS signals is virtually unexplored and may require a different level of theory\cite{goulon2000x}.

Alternative approaches avoiding the multipolar expansion altogether can be envisioned to simplify the numerical calculation of chiral observables and to build different intuitions. 
Rather than using a truncated multipolar expansion and extract the pseudo-scalars measured by a given technique, one can use the complete radiation/matter interaction Hamiltonian\cite{chernyak2015non}. 
The Power-Zienau-Woolley Hamiltonian $H_\text{PZW} = -\int d\bm r \bm P(\bm r)\cdot \bm E(\bm r,t) -\int d\bm r \bm M(\bm r)\cdot \bm B(\bm r,t)$ is based on the full matter polarization and magnetization fields, and can be used as a starting point for a truncated multipolar expansion as discussed above.
Its implementation is not easy since there are no simple closed-form operators corresponding to the classical polarization and magnetization\cite{craig1998molecular}. It is simpler and more physically sound to use the minimal coupling description that is based on the charge and current density operators:
\begin{equation}
H_\text{mc} = -\int d\bm r \bm j(\bm r)\cdot \bm A(\bm r,t) - \frac{e}{2m}\int d\bm r \ \sigma(\bm r) \bm A^2(\bm r,t)
\end{equation}
\noindent where $\bm j$ and $\sigma$ are the current and charge density operators. $\bm A$ is the transverse component of the incoming X-ray pulse.
The expectation values of $\bm j$ and $\sigma$ have an obvious physical meaning and are relatively simple to compute from ab initio packages.
The multipolar expansion of polarization-based signals is thus circumvented, and chiral techniques involving X-ray field with important spatial variations can be considered\cite{rouxel2022molecular}.

For example, the ultrafast X-ray CD signal can be expressed as\cite{mukamel1999principles}:
\begin{multline}
S_\text{CD}(\omega, T) = S_\text{abs}(\omega, T, L) - S_\text{abs}(\omega, T, R)
\label{eq:trcd1}
\end{multline}
\noindent where $\omega$ is the detected spectrum of the transmitted dispersed probe, $T$ is the time delay between the actinic pulse and the probe. The actinic pulse is included implicitly in the wavefunction and its interaction can be calculated perturbatively or numerically.
The frequency-dispersed absorption signal in the minimal coupling picture is given by
\begin{multline}
S_\text{abs}(\omega, T, L/R) = \frac{2}{\hbar}\text{Im}\int d\bold{r} \langle \bold J_{L/R}(\bold r,\omega)\rangle_\Omega \cdot\bold A_{L/R}^*(\bold r,\omega)
\label{eq:trcd2}
\end{multline}
\noindent where the current expectation value is given by
\begin{multline}
\bold J_{L/R}(\bold r,t) = -\frac{i}{\hbar}\int dt' d\bold r' \\
\langle \bold j(\bold r,t)\bold j(\bold r',t') \rangle_\Omega\cdot A_{L/R}(\bold r', t') 
- \langle \bold j(\bold r',t') \bold j(\bold r,t) \rangle_\Omega\cdot A_{L/R}^*(\bold r', t')
\label{eq:trcd3}
\end{multline}
\noindent Here, $\bold j$ is the transition current density and $\bold A_{L/R}$ is the vector potential of the incoming field with left or right circular polarization. $\Omega$ indicates rotational averaging over the molecular ensemble.
The time of arrival $T$ of the CD probe pulse and its central frequency $\omega$ are implicitly contained in the field $\bold A$. 

Eqs.\ref{eq:trcd1}-\ref{eq:trcd3} describe the differential absorption of an X-ray probe using the minimal coupling interaction Hamiltonian. 
This approach offers different intuition and possibilities, and also some new simulation challenges.
The following discussion uses ultrafast X-ray CD as an example, but most of it can be adapted to other polarization-based chiral techniques.

First, the signal is expressed as a spatial overlap between the matter and field tensors. 
This is particularly convenient to include the effect of fields having complex spatial profiles, e.g. an Orbital Angular Momentum (OAM).
X-ray OAM is an independent field degree of freedom from the Spin Angular Momentum (SAM) that can be used conjointly with the OAM.
Signal computations show that combining SAM and OAM could lead to increased chiral sensitivity (higher asymmetry ratio), in an hybrid technique of circular and helical dichroism (CHD)\cite{Mukamel_2019}.
Studies suggest that the longitudinal polarization components that appear when beams are highly focused are the source of the largeer dichroic effect\cite{forbes2021relevance,rouxel2022hard,jiang2023time}
Even when the light polarization is the only degree of freedom used in the dichroic measurement, the local nature of X-ray interactions\cite{ilchen2021site} can make a spatially dependent description more insightful.
Molecular chirality has long been considered as an on or off property (either the molecule is chiral or it is achiral) but X-rays sources now allow to study the spatial structure of the molecular chirality within the probed system.

A drawback of the exact description using current and charge densities is that the required rotational averaging for isotropic molecular ensembles becomes a non-trivial task. 
In the static case and within the multipolar description, the XCD signal is well described by a two-point correlation function of an electric and a magnetic dipoles. This second-rank tensor can be straightforwardly rotationally averaged analytically, leading to the well-known rotatory strength $R_{ij} = \bm \mu_{ji}\cdot \bm m_{ij}$.
In the ultrafast regime, one must include the excited wavefunction into the matter rotational averaging.
Neglecting reorientation during the dynamics (which holds for very short timescales) and considering first order perturbation on the wavefunction, one must then average a 4-rank (3ED/1MD) or a 5-rank tensor (3ED/1EQ). For a 4-rank tensor, there are 4 diagrammatic contributions to the chiral signals and the averaging tensor contains 9 terms, leading to 36 relevant correlation functions.
This approach is tedious, not physically transparent, and becomes inadequate when using propagators allowing for reorientation or using the minimal coupling description.
In this case, rotational averaging is best tackled numerically by calculating the signal on a grid of Euler angles\cite{parrish2019ab} for an initial distribution of molecules and averaging it out.

\begin{figure}[h]
    \centering
    \includegraphics[width=0.9\linewidth]{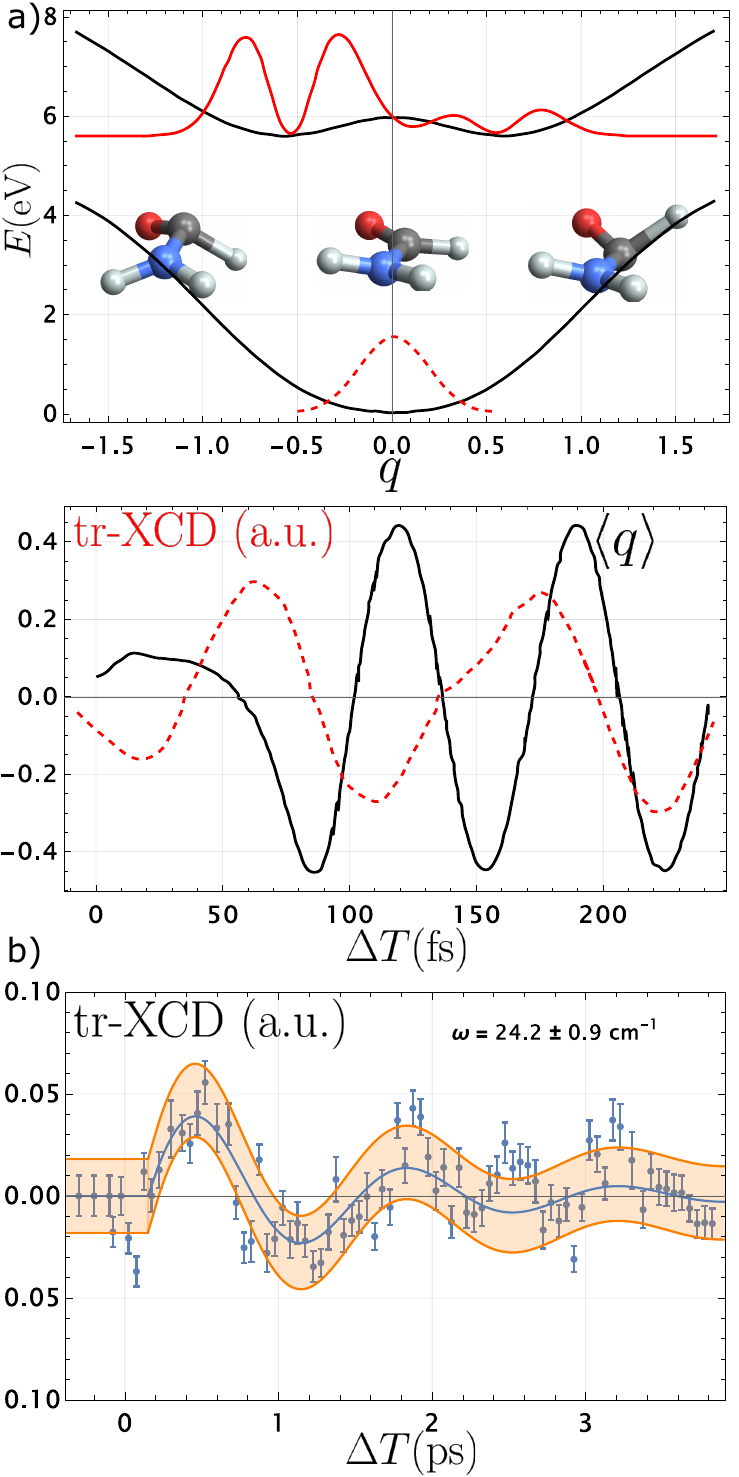}
    \caption{a) The formamide molecule is achiral in its ground state and becomes chiral in its first excited state through pyramidalization. When pumped by circularly-polarized UV light, the nuclear wavepacket oscillates between the two enantiomers (top panel). The tr-XCD signal (middle panel, dashed, red) follows closely the nuclear wavepacket center of mass dynamics (solid, black)\cite{Mukamel_formamide} and is a sensitive probe of the local chiral nuclear dynamics. b) tr-XCD can also be used as a local windows into chiral vibrational dynamics. In the lower panel, low-lying vibrational excitations are pumped by a UV Raman process\cite{mincigrucci2023element} and tr-XCD is used as a local probe of chiral nuclear dynamics.}
    \label{fig:trXCD}
\end{figure}

Third, the temporal convolution that appears in Eq. \ref{eq:trcd3} is a standard feature of time-domain observables when the impulsive limit cannot be used.
It is likely that given the ultrashort nature of some chiral nuclear or electronic dynamics, the impulsive limit cannot be taken in polarization-based FEL dichroic measurements.
For example, ultrafast X-ray CD can be used to monitor the flipping between two enantiomers around a chiral carbon. Typically, it is the lightest group that experience the most motion that can be as small as a single hydrogen atom.
As a result, the chiral nuclear dynamics can be on the timescale of few femtoseconds\cite{Mukamel_formamide}.
Even shorter, ultrafast X-ray dichroic measurements can be used to probe charge migration in chiral molecules.
To achieve chiral sensitivity, one must ensure that the field envelope is indeed shorter that the chiral dynamics, especially in the case in which the molecular bounces between its two enantiomers.
Otherwise, the temporal integration averages them, and the dicroic signal vanishes.
Alternatively, advanced polarization schemes can be employed in which the field envelope is not shorter than the chiral dynamics but its polarization evolves in time in order to follow it.
This can be considered as a form of polarization pulse-shaping.
For example, one could envision an experiment in which the field polarization state is left polarized for half the pulse envelope and then switches to right polarized.
While this is challenging to achieve at an FEL, schemes relying on the combination of polarized radiation from two undulators could achieve non-trivial time-dependent polarizations\cite{Serkez2019}.

Figure \ref{fig:trXCD} summarizes recent results on ultrafast chirality probed by X-rays. In Fig. \ref{fig:trXCD}a, the out-of-plane stretching of formamide\cite{Mukamel_formamide} was probed by pump-probe tr-XCD. In the excited state, the molecule experiences a pyramidalization and becomes chiral.
When pumping the molecule with circular polarized light, an asymmetric wavepacket is generated 
In order to generate an asymmetric nuclear wavepacket in the first excited electronic state, that is the center of mass of the nuclear wavepacket is located on one of the two possible enantiomers and oscillate between them on an ultrafast timescale.
The tr-XCD signal at the carbon K-edge offers a sensitive probe of this time-evolution by following closely this oscillation between the two-enantiomers.
Fig. \ref{fig:trXCD}b displays the tr-XCD signal recorded on ibuprofen dimer pumped by ultrashort UV light and probed at the carbon K-edge\cite{mincigrucci2023element}. Through a Raman process, a large amount of vibrational modes are excited and the ones with low-energies, describing global vibrational modes, are known to be difficult to disentangle.
Using resonant ultrafast X-rays as a windows, one can select which group of carbon atoms are probed by selecting a central frequency and bandwidth exciting only these ones in the pre-edge C K-edge.
The addition of the XCD detection mode further allows to probe only atoms located in parts of the molecule lacking mirror inversion.
For example, we have recently shown that chlorine and fluorine atoms substituted at different locations in a helicene molecule can be used as probes of the local helical state of the molecular helix\cite{freixas2023x}. Helicenes are screw-shape molecules that can racemize by reaching a thermally, globally achiral, excited state of mixed helicity of opposite side of the molecule. From that state, one of the two sides can switch to the opposite helicity and time-resolved XCD resonant with the Cl or F atomic chromophores can be used to monitor on which side this is happening.

While this brief overview has mostly focused on the use of CPL, the probing of molecular chirality can also be achieved using schemes relying on controlled linear polarizations. 
Even-order spectroscopies, for example Sum Frequency Generation of Difference Frequency Generation, are sensitive to chirality in the bulk of isotropic ensembles\cite{ji2006toward}.
A condition on the applicability of the technique is that the triple product of the three involved field polarizations do not vanish, i.e. they are not coplanar. 
This calls for multi-pulse FEL schemes with independent linear polarizations.
Similarly, it has been shown that cross-polarization transient grating experiments is sensitive to molecular chirality\cite{terazima1996transient}. Such experiments involve the crossing of two FEL beams with crossed linear polarizations on a chiral sample and the diffraction of a delayed probe is measured.


\onecolumngrid \newpage \twocolumngrid

\subsection{Towards Time-Resolved Emitter-Site-Selective PECD of Fixed-in-Space Chiral Molecules}
\label{sec:Demekhin}

\begin{center}
Philipp V. Demekhin and Nikolay M. Novikovskiy    
\end{center}

Short-wavelength radiation provided by XFELs governs primary inner-shell ionization and, thus, addresses individual chemical elements (electron emitters) located on specific molecular sites. In many cases, electrons ejected from equivalent emitters experience different chemical shifts and can thus be distinguished due to their kinetic energy. Being created at a specific site of a molecule, an initially well-localized photoelectron wave packet propagated outwards and experiences each and every detail of the molecular potential on its way out. An accompanying multiple scattering of this wave packet on a molecular scaffold results in  characteristic interference and/or diffraction pattern in the  resulting photoelectron angular emission distributions, which thus provide detailed information on the molecular potential itself. In many cases, the photoemission probability depends on the polarization of the ionizing radiation, resulting thereby in a variety of dichroic phenomena. Particularly promising is the fascinating chiroptical phenomenon of forward-backward asymmetry in the photoemission from randomly-oriented chiral molecules with respect to the propagation direction of circularly polarized light  \cite{Ritchie}, known as photoelectron circular dichroism (PECD). 

One usually distinguishes two contributions to PECD: one from a chiral initial electronic state and another from chiral final continuum state of the photoelectron, both naturally imprinted in the electric-dipole transition matrix element. In the case of inner-shell photoionization from an almost-symmetric inner-shell orbital, the former contribution is negligibly small. Thereby, in the inner-shell ionization, PECD is governed solely by the final-state effect, which offers a unique opportunity to inspect chiral asymmetries of the potential of the molecular ion. To date, many experimental and theoretical studies on randomly-oriented chiral molecules in the gas phase confirmed that PECD in  inner-shell ionization \cite{Hergenhahn04, Stener04, Nahon06, Powis08b, Alberti08, Ulrich_2008, Catone12, Pitzer16cpc, Pitzer16jpb, Tia17, ilchen2017emitter, Hartmann19, Nalin21, Fehre21, ilchen2021site, Fehre22a, Fehre22b, Nalin22} is as large as in the outer-shell ionization, where a chiral asymmetry of the initial state can play an additional role. Those studies suggest site-selectivity \cite{Hergenhahn04,Stener04,Alberti08,Catone12,ilchen2017emitter,Fehre22a} and emitter-selectivity \cite{Stener04,ilchen2017emitter,Fehre22a} of inner-shell PECD, as well as its complex dependence on the molecular configuration \cite{Stener04,Pitzer16cpc,Pitzer16jpb,Nalin21} and photoelectron energy \cite{Hergenhahn04,Stener04,Nahon06,Powis08b,Ulrich_2008,Catone12,Nalin21}, indicating also an important role of the core-excited resonances \cite{Alberti08,Catone12,Hartmann19}. Utilizing polarization-controlled FELs in the future may thus be a perspective route to explore the emitter-site-selectivity of inner-shell PECD in a nonlinear regime. 

Any inner-shell ionization is undoubtedly followed by a radiative or Auger decay (or even a cascade of the decay events), which increases the charge state of the residual ion. Such highly-charged molecular ions are typically unstable and undergo an ultrafast dissociation by a Coulomb explosion into several ionic and neutral fragments. Under the assumption of the so-called axial-recoil approximation \cite{Zare72}, which suggests a moderate deterministic rotation of a molecule during its ultrafast fragmentation, the ionic momenta provide access to the spatial orientation of the molecule at the instant of photoionization. Measuring the photoelectron momentum in coincidence with the ionic momenta, allows to access its relative emission angle in the molecular frame of reference \cite{Landers01,Williams12,Fukuzawa19a,Fukuzawa19b}, so-called molecular frame photoelectron angular distributions (MFPADs).

\begin{figure}[b!]
\includegraphics[width=0.48\textwidth]{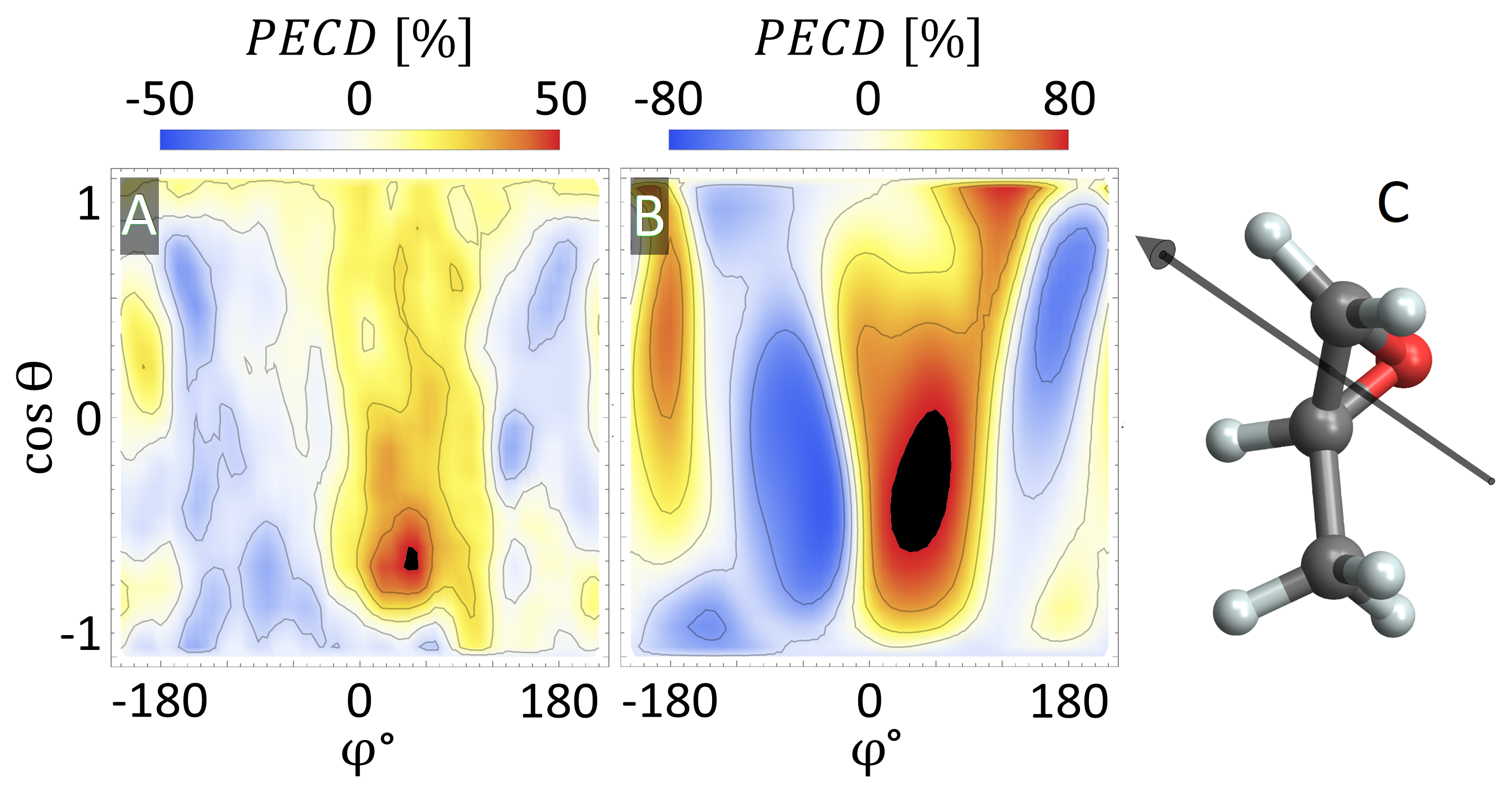}
\caption{Measured (panel A) and computed (panel B) differential PECDs of the O~1s-photoelectrons of S-methyloxirane with the kinetic energy of 11.5~eV as functions of the two photoelectron emission angles for the chosen molecular orientation with respect to the light propagation direction (panel C). At certain photoemission directions (indicated by black color), the measured and computed PECDs exceed the chosen upper/lower limits of the respective asymmetry scales. The figure has been adapted and reprinted with permission from \cite{Fehre21}.}
\label{Fig_4D-PECD_MOx}
\end{figure}

Even achiral molecules, being distinctly oriented in space,  show a number of dichroic phenomena in  photoemission, such as circular dichroism in the angular distribution (CDAD, \cite{Dubs85,Jahnke02}) or an \textit{apparent} PECD \cite{Pier20}, both on the order of 100\%. Such orientation-induced dichroic asymmetries in the photoemission superimpose with the intrinsic dichroic asymmetry present in chiral molecules, resulting in a much stronger PECD for spatially oriented chiral molecules, as well \cite{Tia17,Nalin21,Fehre21,Fehre22a,Fehre22b,Nalin22}. This is because averaging over molecular orientations leads to a loss of information, reducing thereby the PECD substantially \cite{Ritchie}. In particular, fixing already one molecular orientation axis in space increases the PECD by a factor of about ten \cite{Tia17,Nalin21,Fehre22a}, and the differential PECD of a fully fixed-in-space molecule can reach up to 100\% \cite{Tia17,Fehre21,Nalin22}. The latter fact is illustrated in Fig.~\ref{Fig_4D-PECD_MOx}, which depicts a normalized difference of the photoelectron distributions of a fixed-in-space S-methyloxirane molecule recorded for circularly polarized light with positive and negative helicities, $PECD(\theta,\varphi)=[I_{+}(\theta,\varphi)-I_{-}(\theta,\varphi)]/[I_{+}(\theta,\varphi)+I_{-}(\theta,\varphi)]$. The considered  molecular orientation with respect to the light propagation direction is visualized in  panel C. As one can see, for some  photoelectron emission angles  (black spots), the measured PECD in panel A exceeds 50\%, while the computed PECD in panel B is somewhat stronger. Such a strong dichroic asymmetry in the photoemission from fixed-in-space molecules increases the sensitivity for chiral recognition in the gas phase \cite{Fehre21}.

\begin{figure}[!t]
\includegraphics[width=0.42\textwidth]{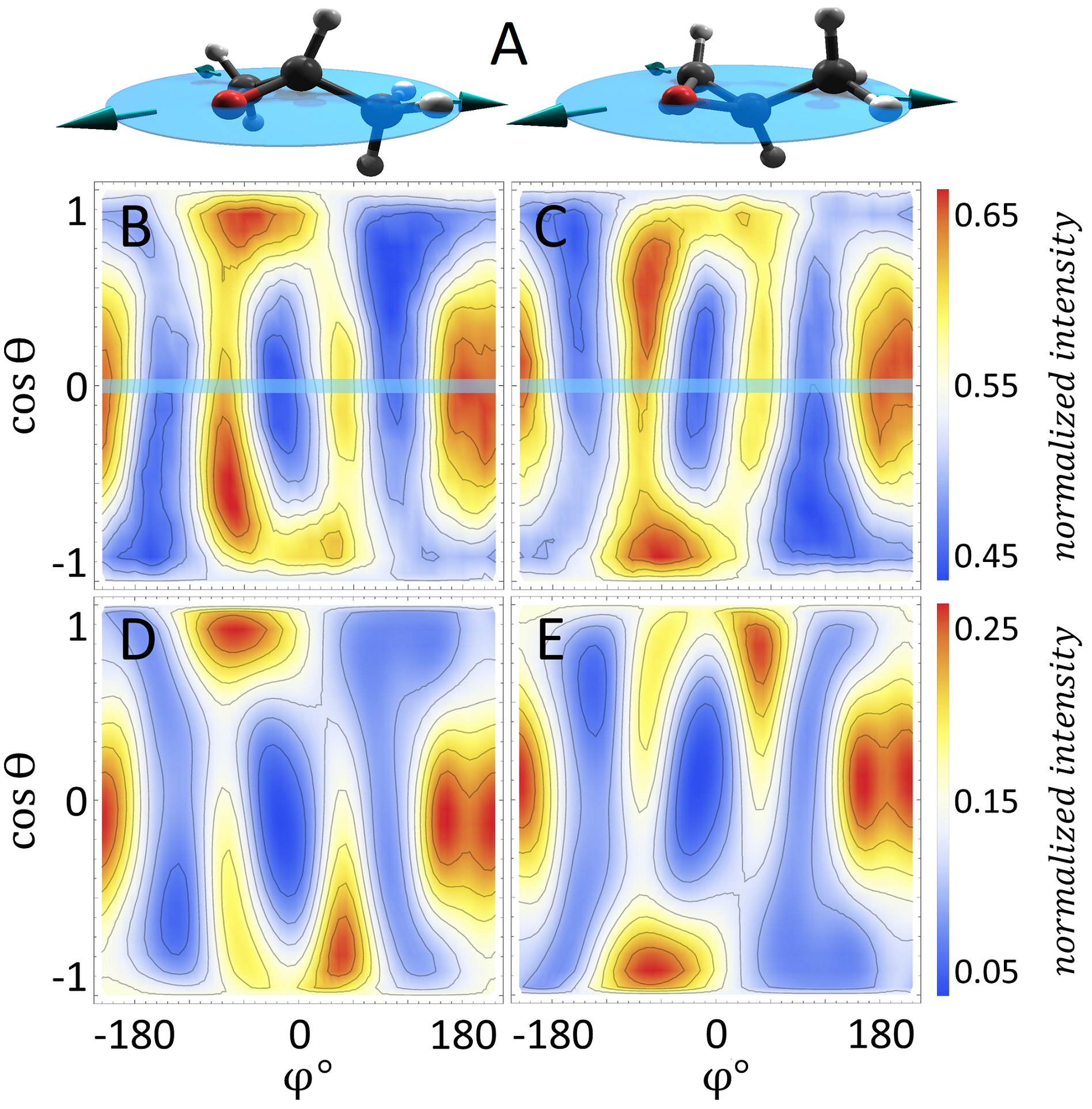}
\caption{A visualization of two  enantiomers of methyloxirane molecule (panel A) and the respective molecular frames of reference  derived from the measured asymptotic momenta of three fragments (turquoise arrows which define cyan plane for each enantiomer). Polarization-averaged angular emission distributions of the O~1s-photoelectrons with the kinetic energy of 11.5~eV, measured (panels B and C) and computed (panels D and E)  for two enantiomers of methyloxirane   in the frame of molecular reference. Switching between two enantiomers reflects the PA-MFPADs with respect to the cyan plane (indicated by the horizontal line in panels B and C). The figure has been adapted from \cite{Fehre22b}.}
\label{Fig_PA-MFPAD_MOx}
\end{figure}

Even polarization-averaged MFPADs are enantio-sensitive \cite{Fehre22b}. This fact is illustrated in Fig.~\ref{Fig_PA-MFPAD_MOx}, which depicts the measured (panels B and C) and computed (panels D and E) PA-MFPADs of two enantiomers of methyloxirane. As one can recognize from panel A, the momenta of three molecular fragments define a plane, and switching between the two enantiomers reflects the molecule with respect to this plane. As a consequence, the respective PA-MFPADs for two enantiomers are just mirror-images of each other with respect to $cos\theta=0$ (the horizontal line in the panels B and C). In the experiment, such PA-MFPADs provide much higher statistics, since they include all photoionization events averaging over all possible directions from which the ionizing light hits the molecule. As a consequence, such PA-MFPADs can be utilized for the determination of the geometrical structure (bond length, angles, etc.) of molecules with enantio-selectivity and an accuracy of about 5\%, even for weak scatterers such as hydrogen atoms \cite{Fehre22b}.

A new type of experiments with polarization-controlled FELs on the  time-resolved PECD of randomly oriented chiral molecule was performed in Ref.~\cite{ilchen2021site} at the AMO beamline of the LCLS at the SLAC National Accelerator Laboratory in the USA. It is based on the X-ray pump -- X-ray probe scheme, where the pump pulse triggers molecular fragmentation dynamics by ionizing the F~1s-shell of a trifluoromethyloxirane molecule. After an ultrafast Auger decay, the doubly-charged molecule can Coulomb-explode into a fluorine ion and the singly-charged mother-fragment. The time-delayed X-ray probe pulse releases an additional F 1s-photoelectron from the mother-fragment, and its PECD is investigated as a function of the time delay between two pulses, i.e., at different internuclear separations between the ions. The PECDs, computed in Ref.~\cite{ilchen2021site} in the SAE approximation for individual combinations of the ejected and ionized fluorine atoms, vary by almost an order of magnitude for rather large elongations of the bond to about 12~a.u. (far beyond the chemical bond breaking), which corresponds to the time-delay of about 150~fs. Such a long-range sensitivity of the PECD can be explained by the long-range influence of the Coulomb potential of the fluorine ion, which is seen by the mother-fragment. However, averaging over possible combinations of the primary and secondary addressed fluorine atoms makes the effect almost independent of the separation, but still observable.

\begin{figure}[!t]
\includegraphics[width=0.48\textwidth]{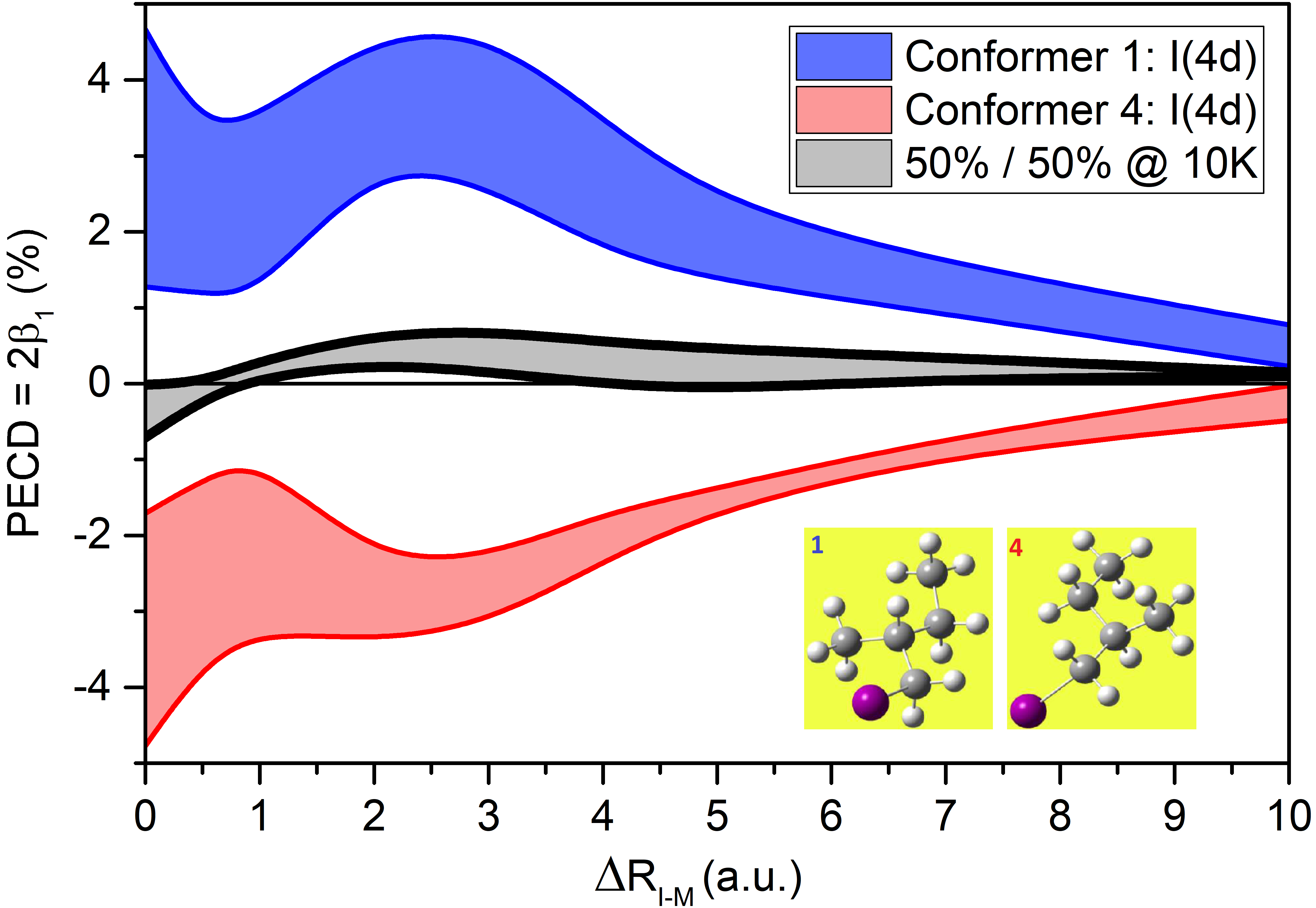}
\caption{Theoretical PECD of the 4d-photoelectrons of the iodine atom, computed for (R)-1-iodo-2-methylbutane at different distances $R_{I-M}=R_{eq}+\Delta R$ between the neutral chiral mother-fragment and neutral iodine atom dissociating along the bond. The shaded areas represent dispersions of the PECDs in the selected kinetic energy interval of 4--8~eV. The theory predicts that averaging over the relevant conformers results in an almost complete cancellation of the effect.}\label{Fig_PECD_IMB}
\end{figure}

In a recent UV pump -- XUV probe experiment \cite{allum2022localized}, performed at the Free-electron LASer in Hamburg (FLASH), a neutral dissociation of a chiral molecule 1-iodo-2-methylbutane was studied by examining the ultrafast evolution of the iodine 4d binding energy. In the experiment, the UV pump pulse excites the molecule in a dissociative state such that a neutral iodine atom is ejected, and the time-delayed XUV pulse probes the 4d-shell of iodine at different separations from the neutral mother-fragment. Unfortunately, the  PECD of the released photoelectrons was not accessible in the experiment \cite{allum2022localized}. A simulation of the time-resolved PECD of the 4d-photoelectrons of the iodine atom, which can be expected in such an experiment, is shown in Fig.~\ref{Fig_PECD_IMB}. These estimates were performed by the Single Center (SC) method and code \cite{SC1,SC2,SC3} in the SAE approximation under assumptions that: (i) the iodine atom dissociates along its initial bond, and (ii) the geometry of the mother-fragment does not adjust in the course of the fragmentation.

A noticeable PECDs of the 4d-photoelectrons of the iodine atom as a function of internuclear separation from the rest of the 1-iodo-2-methylbutane molecule during its neutral dissociation can be seen in Fig.~\ref{Fig_PECD_IMB} for two relevant conformers of the molecule (see insets). The effect persists far beyond the chemical bond breaking up to elongations of the bond by about 10~a.u., which, according to Ref.~\cite{allum2022localized} (see Fig.~10 there), corresponds to a time-delay of about 175~fs. Such a long-range PECD sensitivity can be explained by the large scaffold of the mother fragment, which is seen by the 4d-photoelectron emitted from the iodine atom within a large solid angle even at these separations. Nevertheless,  Figure~\ref{Fig_PECD_IMB} predicts that averaging the PECDs over two conformers, which are relevant at the temperature of 10~K, makes the effect almost undetectable. One, thus, would need to decrease the temperature in a molecular beam to make contribution from the ground-state conformer dominant.

\begin{figure}[b!]
\includegraphics[width=0.48\textwidth]{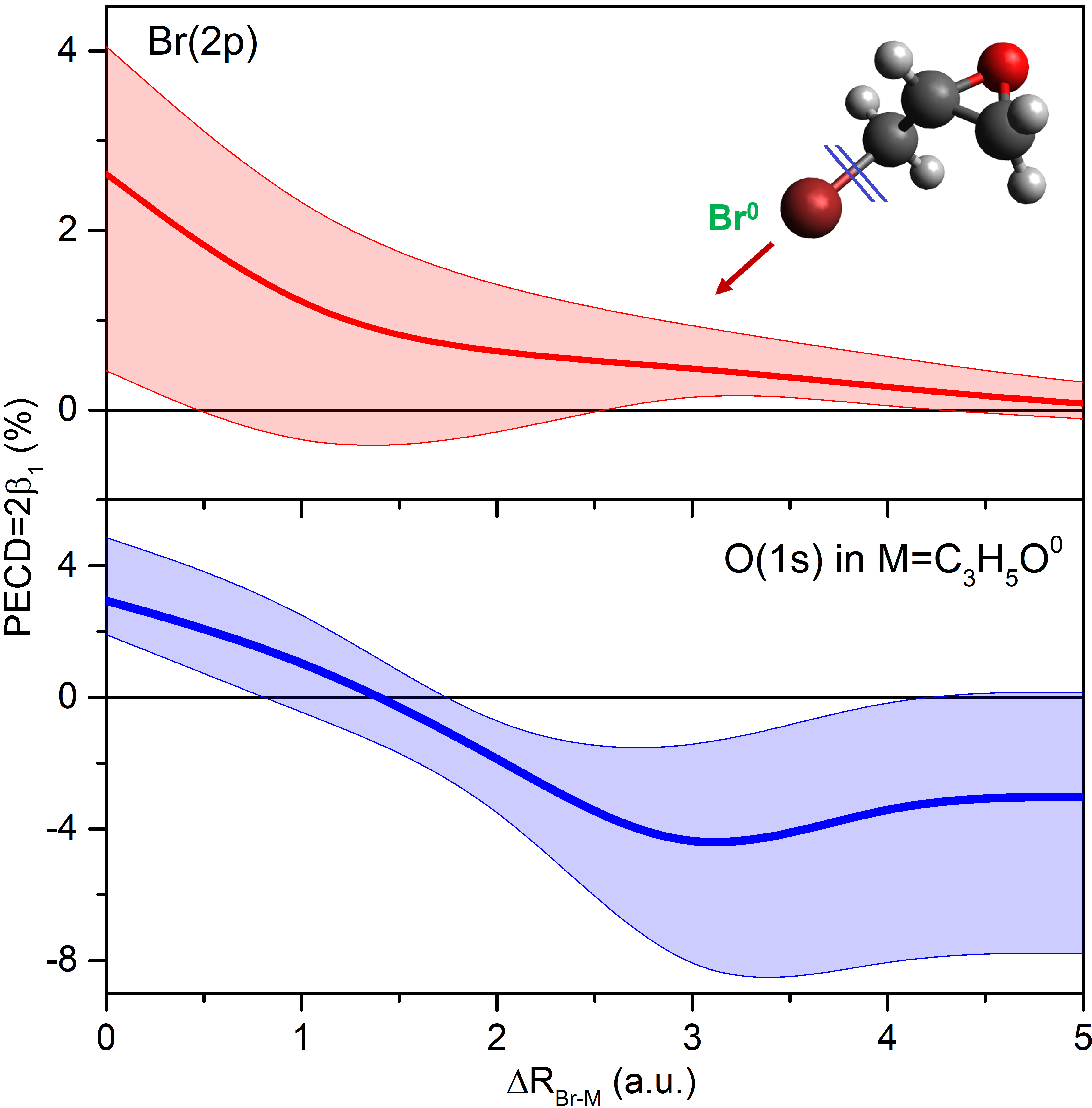}
\caption{Theoretical PECD of the 2p-photoelectrons of the bromine atom (upper panel) and 1s-photoelectrons of the oxygen atom (lower panel), computed for (R)-epibromohydrin at different distances $R_{Br-M}=R_{eq}+\Delta R$ between the neutral chiral mother-fragment and neutral bromine atom dissociating along the bond. The shaded areas represent dispersions of the PECDs in the selected kinetic energy intervals of 4--8~eV for Br~2p- and 5--9~eV for O~1s-photoelectrons. The theory predicts that PECD saturates to its asymptotic values if the bond length is enlarged already by about 5~a.u.}\label{Fig_PECD_BrMOx}
\end{figure}

In order to conduct a successful experiment on time-resolved PECD at a polarization-controlled FEL, it is extremely important to find a suitable candidate-molecule, which is free of the disadvantages discussed above. Since larger chiral molecules appear usually in many  (energetically close-by) conformer configurations, one can look for more compact (and therefore) conformer-free chiral molecules. Figure~\ref{Fig_PECD_BrMOx} depicts results of a theoretical simulation of the time-dependent PECD of epibromohydrin molecules. Similarly to the just discussed process, an UV pump pulse initiates a neutral dissociation of the bromine atom along its bond, and a delayed X-ray pulse probes either the 2p-shell of the bromine atom (upper panel) or the 1s-shell of the oxygen atom in the remaining neutral mother-fragment (lower panel) as a  function of their separation. Simulations were performed  by the SC method \cite{SC1,SC2,SC3} in the SAE approximation under similar assumptions as in Fig.~\ref{Fig_PECD_IMB}. Figure~\ref{Fig_PECD_BrMOx} illustrates a measurable PECDs  which saturate to their asymptotic values (zero for Br 2p-photoelectrons in the upper panel and a constant value for  O 1s-photoelectrons of the chiral mother-fragment in the lower panel) when the respective bond elongates by about 5~a.u. Thus, the time-evolution of the PECDs can be traced here using pump-probe time-delays below 100~fs. Such a short-range sensitivity of the PECDs is a disadvantage of using small chiral molecules, for which photoelectrons emitted from one of the fragment see another fragment only in a relatively small solid angle already at short separations.

Summarizing, studies of time-resolved emitter-site-selective PECD of fixed-in-space chiral molecules in the gas phase and their ultrafast structure determination via photoelectron-diffraction patterns are far-reaching and have a bright perspective for application of FELs with polarization control. Performing, however, robust experiments on the PECD and its temporal  evolution on an ensemble of randomly oriented chiral molecules with the FELs is a very important intermediate milestone.

\onecolumngrid \newpage \twocolumngrid

\section{Summary and Outlook}

Free-electron lasers are evolving to be versatile machines for site specifically investigating polarization dependent phenomena in e.g. chiral and oriented systems in the gas phase via dichroic phenomena. Their ultrashort and ultrabright pulses unprecedentedly allow for creating and interrogating hitherto inaccessible states of (transient) matter, and to investigate electronic and structural properties and dynamics of small systems, with imminent perspectives towards larger molecules such as amino acids and peptides. One of the core pillars of FEL-driven gas-phase science is the fundamental correlation of the physics perspective and the subsequently evolving chemical perspective. Latest technological breakthroughs of FEL-operation such as full polarization control with seeded or attosecond pulses in combination with state-of-the art optical lasers, new sample delivery methods, advanced theoretical modeling and data analysis, as well as sophisticated instrumentation and diagnostics set the stage for a novel field of research with promising prospects for science, industry, and society. Several facilities worldwide have identified this opportunity and have initialized the respective efforts, as presented above in the first chapter of this roadmap. The subsequently discussed opportunities for experimental and theoretical sciences in the light of these developments sketched the state-of-the-art in this young field of research.

\begin{acknowledgements}
This work has been supported by the Bundes\-ministerium f\"ur Bildung und Forschung (BMBF) grant 13K22CHA. M.I. acknowledges funding from the Volks\-wagen Foundation for a Peter-Paul-Ewald Fellowship. 
M.I., P.V.D., N.M.N, and R.D.\ acknowledge support from the Deutsche Forschungsgemeinschaft (DFG)-Project No. 328961117-SFB 1319 ELCH (``Extreme light for sensing and driving molecular chirality''). W.H. and M.I. acknowledge funding of the BMBF-ErUM-Pro project ``TRANSALP" (grant No. 05K22PE3).
P.W.\ acknowledges funding from SLAC under the U.S.\affiliation{} Department of Energy, Office of Science, Office of Basic Energy Sciences under
Contract No.\ DE-AC02-76SF00515.
J.B.\ and J.W.\ acknowledge funding from the DFG under project Nos.\ 429194455 and 465098690. S.B.\ and L.S.\ acknowledge support from DESY (Hamburg, Germany), a member of the Helmholtz Association HGF and the Helmholtz Initiative and Networking Fund. H.D.\ acknowledges the National Natural Science Foundation of China (12125508, 11935020). C.F.\ acknowledges the National Natural Science Foundation of China (12122514 and 11975300).
K.B.\ acknowledges funding from the U.S.\ National Science Foundation under grant No.\ PHY-2110023. N.D. \ acknowledges funding from the U.S.\ National Science Foundation under grant No.\ PHY-2012078. L.A. acknowledges funding from the U.S. National Science Foundation under grant No.~2309133, and from the DOE AMOS program under grant No.~DE-SC0020311.
C.V.\ acknowledges support from the European Union's Horizon 2020 research and innovation program under Grant Agreements No.\ 860553 (SMART-X) and No.\ 654148 (LASERLAB-EUROPE), and from the European COST Action CA18222 (AttoChem). 
F.C.\ acknowledges the Cluster of Excellence ``Advanced Imaging of Matter'' of the Deutsche Forschungsgemeinschaft (DFG) - EXC 2056 - project ID 390715994, the DFG - SFB-925 - Project ID 170620586, and the COST action CA18222 Attochem.
J.R.R.\ acknowledges support from the U.S. Department of Energy, Office of Science, Basic Energy Science, Chemical Sciences, Geosciences and Biosciences Division under Contract No.\ DE-AC02-06CH11357.

\end{acknowledgements}


\bibliographystyle{ieeetr.bst}
\bibliography{bib2}
\end{document}